%% file: paper.tex
\pdfoutput=1
\documentclass[11pt,a4paper]{article}

\usepackage{jheppub}
\usepackage{atlasphysics}
\usepackage{graphicx}
\usepackage[hang,center]{subfigure}
\usepackage{mathptmx}
\usepackage{helvet}
\usepackage{palatino, url,multicol}
\usepackage{appendix}
\usepackage{dcolumn}
\usepackage{paralist}
\usepackage{placeins}
\usepackage{xspace}
\usepackage{multirow}
\usepackage{rotating}
\usepackage{siunitx}
\usepackage{pdflscape}
\usepackage{afterpage}
\PassOptionsToPackage{hyphens}{url}\usepackage{hyperref}
\newcommand{\TheTitle}{Search for direct pair production of the top squark
  in all-hadronic final states in proton--proton collisions at
  $\rts=8\tev$ with the ATLAS detector}

\usepackage{preprintcover}  
\PreprintCoverPaperTitle{\TheTitle}  
\PreprintIdNumber{CERN-PH-EP-2014-112}  
\PreprintCoverAbstract{\input{abstract}}  
\PreprintJournalName{JHEP}

\title{\TheTitle}
\notoc
\author{The ATLAS Collaboration}
\abstract{\input{abstract}}

\begin{document}
\maketitle
\flushbottom

\section{Introduction}
\input{intro}

\section{The ATLAS detector}
\input{detector}

\section{Trigger and data collection}
\input{trigger}

\section{Simulated event samples and SUSY signal modelling}
\input{montecarlo}

\section{Physics object reconstruction}
\input{objecteventsel}

\section{Signal region definitions}
\input{signalregiondefs}

\section{Background estimation}
\input{background}

\section{Systematic uncertainties}
\input{systematics}

\section{Results and interpretation}
\input{results}

\section{Conclusions}
\input{conclusions}


\section*{Acknowledgments}
\input{acknowledgments}

\bibliographystyle{JHEP}
\bibliography{paper}

\onecolumn
\clearpage
\input{atlas_authlist}

\end{document}

%% file: abstract.tex
The results of a search for direct pair production of the scalar partner to the top quark
using an integrated luminosity of \totalluminumnoerr~of proton--proton collision data at $\sqrt{s}=8$~TeV recorded with the ATLAS detector at the LHC are reported. The top squark is assumed to decay via $\stop \to t \ninoone$ or $\stop\to b\chinoonepm \to b  W^{\left(\ast\right)} \ninoone$, where $\ninoone$ ($\chinoonepm$) denotes the lightest neutralino (chargino) in supersymmetric models. The search targets a fully-hadronic final state in events with four or more jets and large missing transverse momentum.  
No significant excess over the Standard Model background prediction is observed,
and exclusion limits are reported in terms of the top squark and
neutralino masses and as a function of the branching fraction of
$\stop \to t \ninoone$. For a branching fraction of $100\%$, top squark masses in the range
$270$--$645\GeV$ are excluded for $\ninoone$ masses below $30\GeV$.
For a branching fraction of $50\%$ to either $\stop \to t \ninoone$ or
$\stop\to b\chinoonepm$,  and assuming the $\chinoonepm$
mass to be twice the $\ninoone$ mass, top squark masses in the range
$250$--$550\GeV$ are excluded for $\ninoone$ masses below $60\GeV$.

%% file: intro.tex
The recent observation of the Standard Model (SM) Higgs
boson~\cite{:2012gk,:2012gu}
has brought renewed attention to the
gauge hierarchy
problem~\cite{Weinberg:1975gm,Gildener:1976ai,Weinberg:1979bn,Susskind:1978ms}.
However,
the existence (and mass) of this fundamental scalar boson does not resolve
the tension between the electroweak and Planck scales.
Supersymmetry
(SUSY)~\cite{Miyazawa:1966,Ramond:1971gb,Golfand:1971iw,Neveu:1971rx,
  Neveu:1971iv,Gervais:1971ji,Volkov:1973ix,Wess:1973kz,Wess:1974tw} 
 provides an extension of the SM which can resolve the hierarchy
 problem~\cite{Dimopoulos:1981zb,Witten:1981nf,Dine:1981za,
   Dimopoulos:1981au,Sakai:1981gr,Kaul:1981hi}
 by introducing supersymmetric partners of  the known bosons and fermions.
The dominant contribution to the divergence of the Higgs boson mass
arises from loop diagrams involving the top quark; these can be largely
cancelled if a scalar partner of the top quark (top squark)
exists and has a mass below $\sim 1\TeV$~\cite{Barbieri:1987fn,deCarlos:1993yy}.
Furthermore, a compelling by-product of $R$-parity conserving SUSY
models~\cite{Fayet:1976et,Fayet:1977yc,Farrar:1978xj,
  Fayet:1979sa,Dimopoulos:1981zb}
is a weakly interacting, stable, lightest supersymmetric particle
(LSP): a possible candidate
for dark matter.

\begin{figure}[!b]
\begin{center}
\subfigure[]{\includegraphics[width=0.35\textwidth,trim=0 0 -20 0]{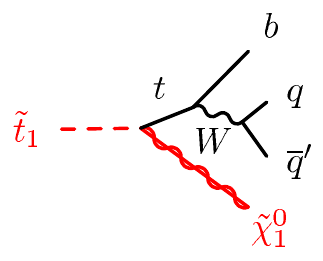}}
\subfigure[]{\includegraphics[width=0.35\textwidth,trim=-20 0 0 0]{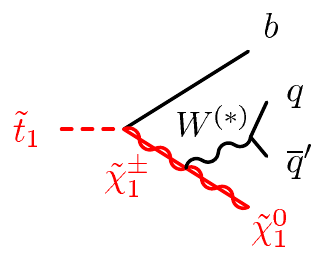}}
\caption{\label{fig:feynman} Feynman diagrams illustrating the $\tone$ decay modes considered: (a)~$\tone \rightarrow t \ninoone$ and (b)~$\tone \rightarrow b \chinoonepm\rightarrow b  W^{\left(\ast\right)} \ninoone$.}
\end{center}
\end{figure}

In SUSY models, there are two scalar partners for the top quark
(denoted $\tleft$ and $\tright$),
corresponding to left-handed and right-handed top quarks.  Due to the
large Yukawa coupling of the top quark, there can be significant mixing between
$\tleft$ and $\tright$, leading to a large splitting between the
two mass eigenstates, denoted $\tone$ and $\ttwo$, where $\tone$ refers
to the lighter of the two states.
To leading order the direct top squark production cross section is given in perturbative Quantum Chromodynamics by gluon-gluon and $\qqbar$ fusion and depends only on the $\tone$ mass~\cite{Beenakker:1997ut,Beenakker:2010nq,Beenakker:2011fu}.
The decay of the top squark depends on the left-right admixture as well
as on the masses and mixing parameters of the fermionic partners of the
electroweak and Higgs bosons (collectively known as charginos,
$\tilde{\chi}_{i}^{\pm}$, $i=1,2$  and
neutralinos,  $\tilde{\chi}_{i}^{0}$, $i=1$--$4$)
to which the top squark can decay.  No
assumption is made on the LSP mass in this paper.  While the LSP is
assumed to be the \ninoone\  it could, for example, be a very
light gravitino (the fermionic partner of the graviton),
which would evade existing limits on the neutralino mass. 
In addition, $\tone$ is
assumed to be heavier than the top quark and to decay via either
$\tone \rightarrow t \ninoone$ or $\tone \rightarrow b \chinoonepm
\rightarrow b  W^{\left(\ast\right)} \ninoone$, as illustrated in figure~\ref{fig:feynman}.

This paper presents the results of a search for direct pair
production of $\tone$ using a dataset corresponding to an integrated luminosity of \totalluminoerr. These data were
collected by the ATLAS detector at the Large Hadron Collider (LHC)
through proton--proton collisions at a centre-of-mass energy of
$\sqrt{s} = 8\,\TeV$.   Only fully hadronic final states are
considered.  The undetected LSP, $\ninoone$, leads to missing transverse
momentum (\ptmiss, whose magnitude is referred to as \met).
If both top squarks decay via $\tone \rightarrow t
\ninoone$, the experimental signature is  a pair of reconstructed
hadronic top quarks plus significant \met.  The $b \chinoonepm \rightarrow b  W^{\left(\ast\right)} \ninoone$ decay
mode leads to a similar final-state topology, apart from the mass of
the resulting hadronic system.
The results are presented as a
function of the branching fraction $B\left(\tone\to t\ninoone\right)$.

Previous searches for $R$-parity-conserving direct top squark production at centre-of-mass energies of $7$
and $8\,\TeV$ have been reported by the
ATLAS~\cite{Aad:2012cz,:2012si,Aad:2012tx,Aad:2012uu,Aad:2012yr,Aad:2013ija,Aad:2014mha}
and
CMS~\cite{Chatrchyan:2012uea,Chatrchyan:2012wa,Chatrchyan:2013lya,Chatrchyan:2013xna,Chatrchyan:2013mya,Khachatryan:2014doa}
collaborations. The previous ATLAS search in the all-hadronic channel \cite{:2012si}
considered only the nominal ``fully resolved" experimental signature of
six distinct
jets (two from the bottom quarks and two from each of the
\Wboson\ decays) and significant missing transverse momentum from the
LSPs.  In this paper, the experimental sensitivity to this signature
is enhanced by
considering in addition ``partially resolved" events with four or five jets in the
final state, which can occur if one or more jets are below the
reconstruction threshold or if the decay products of Lorentz-boosted
top quarks are sufficiently collimated. Furthermore, the sensitivity
to the $\tone\to b\chinoonepm$ decays is augmented by including events
with exactly five jets in the final state (targeting final
states where one of the jets from the $W^{(\ast)}$ decay has low
transverse momentum, \pt).

%% file: detector.tex
The ATLAS detector~\cite{DetectorPaper:2008} consists of inner tracking devices
surrounded by a superconducting solenoid, electromagnetic
and hadronic calorimeters and a muon spectrometer
in a toroidal magnetic field. The inner detector,
in combination with the 2~T field from the solenoid,
provides precise tracking of charged particles for $|\eta| <
2.5.$\footnote{ATLAS uses a right-handed coordinate system with the
  $z$-axis along the beam pipe.  The $x$-axis points to the centre of
  the LHC ring and the $y$-axis points upward.  The azimuthal angle $\phi$ is measured 
around the beam axis and the polar angle $\theta$ is the
angle from the beam axis. The pseudorapidity is defined as $\eta = -\ln \tan(\theta/2)$. The distance~$\mathrm{\Delta} R$ in
the $\eta$--$\phi$ space is defined as $\mathrm{\Delta} R = \sqrt{(\mathrm{\Delta}\eta)^2+(\mathrm{\Delta}\phi)^2}$.}
 It consists of a silicon pixel detector, a silicon strip
detector and a straw-tube tracker that also provides transition
radiation measurements for electron identification.
A high-granularity electromagnetic calorimeter system, with acceptance covering $|\eta| < 3.2$, uses liquid argon (LAr) as the active medium.  A scintillator-tile calorimeter provides hadronic coverage for $|\eta| < 1.7$. The end-cap and forward regions, spanning $1.5 < |\eta| < 4.9$, are instrumented with LAr calorimeters for both electromagnetic and hadronic measurements. The muon spectrometer has separate trigger and high-precision tracking chambers which provide trigger coverage for $|\eta| < 2.4$ and muon identification and momentum measurements for $|\eta| < 2.7$.

%% file: trigger.tex
The data were collected from March to December 2012 at a $pp$ centre-of-mass energy of $8\tev$ using several triggers.  For the primary search region, a missing transverse momentum trigger was used, which bases the bulk of its rejection on
the vector sum of transverse energies
deposited in projective trigger towers (each with a size of approximately
$\Delta\eta \times \Delta\phi \sim 0.1 \times 0.1$~for $|\eta|<2.5$; these are larger and less regular in the more forward regions).  A more
refined calculation based on the vector sum of all calorimeter cells
above an energy threshold is made at a later stage in the trigger processing.
The trigger required \met~$>$~80 \GeV, and is fully efficient for
offline calibrated $\met > 150$~\GeV\ in signal-like events. An
integrated luminosity of \totallumi\ was collected using this trigger.
The luminosity uncertainty is derived,
following the same methodology as that detailed in
ref.~\cite{Aad:2013ucp}, from a preliminary calibration of the
luminosity scale obtained from beam-separation scans performed in
November 2012.

Data samples enriched in the major sources of background were
collected with electron or muon triggers, yielding an integrated luminosity of \totallumileptrig.  The electron trigger
selects events based on the presence of clusters of energy in  the electromagnetic calorimeter, with a
shower shape consistent with that of an electron, a matching track
in the tracking system, and a transverse energy (\et) threshold of $24\,\GeV$.  In order to recover some
of the efficiency for
high-\pt~electrons, events were also collected with a single-electron
trigger with looser requirements, but with the \et~threshold set to $60\,\GeV$.
The muon trigger
selects events containing one or more muon candidates based on tracks
identified in the muon spectrometer and inner detector.  For the
single-muon trigger, the \pt~threshold was 24~\GeV.
To recover some of the efficiency for higher-\pt~muons, events were also
collected with a single-muon trigger with a \pt~threshold of
36~\GeV~but with otherwise looser requirements.

Triggers based on the presence of high-\pt\ jets were used to collect data samples for the estimation of
the multijet and all-hadronic \ttbar~background.
The jet \pt~thresholds ranged from $55$ to $460\,\GeV$.  In order
to stay within the bandwidth limits of the trigger system, only  a
fraction of events passing these triggers were recorded to permanent
storage.

%% file: montecarlo.tex
\label{sec:montecarlo}

Samples of simulated events are used for the description of the background and
to model the SUSY signal.  
Top quark pair production where at least one of
the top quarks decays to a lepton is simulated with \powhegbox~\cite{Frixione:2007vw}.  To improve the agreement between data and
simulation, \ttbar\ events are reweighted based on the \pt\ of the
\ttbar\ system; the weights are extracted from the ATLAS measurement
of the \ttbar\ differential cross section at 7 \TeV, following the
methods of ref.~\cite{Aad:2012hg} but updated to the full 7 \TeV\  dataset.
\powhegbox\ is used to simulate single-top production in the $s$- and $\Wboson t$-channels, while \acer~\cite{kersevan:2004yg} is used for the $t$-channel. \sherpa~\cite{Gleisberg:2008ta} is used for \Wjets\ and \Zjets\ 
production with up to four additional partons (including heavy-flavour jets) as well as for diboson ($\Wboson\Wboson$,
$\Zboson\Zboson$ and $\Wboson\Zboson$) production.
\madgraph~\cite{Alwall:2011uj} generation with up to two additional partons is used for $\ttbar+\Wboson$ and
$\ttbar+\Zboson$ production.  

The underlying-event model is the ATLAS AUET2B tune
~\cite{ATL-PHYS-PUB-2011-009} of \pythia~\cite{Sjostrand:2006za}
except for the \ttbar\ and single-top
samples where  the Perugia 2011 C tune~\cite{PhysRevD.82.074018} is
used.  The parton distribution function (PDF) sets used for the SM background are CT10~\cite{Lai:2010vv} for the \powhegbox\ \ttbar\ and \sherpa\ samples, and CTEQ6L1~\cite{Pumplin:2002vw} for the \madgraph, \powhegbox\ single-top, and \acer\ samples.  
  
For the initial comparison with data, all SM background cross sections are
normalized to the results of higher-order calculations when available.
The theoretical cross sections for \Wjets\ and \Zjets\
are calculated with DYNNLO~\cite{Catani:2009sm} with the
MSTW 2008 NNLO~\cite{Martin:2009iq} PDF set.  The same ratio of the next-to-next-leading-order (NNLO)
to leading-order cross sections is applied
to the production of $\Wboson/\Zboson$ in
association with heavy-flavour jets.
The inclusive \ttbar~cross section is calculated at NNLO, including resummation of next-to-next-to-leading-logarithmic (NNLL) soft gluon terms~\cite{Cacciari:2011hy}, with Top++~\cite{Czakon:2011xx} using MSTW 2008 NNLO PDFs.
The production cross sections of \ttbar\ in association with $\Wboson/\Zboson$ are normalized to 
NLO cross sections~\cite{Campbell:2012dh,Garzelli:2011is}.
Approximate NLO+NNLL 
calculations are used for single-top production cross sections~\cite{Kidonakis:2010tc,Kidonakis:2010ux,Kidonakis:2011wy}.
  For the diboson cross sections,
MCFM~\cite{Campbell:2011bn} with
MSTW 2008 NLO PDFs is used.

The signal samples are generated with three different configurations:
(1) both top squarks decay via $\tone \rightarrow t \ninoone$, (2) one
top squark decays via $\tone \rightarrow t \ninoone$ and the other via
$\tone \rightarrow b \chinoonepm \rightarrow b  W^{\left(\ast\right)}
\ninoone$, and (3) both top squarks decay via
$\tone \rightarrow b \chinoonepm \rightarrow b  W^{\left(\ast\right)}
\ninoone$.  With appropriate weighting of these configurations,
the analysis sensitivity as a function of the branching fraction to 
$t \ninoone$ can be assessed.  In the samples with a decay to
$b\chinoonepm \rightarrow b  W^{\left(\ast\right)}
\ninoone$, the mass of $\chinoonepm$ is chosen to be twice
that of $\ninoone$ (motivated by models of gaugino unification). The
signal samples are generated in a grid across the plane of top squark
and $\ninoone$ masses with a grid spacing of $50 \GeV$ across most of
the plane.   The top squark mass ranges from $200$ to $750 \GeV$.
In the samples for which both top squarks decay via $\tone \rightarrow t
\ninoone$, the $\ninoone$ mass ranges from $1 \GeV$ up to approximately
$5 \GeV$ below the kinematic
limit.  For the samples involving
$\tone \rightarrow b \chinoonepm \rightarrow b  W^{\left(\ast\right)}
\ninoone$ decays, the chargino mass ranges from $100 \GeV$ (taking into
account the LEP limits~\cite{LEPCharginoLimit}
on the lightest chargino mass) up to approximately
$10 \GeV$ below the $\tone$ mass.

The signal samples for the scenario where both top squarks decay to a
top quark and a neutralino are generated using
\herwigpp~\cite{Bahr:2008pv}.
The neutralino is fixed to be a pure
bino, enhancing the decay of the $\tright$ component of $\tone$ to a
right-handed top quark.
Hadronic decays are expected to be less sensitive to such
polarization effects; a subset of signal samples generated
with top squarks
corresponding to the left-handed top quark yielded a $< 5\%$ increase in
the signal acceptance. The increase in the acceptance due to the
polarization is neglected in this paper.  The signal samples with mixed decays
are generated using \madgraph.
The differences between \herwigpp\ and \madgraph\ were evaluated for a few
samples and found to be negligible.
In these mixed decay samples, the chargino is fixed to be a pure wino,
and thus top squark decays into $b\chinoonepm$ originate from the
$\tleft$ component of the top squark.
The signal samples for which both squarks decay to
$b\chinoonepm$ are generated with \madgraph\, where once again the
chargino is fixed to be a pure wino.  The PDF set used for all signal
samples is CTEQ6L1.

All signal samples are normalized to cross sections
calculated to NLO in the strong coupling constant, adding the
resummation of soft gluon emission at next-to-leading-logarithmic
accuracy (NLO+
NLL)~\cite{Beenakker:1997ut,Beenakker:2010nq,Beenakker:2011fu}. The
nominal cross section and its uncertainty are taken from an envelope
of cross-section predictions using different PDF sets and
factorization and renormalization scales, as described
in ref.~\cite{Kramer:2012bx}.  The production cross section ranges
from approximately 2 pb to 0.008 pb for a top squark mass of $300\,\GeV$ to $700\,\GeV$ respectively.

The detector simulation
\cite{atlassimulation} is performed using GEANT4~\cite{geant4} or
a fast simulation framework where the showers in the electromagnetic
and hadronic calorimeters are simulated with
a parameterized description~\cite{ATL-PHYS-PUB-2010-013} and the rest
of the detector is simulated with GEANT4. The fast simulation was validated against full GEANT4 simulation for several signal points.
All samples are produced with
a varying number of simulated minimum-bias interactions overlaid on the
hard-scattering event to account for
multiple $pp$ interactions in
the same bunch crossing (pileup). The simulation is reweighted to match the distribution in data, which varies between approximately 10 and 30 interactions in each bunch crossing for this dataset. The overlay
also treats the impact of pileup from bunch crossings
other than the one in which the event occurred.
Corrections are applied to the simulated samples to account for
differences between data and simulation for the lepton
trigger and reconstruction efficiencies, momentum scale and
resolution, and for the efficiency of identifying
jets originating from the fragmentation of $b$-quarks,  together with the probability
for mis-tagging light-flavour and charm quarks.

%% file: objecteventsel.tex
\label{sec:objrec}

The reconstructed primary vertex
\cite{PV} is
required to be consistent with the luminous region and to
have at least five associated tracks with $\pt > 400\mev$;
when more than one such vertex is found, the vertex with
the largest summed $\pt^{2}$ of the associated tracks is chosen.

Jets are constructed from three-dimensional clusters of noise-suppressed calorimeter cells~\cite{topoclusters} using the \antikt\ algorithm~\cite{Cacciari:2008gp,Cacciari:2005hq,Cacciari:2011ma} with a distance parameter $R = 0.4$ and calibrated with a local cluster weighting algorithm~\cite{Issever:2004qh}. An area-based correction is applied for energy from additional proton--proton collisions based on an estimate of the pileup activity in a given event using the method proposed in ref.~\cite{Cacciari:2007fd}. Jets are calibrated~\cite{Aad:2011he} and required to have $\pt > 20\,\GeV$ and $|\eta| <4.5$. Events containing jets arising from detector noise, cosmic-ray muons, or other non-collision sources are removed from consideration~\cite{Aad:2011he}. Once the \MET\ is computed and any ambiguity with electrons or muons is resolved (as described below), signal jets are required to have $\pt > 35 \GeV$ and $|\eta|<2.8$.
Jets containing a $b$-quark and within the acceptance of the inner detector ($|\eta|<2.5$) are identified with an algorithm that
exploits both the track impact parameters and secondary vertex
information~\cite{ATLAS-CONF-2012-043}; this algorithm is based on a
neural network using the output weights of the IP3D, JetFitter+IP3D,
and SV1 algorithms (defined in refs.~\cite{ATLAS-CONF-2011-089,ATLAS-CONF-2011-102}).  The
identification of these ``$b$-tagged jets" has an average efficiency
of $70\%$ for jets originating from the fragmentation of a $b$-quark
in simulated \ttbar~events,
a rejection factor of approximately 150 for light-quark and gluon jets
(depending on the \pt~of the jet), and a rejection factor of
approximately 5 for charm jets. 

Electrons, which are reconstructed from energy clusters in the electromagnetic
calorimeter matched to a track in the inner detector~\cite{Aad:2011mk}, are required to have $|\eta| < 2.47$, $\pt>10\,\GeV$, and must pass a variant of the ``loose'' selection defined in ref.~\cite{Aad:2011mk} that was re-optimized
for 2012 data. In the case where the separation between an electron candidate and a non-$b$-tagged jet is $\Delta R < 0.2$, the object is considered to be an electron. If the separation between an electron candidate and any jet satisfies $0.2 < \Delta R < 0.4$, or if the separation between an electron candidate and a $b$-tagged jet is $\Delta R < 0.2$, the electron is not counted. Muons, which are identified either as a combined track in the muon spectrometer
and inner detector systems, or as an inner detector
track matched with a  muon spectrometer segment
\cite{ATLAS-CONF-2011-021,ATLAS-CONF-2011-063}, are required to have
$|\eta| < 2.4$ and $\pt>10\,\GeV$. If the separation between a muon and any jet is $\Delta R < 0.4$, the muon is not counted.

The \ptmiss\ is the negative vector sum
of the \pT\ of the clusters of calorimeter cells, which are calibrated
according to their associated reconstructed object (e.g. preselected
jets and electrons), and the \pT\ of preselected muons.
The missing transverse momentum from the tracking system (denoted as \ptmisstrk, with magnitude \mettrk)
is computed from the vector sum of the reconstructed inner detector tracks with $\pt > 500\MeV$, $|\eta|<2.5$, in association with the primary vertex in the event. 

The requirements on electrons and muons are tightened for the
selection of events in background control regions (described in
section~\ref{sec:background}) containing leptons.
Electrons are required to pass a variant of the ``tight'' selection
of ref.~\cite{Aad:2011mk} re-optimized for 2012 data, and are required to satisfy track- and calorimeter-based isolation criteria. The scalar sum of the \pt\ of tracks within a cone of size
$\Delta R = 0.3$ (``track isolation'') around the electron (excluding the electron itself) is required to be
less than $16\%$ of the electron \pt.  The scalar sum  of the \et~of pileup-corrected calorimeter energy deposits within a cone of
size $\Delta R = 0.3$ (``calorimeter isolation'') around the electron (again, excluding the electron itself) is required to be less than
$18\%$ of the electron \pt.  The impact parameter of the electron in the transverse plane with respect to the reconstructed event primary
vertex ($|d_0|$) is required to be less than five times the impact
parameter uncertainty ($\sigma_{d0}$). The impact parameter along the
beam direction, $\left|z_0 \times \sin\theta\right|$, is required to be less than $0.4$~mm. Further
isolation criteria on reconstructed muons are also imposed: both the
track and calorimeter isolation are required to be less than $12\%$ of
the muon  \pt. In addition, the requirements $|d_0| < 3\sigma_{d0}$ and
$\left|z_0 \times \sin\theta\right| <0.4$~mm are imposed for muon candidates.  The lepton \pt~requirements vary by background control region, as summarized in tables~\ref{tab:CRA}$-$\ref{tab:CRC} in section~\ref{sec:background}.

%% file: signalregiondefs.tex
\label{sec:signalregiondefs}

The search for direct top squark pair production in the all-hadronic
channel has a nominal experimental signature of six distinct jets (two
of which originate from $b$-quarks), no reconstructed electrons or
muons, and significant \met\ from the LSPs. The stringent requirement
of $\met>150\GeV$ needed to satisfy the trigger rejects the vast
majority of background from multijet and all-hadronic top quark
events. Major background contributions include \ttbar\ and \Wjets\ events where
one \Wboson\ decays via a low-momentum or mis-reconstructed lepton
plus a neutrino; after the event selection described later in the
text, approximately 40\% of the \ttbar\ background arises from 
a \Wboson\ decaying to a $\tau$ lepton that decays hadronically, while the remainder arises
equally from \Wboson\ decays to an electron, muon or
leptonically decaying $\tau$.
Other important background contributions are
\Zjets\ and \ttbar+\Zboson\ events where the
\Zboson\ decays via neutrinos that escape detection, and single-top
events.

This nominal signature, which is labelled ``fully resolved" since all
of the top squark decay products are individually reconstructed, is sensitive to a
wide range of top squark and LSP masses in both the $\stop \to
t\ninoone$ and $\stop \to b \chinoonepm, \chinoonepm
\to\Wboson^{\left(*\right)}\ninoone$ decay modes. The experimental
sensitivity to the $\stop \to t\ninoone$ decay mode, especially for
high top squark masses, is enhanced by also considering a second
category of ``partially resolved" events with particularly high $\met$
and four or five reconstructed jets. This final state can occur
if the top quarks are sufficiently Lorentz-boosted such that their
decay products merge, and/or if one
or more top decay products is below the reconstruction threshold.
The consideration of a third category of events
with exactly five jets in the final state but a less stringent $\met$
requirement augments the sensitivity to $\stop\to b\chinoonepm$
decays, particularly where one of the jets from the decay of the
$\Wboson^{\left(*\right)}$ has low \pT\ and is not reconstructed.

\begin{figure}[tb]
\center
\includegraphics[width=0.60\textwidth,trim=0 20 0 0]{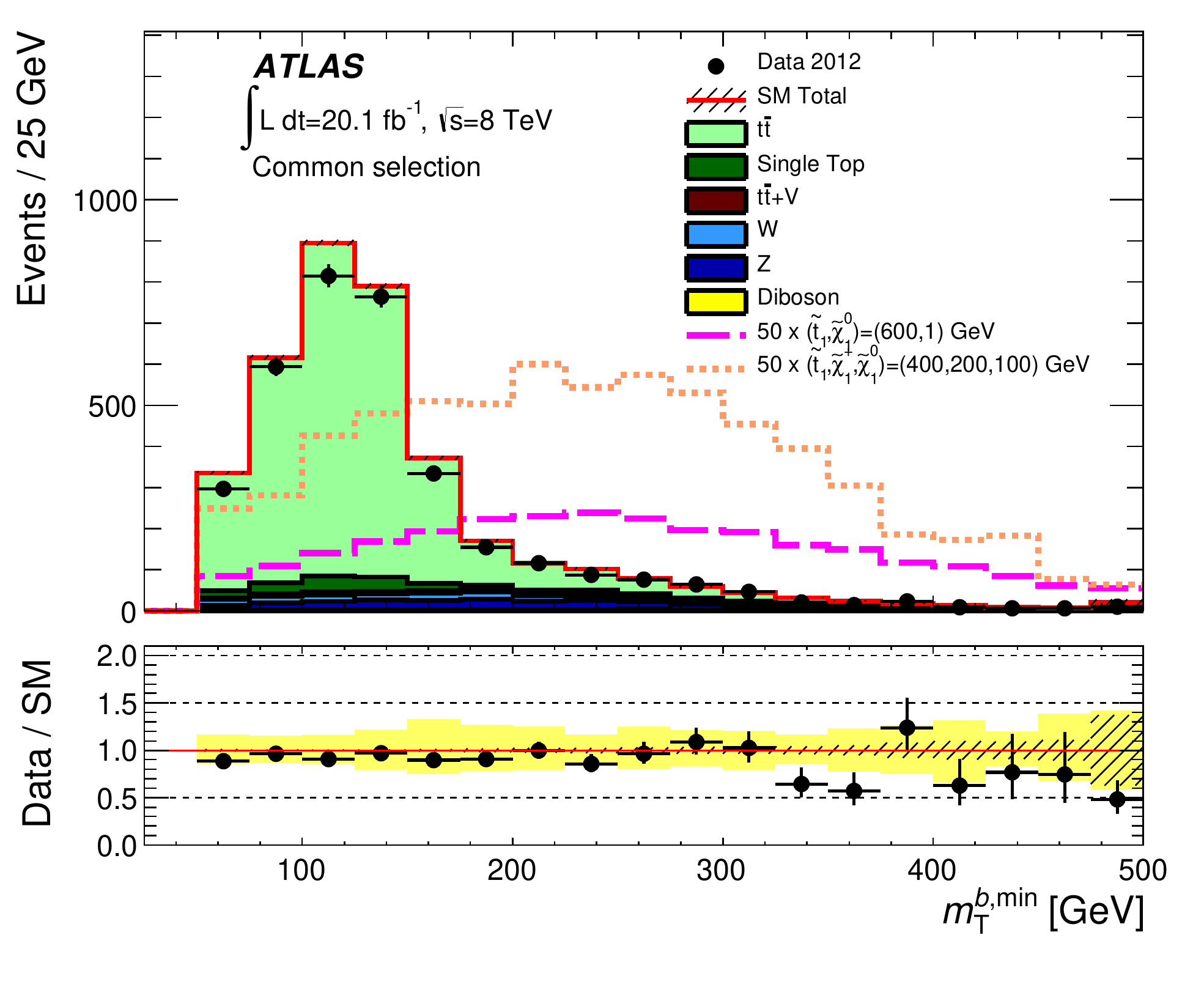}
\caption{
  The distribution of \mtbmetmindphi\ in events with at least four jets that pass the common selection requirements described in the text (see also table~\ref{tab:SRcommon}), excluding the requirement on \mtbmetmindphi.
  The stacked histograms show the SM expectation from simulation compared to the data (points). Simulated signal samples where $m_{\stop}=600$~GeV, $m_{\ninoone}=1$~GeV (pink dashed line) and $m_{\stop} = 400\GeV, m_{\chinoonepm} = 200\GeV, m_{\ninoone} = 100\GeV$ (orange dotted line) are overlaid; the expected number of signal events is multiplied by a factor of 50 for improved visibility. The ``Data/SM'' plot shows the ratio of data events to the total
  Standard Model expectation.  The rightmost bin includes all overflows.
  The hatched uncertainty band around  the
  Standard Model expectation shows the statistical uncertainty and
  the yellow band (shown only for the ``Data/SM'' plot)
  shows the combination of statistical and experimental systematic
  uncertainties.}
\label{fig:mTpreselection}
\end{figure}

All three categories of events share common selection criteria including $\met > 150$~\GeV, as summarized in table~\ref{tab:SRcommon}. The two highest-\pT\ signal jets are required to have $\pt > 80\GeV$ and
two of the signal jets must be $b$-tagged. Events containing reconstructed electrons or muons with $\pT>10\GeV$ are vetoed (thus, $N_{\rm{lep}} = 0$). 
Events with \met\ arising from mis-measured jets are rejected by requiring an angular separation between the azimuthal angle ($\phi$) of the \met\ and any of the three highest-\pt~jets in the event: $\dphijetmet > \pi/5$~radians. Further reduction of such events is achieved by requiring the \ptmisstrk\ to be aligned in $\phi$ with respect to the \ptmiss\ calculated from the calorimeter system: $\dphimettrk<\pi/3$~radians. A substantial rejection of \ttbar\ background is achieved by requiring that the transverse mass ($\mT$) calculated from the \met~and the $b$-tagged jet closest in $\phi$ to the \ptmiss\ direction exceeds the top mass, as illustrated in figure~\ref{fig:mTpreselection}:

\begin{equation}
\mtbmetmindphi\ = \sqrt{2\,\ptb\,\met \left[1-\cos{\Delta\phi\left(\vecptb,\ptmiss\right)}\right]} > 175\,\GeV.
\end{equation}
In this and subsequent figures displaying simulated signal distributions, the signal samples represent $\tone\tonebar$ production where both top squarks decay via $\tone \to t \ninoone$ for $m_{\tone} = 600\GeV$ and $m_{\ninoone}=1\GeV$, or $\tone\tonebar$ production where one top squark decays via $\tone \to t \ninoone$ and the other decays via $\tone  \rightarrow b \chinoonepm$ in each event, for $m_{\tone} = 400\GeV, m_{\chinoonepm} = 200\GeV$, and $m_{\ninoone} = 100\GeV$.

Beyond these common requirements, the three categories of events are
further subdivided and optimized individually to target the neutralino
or chargino decay modes and particular top squark mass ranges. The
most powerful discriminating variable is the \MET\ resulting from the
undetected LSPs; signal regions with higher (lower) \MET\ requirements
have increased sensitivity to potential signals where the difference
in mass between the top squark and the LSP is large (small). The
selection criteria obtained from the simulation-based optimization procedure are described in the following.

\subsection{Fully resolved signal region (SRA)}
\label{sec:SRA}

This first category of events (SRA) encompasses the nominal signature
of six jets plus \MET. Events with additional jets from initial- or
final-state radiation are also accepted. These events are then divided into four signal regions (SRA$1$--$4$) with increasing \MET; the specific requirements are summarized in table~\ref{tab:SRA}. 

Since all jets from the top quark decays are fully resolved, the two
top candidates are reconstructed from signal jets according to the following
algorithm. The two jets with the highest $b$-tagging weight
are selected first.  From the remaining
jets in the event, the two closest jets in the $\eta$--$\phi$ plane are
combined to form a $\Wboson$ boson candidate; this candidate is then
combined with the $b$-tagged jet closest in the $\eta$--$\phi$ plane to form the first top candidate with mass $\mtopcandzero$.  A
second $\Wboson$ boson candidate is formed by repeating the procedure on the
remaining jets; this candidate is then combined with the second of the
selected $b$-tagged jets to form the second top candidate with mass $\mtopcandone$. The mass requirements on each top candidate are rather loose to ensure high signal efficiency.

Two additional discriminating quantities are introduced to reject the
dominant \ttbar\ background in these signal regions. The first
quantity, $\min[\mtjetimet]$, is the minimum value of the transverse
mass calculated from each of the signal jets and the \ptmiss, and
ensures that the final-state jets are well separated from the
\ptmiss. The second quantity is a ``$\tau$ veto'' in which events that
contain a non-$b$-tagged jet within $|\eta| < 2.5$ with $\le 4$
associated tracks with $\pT>500\mev$, and where the $\Delta \phi$ between the jet and the \ptmiss~is less than $\pi/5$ radians, are vetoed since they are likely to have originated from a $W\rightarrow\tau\nu$ decay.

\begin{table}[p]
\footnotesize
\caption{Selection criteria common to all signal regions.}
\begin{center}
\begin{tabular}{l|c} \hline\hline
Trigger & \met \\ \hline
$N_{\rm{lep}}$ & 0 \\ \hline
$b$-tagged jets & $\ge2$ \\ \hline
$\met$ & $> 150\GeV$ \\ \hline
 & \\ [-2.5ex]
$\dphijetmet$ & $> \pi/5$ \\ 
 & \\ [-2.5ex] \hline
 & \\ [-2.5ex]
$\dphimettrk$ & $<\pi/3$ \\ 
 & \\ [-2.5ex] \hline
$\mtbmetmindphi$ & $> 175 \gev$ \\ \hline \hline
\end{tabular}
\end{center}
\label{tab:SRcommon}

\vskip 0.1in

\caption{Selection criteria for SRA, the fully resolved topology, with $\ge 6$ \antikt\ $R=0.4$ jets.}
\begin{center}
\begin{tabular}{l|c|c|c|c} \hline\hline
 & SRA1 & SRA2 & SRA3 & SRA4  \\ \hline \hline
\antikt\ $R=0.4$ jets & \multicolumn{4}{c}{$\ge 6,~\pt>80,80,35,35,35,35 \gev$}\\ \hline
\mtopcandzero & \multicolumn{2}{c|}{$<225 \gev$} & \multicolumn{2}{c}{[50,250] \gev} \\ \hline
\mtopcandone & \multicolumn{2}{c|}{$<250 \gev$} & \multicolumn{2}{c}{[50,400] \gev} \\ \hline
$\min[\mtjetimet]$ & \multicolumn{2}{c|}{--} & \multicolumn{2}{c}{$>50 \gev$}\\ \hline
$\tau$ veto & \multicolumn{4}{c}{yes} \\ \hline
\met & $> 150 \gev$ & $> 250 \gev$ & $> 300 \gev$ & $> 350 \gev$ \\ \hline
\hline
\end{tabular}
\end{center}
\label{tab:SRA}

\vskip 0.1in

\caption{Selection criteria for SRB, the partially resolved topology, with four or five \antikt\ $R=0.4$ jets, reclustered into \antikt\ $R=1.2$ and $R=0.8$ jets.}
\begin{center}
\begin{tabular}{l|c|c} \hline\hline
 & SRB1 & SRB2   \\ \hline \hline
\antikt\ $R=0.4$ jets & 4 or 5, $\pt > 80,80,35,35,(35) \gev$ & 5, $\pt >
100,100,35,35,35 \gev$ \\ \hline
\topmassasym & $< 0.5$ & $> 0.5$ \\ \hline
\ptantikttwelvezero & -- & $> 350 \gev$ \\ \hline
\mantikttwelvezero & $> 80 \gev$ & $[140,500] \gev$  \\ \hline
\mantikttwelveone & $[60,200] \gev$ & -- \\ \hline
\mantikteightzero & $> 50 \gev$ & $[70,300] \gev$ \\ \hline
\mtuntagmetmindphi  & $> 175 \gev$ & $ > 125 \gev$ \\ \hline
\mtlowestptmet & $> 280 \gev$ for 4-jet case & -- \\ \hline
\htsig & -- & $> 17\sqrt{\gev}$ \\ \hline
\met & $> 325 \gev$ & $> 400\gev$ \\ \hline
\hline
\end{tabular}
\end{center}
\label{tab:SRB}

\vskip 0.1in

\caption{Selection criteria for SRC, targeting the scenario in which one top squark decays via $\stop \to b \chinoonepm$, with five \antikt\ $R=0.4$ jets.}
\begin{center}
\begin{tabular}{l|c|c|c} \hline\hline
 & SRC1 & SRC2 & SRC3   \\ \hline \hline
\antikt\ $R=0.4$ jets & \multicolumn{3}{c}{5, $\pt > 80,80,35,35,35 \gev$} \\ \hline
\dphibb & \multicolumn{3}{c}{$> 0.2\pi$} \\ \hline
\mtbmetmindphi & $> 185 \gev$ &$> 200 \gev$ &$> 200 \gev$\\ \hline
\mtbmetmaxdphi & $> 205 \gev$ &$ > 290 \gev$ &$ > 325 \gev$\\ \hline
$\tau$ veto & \multicolumn{3}{c}{yes} \\ \hline
\met & $> 160 \gev$ & $> 160\gev$ & $> 215\gev$ \\ \hline
\hline
\end{tabular}
\end{center}
\label{tab:SRC}
\end{table}

\afterpage{\clearpage}

\subsection{\boldmath Partially resolved signal region targeting $\stop \to t\ninoone$ decays (SRB)}

An alternative top reconstruction algorithm is applied to events with four or five jets in the final state (SRB).  The \antikt\ clustering algorithm~\cite{Cacciari:2008gp} is applied to $R=0.4$ signal jets, using reclustered distance parameters of $R=0.8$ and $R=1.2$.  Nominally, the all-hadronic decay products of a top quark can be reconstructed as three distinct jets, each with a distance parameter of $R=0.4$. The transverse shape of these jets are typically circular with a radius equal to this distance parameter, but when two of the jets are less than $2R$ apart in $\eta$--$\phi$ space, the one-to-one correspondence of a jet with a top daughter is violated. To some extent, this can be tolerated in the fully resolved scenario without an appreciable efficiency loss, but as the jets become closer together the majority of the \pT\ is attributed to one of the two jets, and the lower-\pT\ jet may drop below the minimum \pT\ requirement. In SRB, at least two reclustered $R=1.2$ jets are required, and selection criteria are employed based on the masses of these top candidates: \mantikttwelvezero $\left(\mantikttwelveone\right)$ is the mass of the highest-\pT\ (second-highest-\pt) \antikt\ $R=1.2$ reclustered jet. Requirements are also placed on the \pT\ of the highest-\pt\ \antikt\ $R=1.2$ reclustered jet (\ptantikttwelvezero) and the mass of the highest-\pt\ \antikt\ $R=0.8$ reclustered jet (\mantikteightzero); this latter requirement rejects background without hadronic $\Wboson$ candidates.

For signal regions employing reclustered jets, it is useful to categorize events based on 
the top mass asymmetry \topmassasym, defined as:

\begin{equation}
\topmassasym = \frac{|\mantikttwelvezero - \mantikttwelveone|}{\mantikttwelvezero + \mantikttwelveone}.
\label{eqn:topmassasym}
\end{equation}

\noindent Events where $\topmassasym < 0.5$ tend to be well balanced, with two well reconstructed top candidates (SRB1). In events where $\topmassasym > 0.5$ the top candidates tend to be overlapping or otherwise less well reconstructed (SRB2). Since the two categories of events have markedly different signal-to-background ratios, the selections for these two categories of events are optimized separately.

The background in SRB, which is dominated by $\Zjets$ and $\Wjets$, is suppressed by requirements on the transverse mass of the \ptmiss\ and the non-$b$-tagged jet closest in $\Delta \phi$ to the \ptmiss, \mtuntagmetmindphi, and the transverse mass of the fourth jet (in \pT\ order) and the \ptmiss, \mtlowestptmet; this latter variable is used only in four-jet events in SRB1. In SRB2, where the top candidates tend to be less well-reconstructed and background rejection is especially challenging, only five-jet events are considered and an additional requirement is made on a measure of the \MET\ significance: \htsig, where \HT\ is the scalar sum of the \pt\ of all five jets in the event. The full set of selection requirements for SRB are detailed in table~\ref{tab:SRB}; the logical OR of SRB1 and SRB2 is considered in the final likelihood fit.

\subsection{\boldmath Signal region targeting $\stop\to b\chinoonepm$ decays (SRC)}

For potential signal events where at least one top squark decays via
$\stop\to b\chinoonepm$,
$\chinoonepm\to\Wboson^{\left(*\right)}\ninoone$, requiring exactly
five jets in the final state enhances the sensitivity to smaller
values of $m_{\chinoonepm} - m_{\ninoone}$ at the expense of increased
background. Additional discrimination against background where the
two $b$-tagged jets come from a gluon emission is provided by a
requirement on \dphibb, the azimuthal angle between the two highest-\pt\ $b$-tagged jets. To reduce the \ttbar\ background, tighter
requirements on \mtbmetmindphi\ are employed. The quantity
\mtbmetmaxdphi\ is also used, which is analogous to
\mtbmetmindphi\ except that the transverse mass is computed with the
$b$-tagged jet that has the largest $\Delta\phi$ with respect to the
\ptmiss\ direction. These two criteria, along with the \MET, are
tightened in subsequent SRC$1$--$3$ sub-regions as summarized in
table~\ref{tab:SRC}. Finally, the same $\tau$ veto
as described in section~\ref{sec:SRA} is applied.

%% file: background.tex
\label{sec:background}

The main background contributions in SRA and SRC arise
from \ttbar~production where one top quark decays semileptonically and the
lepton (particularly  a hadronically decaying $\tau$ lepton) is either not identified or reconstructed as a jet,
$Z(\rightarrow \nu \overline{\nu})$ plus heavy-flavour
  jets, and the irreducible background from
  \ttbar+$Z(\rightarrow \nu \overline{\nu})$.  In SRB,
  an important contribution comes from $\Wboson$ plus heavy-flavour jets,
  where again the $\Wboson$ decays semileptonically and the lepton is reconstructed as a jet or not
  identified.   Other
background processes considered are multijets, single top,
$\ttbar+\Wboson$, and diboson production.

With the exception of all-hadronic \ttbar\ and multijet production,
all background contributions are estimated primarily from simulation.
Control regions (CRs) are used to adjust the
normalization of the simulated background contributions in each signal region from
semileptonic \ttbar, $Z \to \nu\overline{\nu}$ plus heavy-flavour jets and, in the case of SRB,
$\Wboson$ plus heavy-flavour jets. The all-hadronic \ttbar\ and multijet contributions are estimated from data alone in a multijet control region. The control regions are
designed to be orthogonal to the signal regions while enhancing a particular source of background; they are used to
normalize the simulation for that background to data. The control regions are chosen to be kinematically close to the corresponding signal region, to minimize the systematic uncertainty associated with extrapolating the background yield from the control region to the signal region, but also to have enough data events to avoid a large statistical uncertainty in the background estimate. In addition, control region selections are chosen to minimize potential contamination from signal in the scenarios considered.
As the control regions  are not
always pure in the process of interest, simulation is used to estimate
the cross contamination between control regions; the normalization factors and the cross
contamination are determined simultaneously for all regions using a
fit described  in section~\ref{sec:fit}.

The selection requirements  for all control regions used in this
analysis are summarized in tables~\ref{tab:CRA}, \ref{tab:CRB}, and
\ref{tab:CRC} for SRA, SRB, and SRC, respectively.

\begin{table}[!htb]
\caption{Selection criteria for control regions associated with SRA.  Only the requirements that differ from the common selection in table~\ref{tab:SRcommon} and those in table~\ref{tab:SRA} are listed; ``same" indicates the same selection as the signal region.}
\begin{center}
\footnotesize
\begin{tabular}{l|c|c|c} \hline\hline
             & {\bf\boldmath \ttbar\ CR} & {\bf\boldmath \Zjets\ CR} & {\bf Multijet CR} \\ \hline \hline
Trigger & electron (muon) & electron (muon) & same \\ \hline
$N_{\rm{lep}}$ & 1 & 2 & same  \\ \hline
\ptl\ & $>35 (35) \gev$ & $>25 (25) \gev$ & -- \\ \hline
\ptltwo\ & same & $> 10 (10) \gev$  & same \\ \hline
\mll\ & -- & $[86,96] \gev$ & -- \\ \hline
\mettrk & -- & -- & same\\ \hline
 & & & \\ [-2.5ex]
$\dphimettrk$ & -- & -- & -- \\ 
 & & & \\ [-2.5ex] \hline
 & & & \\ [-2.5ex]
$\dphijetmet$ & $> \pi/10$ & -- & $< 0.1$ \\ 
 & & & \\ [-2.5ex]\hline
\mtbmetmindphi & $> 125 \gev$ & -- & -- \\ \hline
\mtlepmet & $[40,120] \gev$ & -- & -- \\ \hline
$\min[\mtjetimet]$ & -- & -- & -- \\ \hline
$\mtopcandzero$ or $\mtopcandone$ & $< 600 \gev$ & -- & -- \\ \hline
\met & $> 150 \gev$ & $ < 50 \gev$ & $> 150 \gev$ \\ \hline
\METprime & -- & $>70 \gev$ & -- \\ \hline
\hline
\end{tabular}
\end{center}
\label{tab:CRA}
\end{table}

\begin{table}[p]
\caption{Selection criteria for control regions associated with SRB.  Only the requirements that differ from the common selection in table~\ref{tab:SRcommon} and those in table~\ref{tab:SRB} are listed; ``same" indicates the same selection as the signal region.}
\begin{center}
\footnotesize
\begin{tabular}{l|c|c|c|c} \hline\hline
             & {\bf\protect\boldmath \ttbar\ CR} & {\bf\boldmath $\Wboson$+jets CR} & {\bf\boldmath $\Zboson$+jets CR}  & {\bf Multijet
    CR}  \\ \hline \hline
Trigger & electron (muon) & electron (muon)& electron (muon) & same \\ \hline
$N_{\rm{lep}}$ & 1 & 1 & 2 & same\\ \hline
\ptl\ & $>35 (35) \gev$ & $> 35 (35) \gev$ & $>25 (25) \gev$ & --  \\ \hline
\ptltwo\ & same & same & $>10 (10) \gev$ & same  \\ \hline
\mll\ & -- & -- & $[86,96] \gev$ & -- \\ \hline
\antikt\ $R=0.4$ jets & [4,5] & [4,5] & 5 & same \\ \hline
\ptj & \multicolumn{3}{c|}{$> 80,80,35,35,(35) \gev$} & same \\ \hline
$N_{b\textrm{-jet}}$ & same & 1 & same & same  \\ \hline
\mettrk & -- & -- & -- & same  \\ \hline
& & & & \\ [-2.5ex]
$\dphimettrk$ & -- & -- & -- & --   \\
& & & & \\ [-2.5ex] \hline
& & & & \\ [-2.5ex]
$\dphijetmet$ & $> \pi/10$ & $> \pi/10$ & -- & $< 0.1$  \\
& & & & \\ [-2.5ex] \hline
\mtbmetmindphi & -- & -- & -- & -- \\ \hline
\mtlepmet & $[40,120] \gev$ & $[40,120] \gev$ & -- & --  \\ \hline
\met & $> 150 \gev$ & $>150 \gev$ & $ < 50 \gev$ & $> 150 \gev$ \\ \hline
\METprime & -- & -- & $>70 \gev$ & -- \\ \hline
$\ptantikttwelvezero$ & --  & -- & -- & -- \\ \hline
\mtuntagmetmindphi  & -- & -- & -- & -- \\ \hline
\mtlowestptmet & -- & -- & -- & --
\\ \hline
\topmassasym & $<0.5$ for 4-jet case & $< 0.5$ & -- & --  \\ \hline
\mantikttwelvezero & -- & $< 40 \gev$ & -- & -- \\ \hline
\mantikttwelveone & --  & -- & -- & -- \\ \hline
\mantikteightzero & -- & -- & -- & -- \\ \hline
\htsig & -- & -- & -- & -- \\ \hline
\hline
\end{tabular}
\end{center}
\label{tab:CRB}

\vskip 0.10in

\caption{Selection criteria for control regions associated with SRC.  Only the requirements that differ from the common selection in table~\ref{tab:SRcommon} and those in table~\ref{tab:SRC} are listed; ``same" indicates the same selection as the signal region.}
\begin{center}
\footnotesize
\begin{tabular}{l|c|c|c} \hline\hline
             & {\boldmath \ttbar} {\bf CR} & {\bf\boldmath \Zjets\ CR} & {\bf Multijet CR} \\ \hline \hline
Trigger & electron (muon) & electron (muon) & same \\ \hline
$N_{\rm{lep}}$ & 1 & 2 & same \\ \hline
\ptl\ & $>35 (35) \gev$ & $>25 (25) \gev$ & -- \\ \hline
\ptltwo\ & same & $> 10 (10) \gev$ & same \\ \hline
\mll\ & -- & $[86,96] \gev$ & -- \\ \hline
\mettrk & -- & -- & same \\ \hline
 & & & \\ [-2.5ex]
$\dphimettrk$ & -- & -- & -- \\ 
 & & & \\ [-2.5ex] \hline
 & & & \\ [-2.5ex]
$\dphijetmet$ & $> \pi/10$ & -- & $< 0.1$ \\ 
 & & & \\ [-2.5ex] \hline
\dphibb\ & same & -- & -- \\ \hline
\mtbmetmindphi & $> 150 \gev$ & -- & -- \\ \hline
\mtbmetmaxdphi & $> 125 \gev$ & -- & -- \\ \hline
\mtlepmet & $[40,120] \gev$ & -- & -- \\ \hline
\met & $> 100 \gev$ & $ < 50 \gev$ & $> 150 \gev$ \\ \hline
\METprime & -- & $>70 \gev$ & -- \\ \hline
\hline
\end{tabular}
\end{center}
\label{tab:CRC}
\end{table}

\subsection{\boldmath \ttbar\ background}

\begin{figure}[tbp]
\begin{center}
\subfigure[]{\includegraphics[width=0.45\textwidth,trim=0 40 0 0]{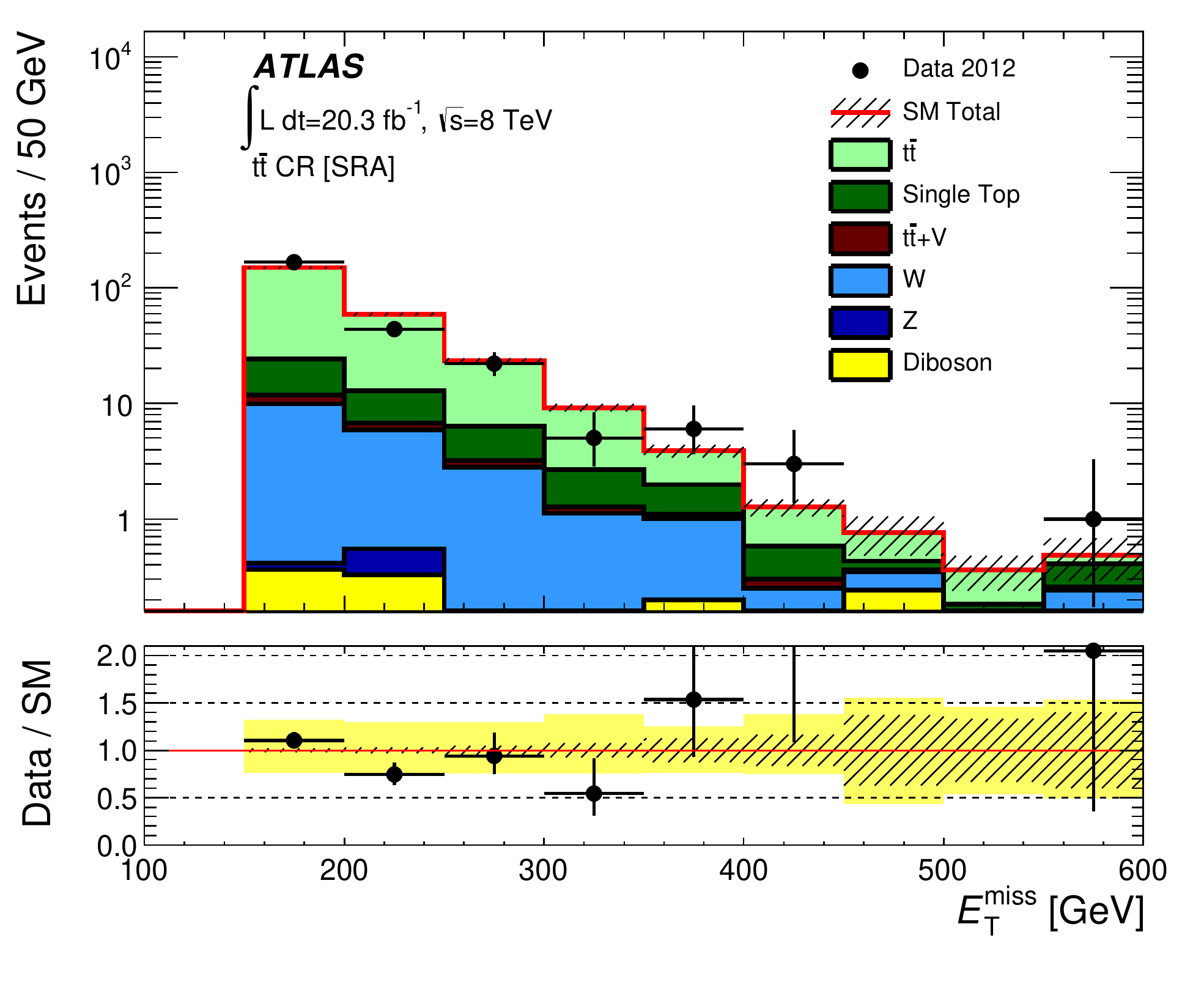}}
\subfigure[]{\includegraphics[width=0.45\textwidth,trim=0 40 0 0]{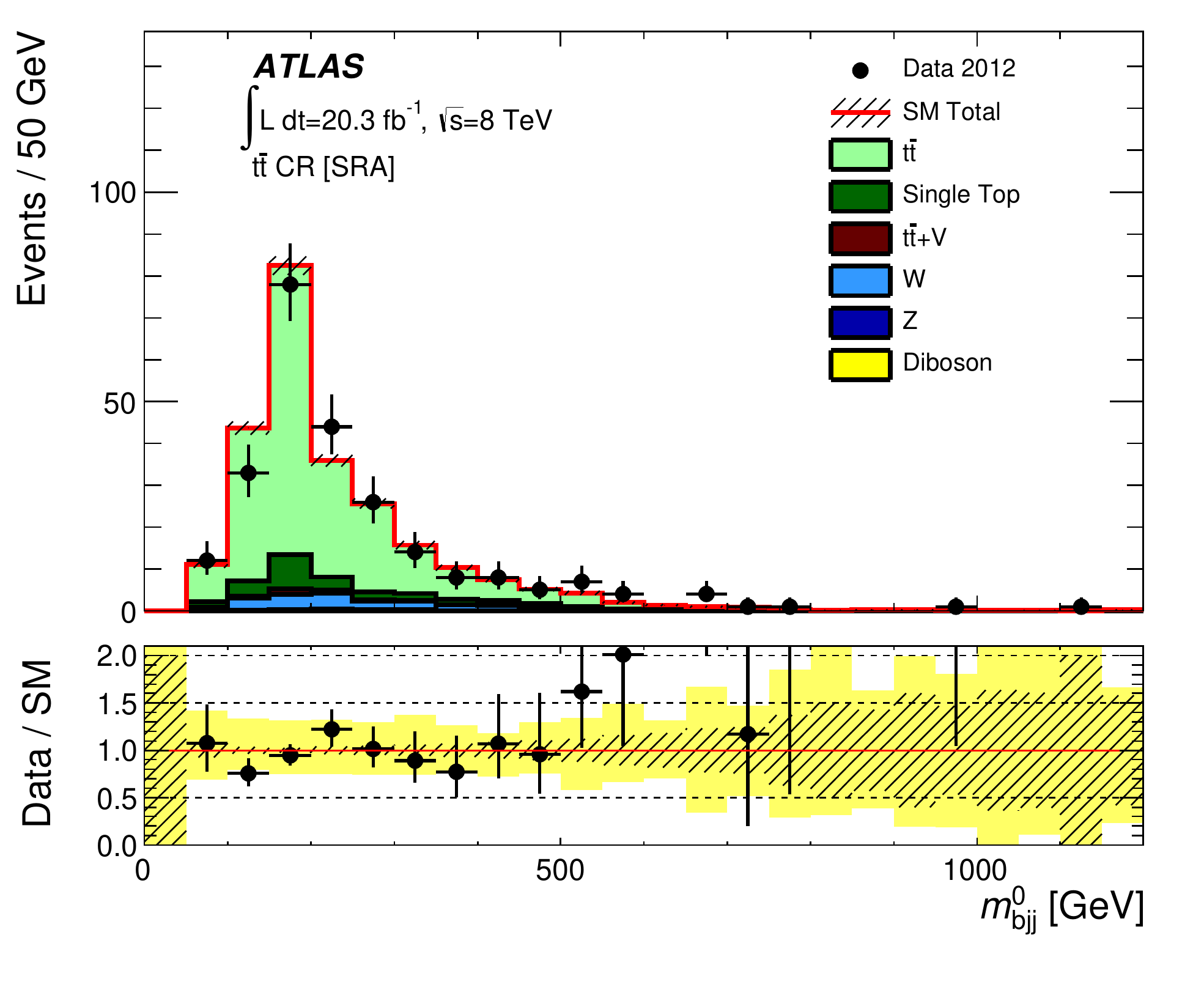}}
\subfigure[]{\includegraphics[width=0.45\textwidth,trim=0 40 0 0]{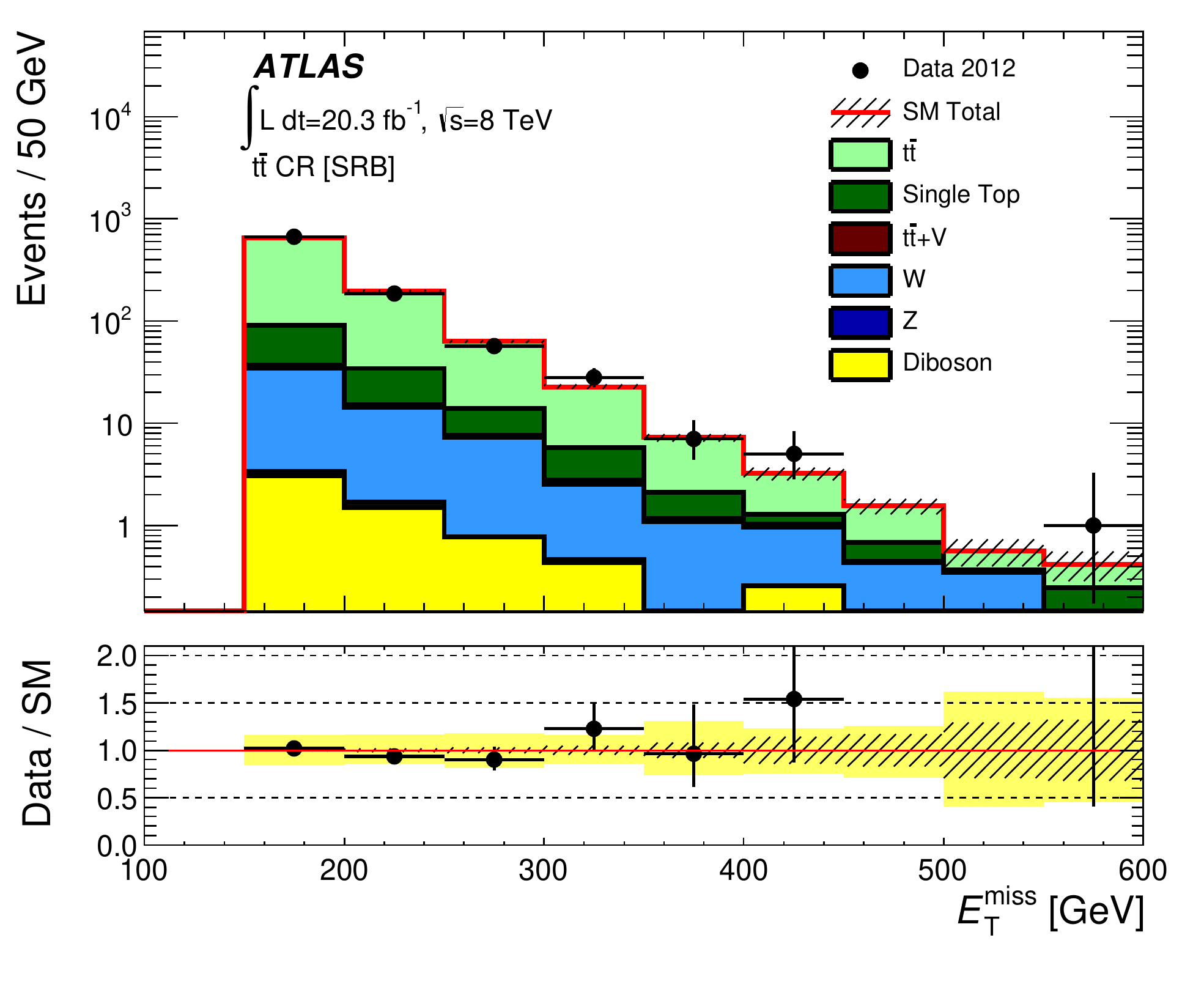}}
\subfigure[]{\includegraphics[width=0.45\textwidth,trim=0 40 0 0]{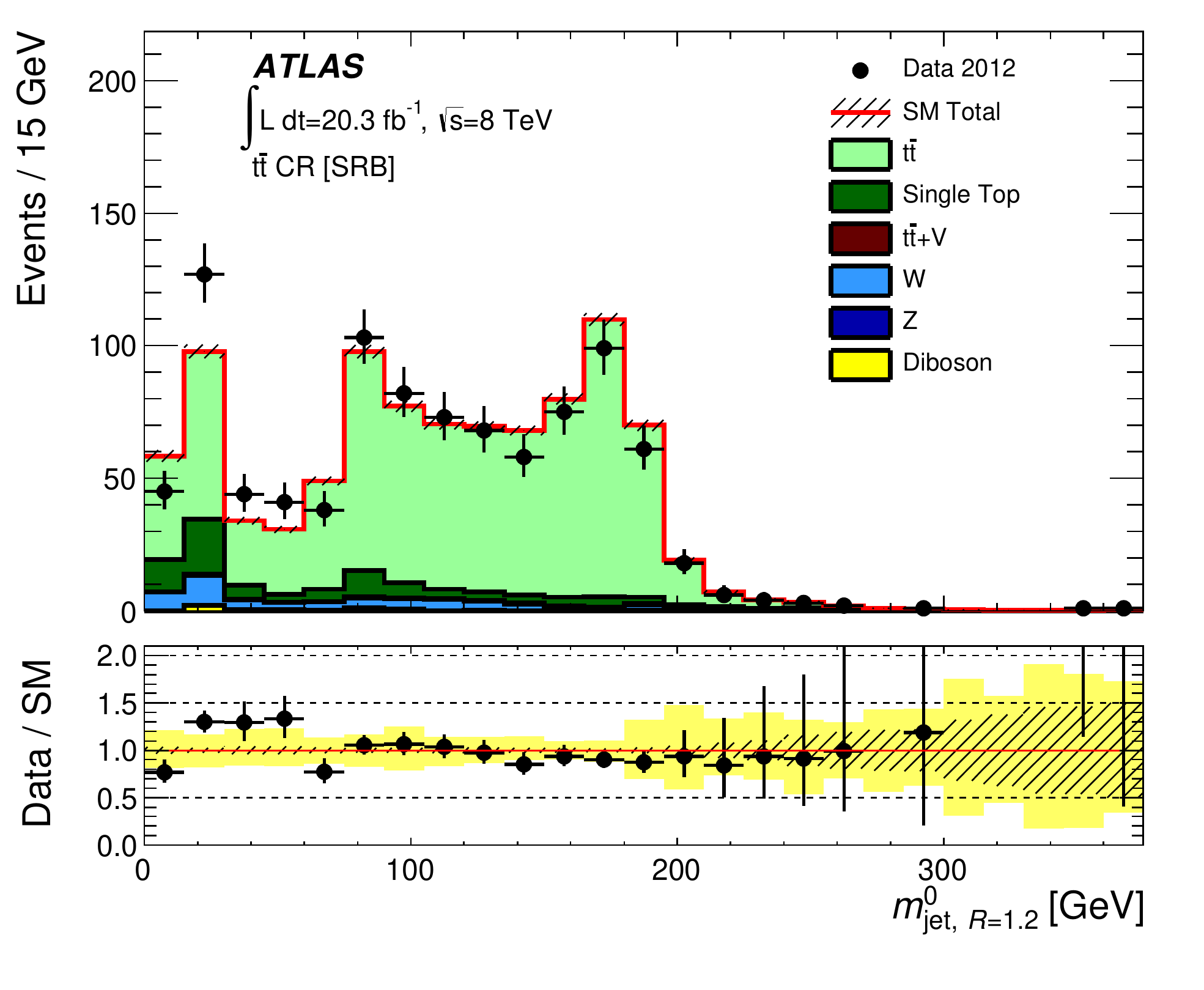}}
\subfigure[]{\includegraphics[width=0.45\textwidth,trim=0 40 0 0]{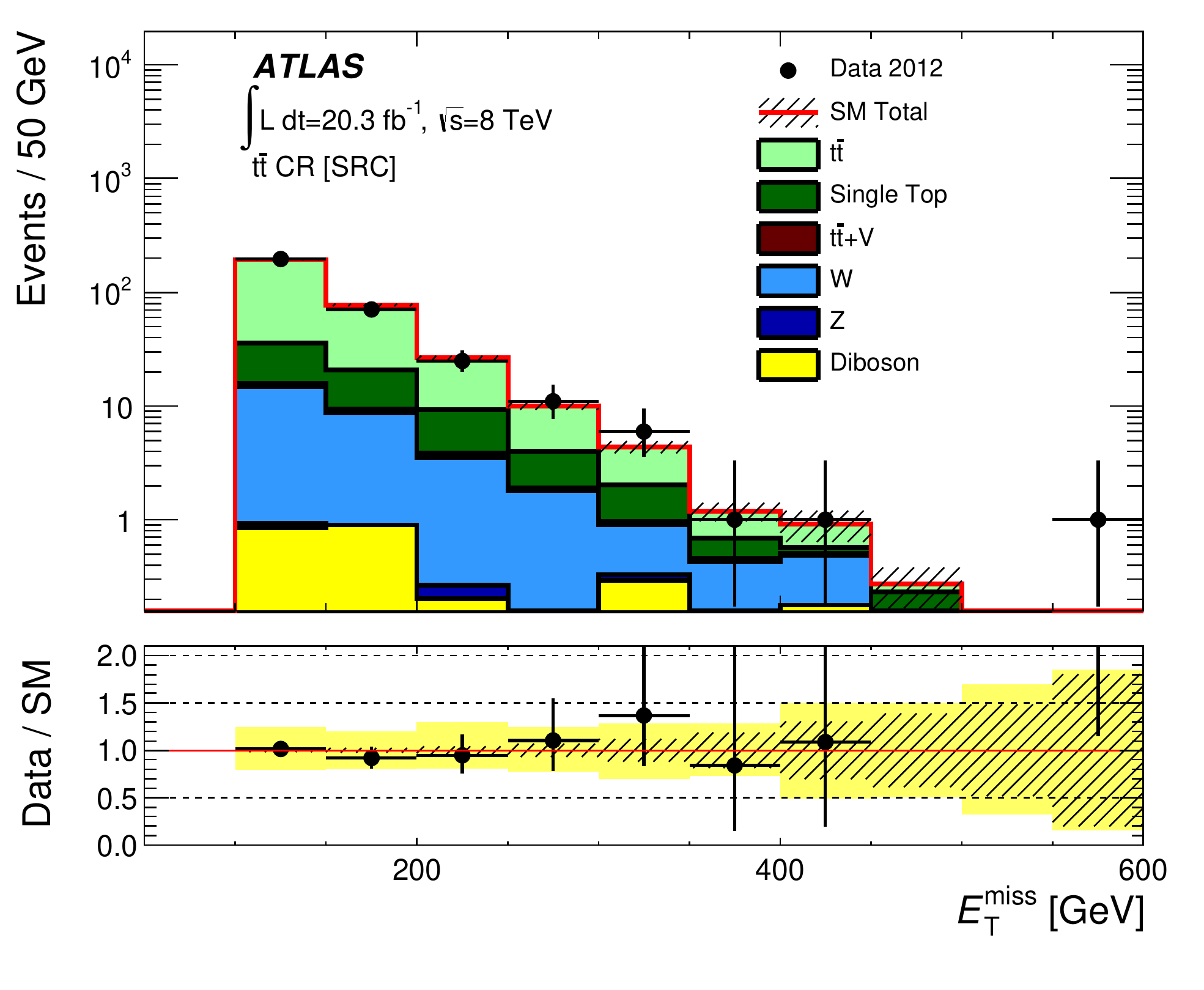}}
\subfigure[]{\includegraphics[width=0.45\textwidth,trim=0 40 0 0]{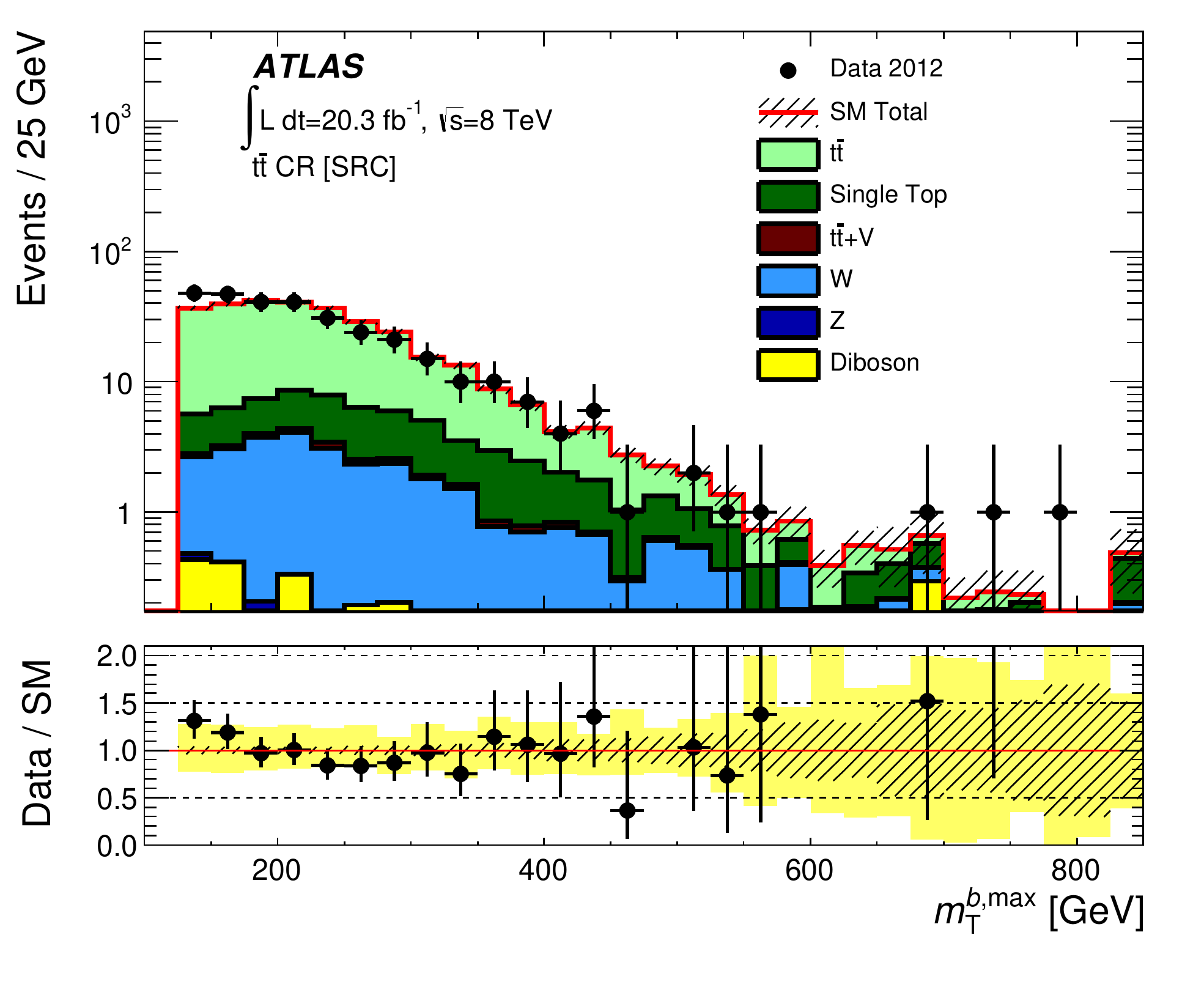}}
\caption{
  Distributions in the semileptonic \ttbar~control region of (a) \met\ and (b) \mtopcandzero\ for SRA, (c) \met\ and (d) \mantikttwelvezero\ for SRB, and (e) \met\ and (f) \mtbmetmaxdphi\ for SRC after the application of all selection requirements.
  All kinematic quantities were
  recalculated after treating the lepton as a jet.
  The stacked histograms show the Standard Model expectation,
  normalized using the factors summarized in table~\ref{tb:norm_after_fit}.
  The ``Data/SM'' plots show the ratio of data events to the total
  Standard Model expectation.  The rightmost bin includes all overflows.
  The hatched uncertainty band around  the
  Standard Model expectation shows the statistical uncertainty and
  the yellow band (shown only for the ``Data/SM'' plots)
  shows the combination of statistical and detector-related systematic
  uncertainties.
}
\label{fig:CRTop}
\end{center}
\end{figure}

The control region for the semileptonic \ttbar~background is defined
with requirements similar to those described in
section~\ref{sec:signalregiondefs} for the top squark signal candidates;
however, in order to enhance the contribution from the semileptonic
\ttbar\ process,  the events are required to be based on the
single-electron or single-muon trigger and to contain a single isolated electron or
muon as described in section~\ref{sec:objrec}.
The transverse mass of the lepton and \met\ is required to be
close to the $\Wboson$ mass, namely between $40$ and $120\,\GeV$.
Events with an additional isolated
electron or muon with transverse momentum greater than $10\,\GeV$ are rejected.
The identified lepton is then treated as a
non-$b$-tagged jet before imposing the jet and $b$-tagged jet
multiplicity requirements.  Several signal region requirements are relaxed (or not
applied at all) in order to have enough events, while keeping
systematic uncertainties related to extrapolating the background yield from the control region to the signal region under control.
Figure~\ref{fig:CRTop} compares several distributions in data and simulation in the
semileptonic \ttbar\ control region for each signal region; the background expectations are normalized using the factors summarized in table~\ref{tb:norm_after_fit}.

\afterpage{\clearpage}

\subsection{\boldmath \Wjets\ background}

For SRB only, a control region is defined for the normalization of the
\Wjets\ background.  The control region requirements are similar
to those in the \ttbar\ control region for SRB but a number of requirements are
changed in order to enhance the $\Wboson$ plus heavy-flavour jets process
over \ttbar.  The top mass asymmetry in eq.~\ref{eqn:topmassasym} is
restricted to the region $\topmassasym < 0.5$, the number of
$b$-tagged jets is required to be exactly one, and the mass of the
leading \antikt\ $R=1.2$ reclustered jet is required to be less than
$40\,\GeV$.  The full list of requirements can be found in table~\ref{tab:CRB}.  The lepton is treated as a non-$b$-tagged jet (in a similar fashion as the \ttbar\ control region) before imposing the jet and $b$-tagged jet multiplicity requirements. Figure~\ref{fig:CRW}
shows the \met\ and \mtbmetmindphi\ distributions in this control region; the background expectations are normalized using the factors summarized in table~\ref{tb:norm_after_fit}.

\subsection{\boldmath \Zjets\ background}
\label{sec:Zjets}

\begin{figure}[p]
\begin{center}
\subfigure[]{\includegraphics[width=0.48\textwidth,trim=0 40 0 10]{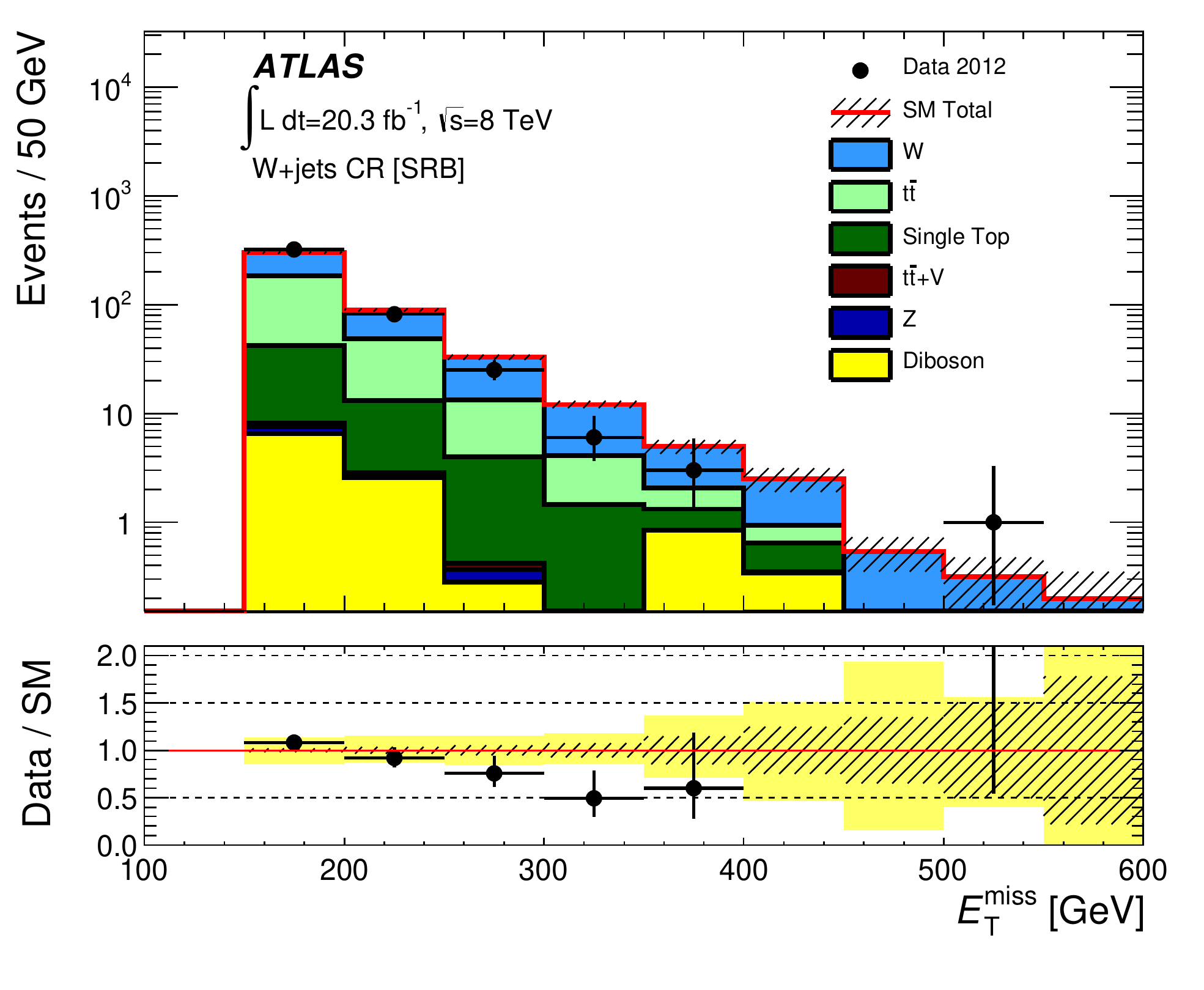}}
\subfigure[]{\includegraphics[width=0.48\textwidth,trim=0 40 0 10]{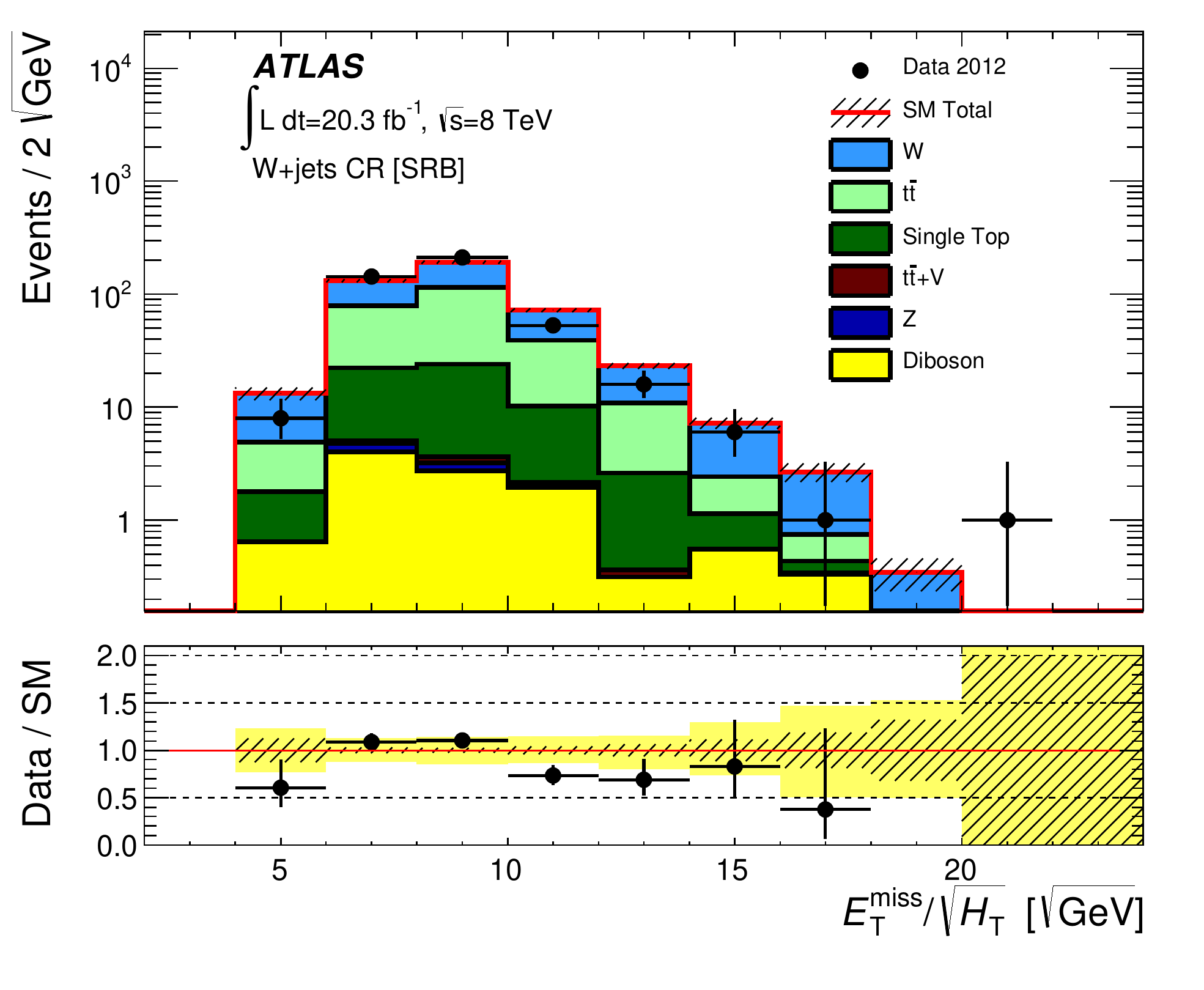}}
\caption{The (a) \met\ and (b) \htsig\ distributions in the \Wjets\ control region after all
 selection requirements. All kinematic quantities were
  recalculated after treating the lepton as a jet.
  The stacked histograms show the Standard Model expectation,
  normalized using the factors summarized in table~\ref{tb:norm_after_fit}.
  The ``Data/SM'' plots show the ratio of data events to the total
  Standard Model expectation.  The rightmost bin includes all overflows.
  The hatched uncertainty band around  the
  Standard Model expectation shows the statistical uncertainty and
  the yellow band (shown only for the ``Data/SM'' plots)
  shows the combination of statistical and detector-related systematic
  uncertainties.}
\label{fig:CRW}
\end{center}
\vskip 0.1in
\begin{center}
\subfigure[\label{fig:zCRa}]{\includegraphics[width=0.49\textwidth,trim=0 40 0 10]{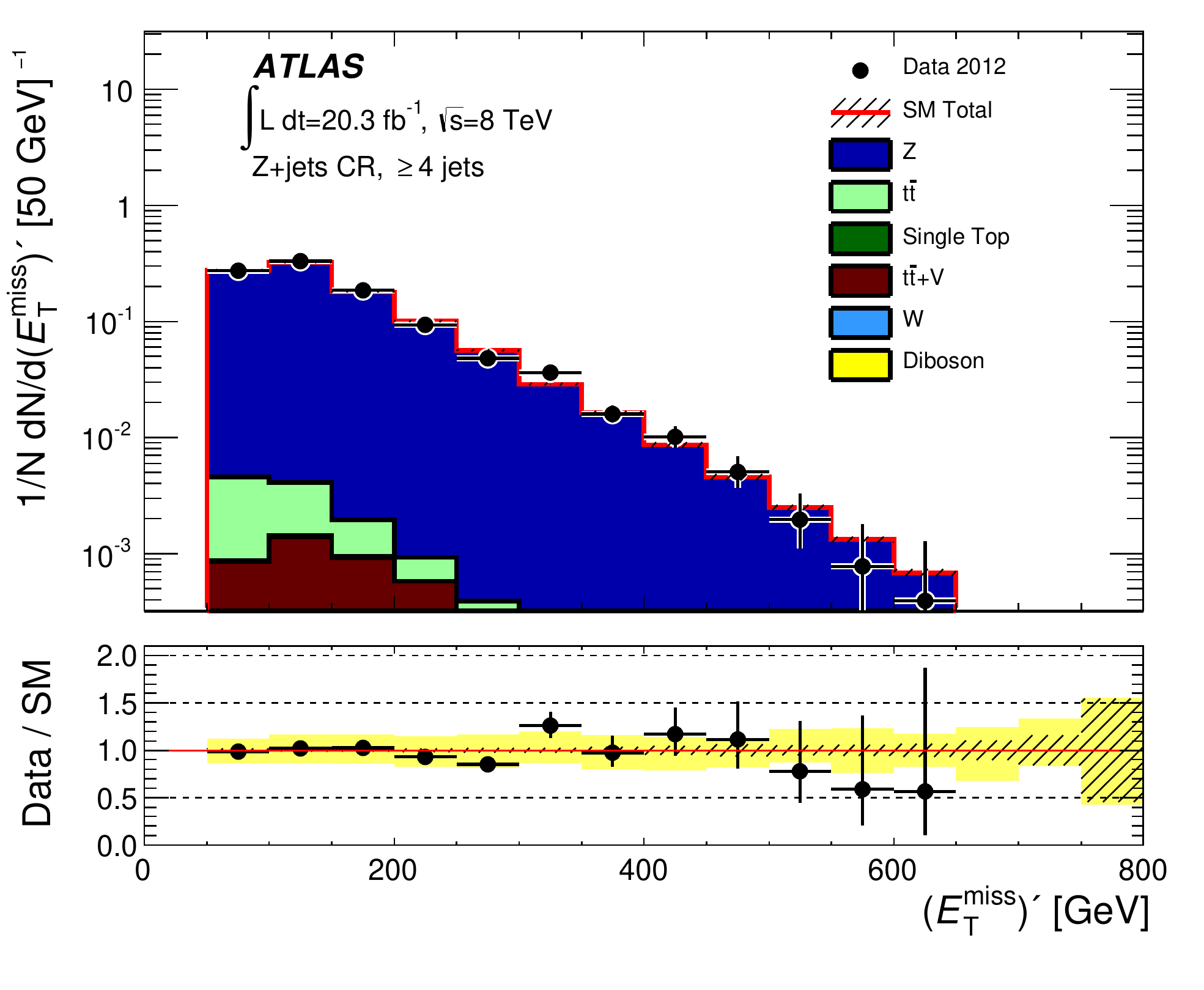}}%
\subfigure[\label{fig:zCRb}]{\includegraphics[width=0.49\textwidth,trim=0 40 0 10]{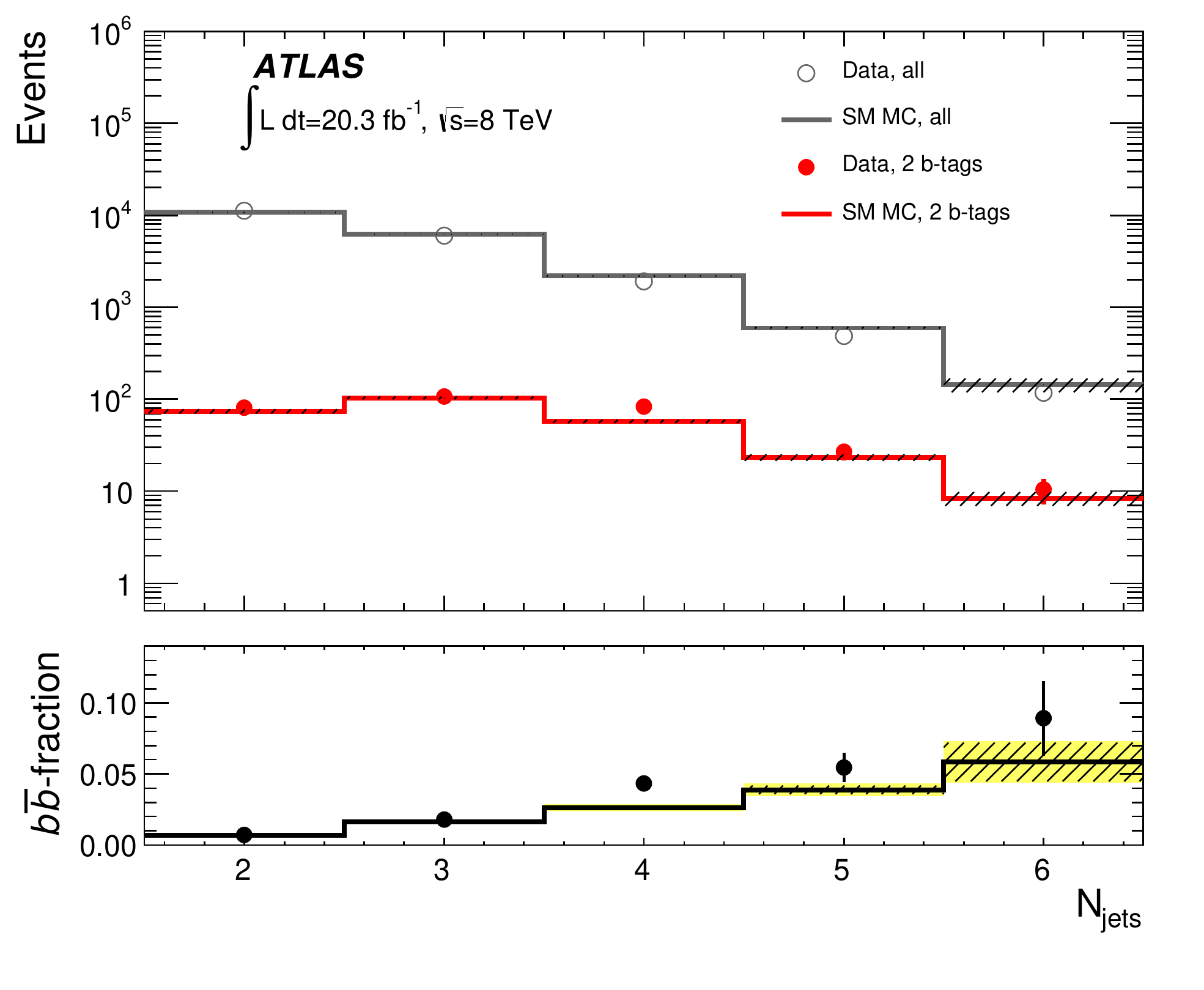}}
\caption{
  (a) The \METprime\ distribution in the \Zjets\ control region after all
  selection requirements for $\ge 4$ jets, normalized to unit area.
  The stacked histograms show the Standard Model
  expectations. The ``Data/SM'' plot shows the ratio of data events to the total
  Standard Model expectation.  The rightmost bin includes all overflows.
  The hatched uncertainty band around  the
  Standard Model expectation shows the statistical uncertainty and
  the yellow band (shown only for the ``Data/SM'' plots)
  shows the combination of statistical and detector-related systematic
  uncertainties. (b) The number of events in data and simulation as a function of jet multiplicity. The open (solid) points show all events (events with two or more $b$-tagged jets) in data, while the grey (red) line indicates the SM expectation for all events (events with two or more $b$-tagged jets). The $\bbbar$-fraction in data (simulation) is shown in the bottom panel, as indicated by the points (line). The hatched areas indicate MC statistical uncertainties and the yellow band (shown only for the $\bbbar$-fraction) includes the $b$-tagging systematic uncertainty. The rightmost bin includes all overflows. 
}
\label{fig:zCR}
\end{center}
\end{figure}

The control region for $Z(\rightarrow \nu\overline{\nu})$ plus heavy-flavour jets
background is based on a sample of
$Z (\rightarrow \ell\ell)$+jets events (where $\ell$ denotes
either an electron or muon).  The events are collected with the
single-lepton triggers.  Exactly two oppositely charged electrons or muons
are required; the higher-\pT\ lepton must satisfy $\ptl > 25\,\GeV$ and the lower-\pT\ lepton must satisfy $\ptltwo > 10\,\GeV$. The invariant mass of the dilepton pair (\mll) is
required to be between 86 and 96~\GeV.   To reduce the
\ttbar~contamination, the events are required to have $\met < 50$~\GeV.
The reconstructed dileptons are then removed from the event to mimic the $Z(\rightarrow \nu\overline{\nu})$ decay and the vector sum of their momenta is added to
the \ptmiss; the events are required to have a recalculated $\METprime > 70$~\GeV.
No requirements are made on the number of $b$-tagged jets.  Monte
Carlo studies indicate that the \met\ in $Z$+jets events (with
$Z(\rightarrow \nu\overline{\nu})$) is completely dominated by the
neutrinos from the $Z$ decay, independent of the presence of $b$-tagged jets.
The shape
of the $\METprime$ distribution, comparing data to simulation,
is shown in figure~\ref{fig:zCRa}; the normalization of the simulation
for the estimation of the background is described below.

The fraction of events containing two or
more $b$-tagged jets (henceforth denoted the $\bbbar$-fraction)
is found in simulation and data to scale linearly with the
number of jets in the event as shown in figure~\ref{fig:zCRb}, as expected when the primary source of
$b$-tagged jet pairs is gluon radiation followed by
splitting.  The number of events in the \Zjets\ control region
is corrected in each jet multiplicity bin in simulation and data by
the $\bbbar$-fraction. The $\bbbar$-fraction is fit to a
linear function of jet multiplicity, starting at a jet multiplicity of
two, in order to improve the statistical accuracy for high jet-multiplicity $\Zboson$ events.  After correcting
for the $\bbbar$-fraction, the \Zjets\ simulation is
normalized to the data in the control region.

\clearpage

\subsection{\boldmath Multijet background}

\begin{figure}[b]
\begin{center}
\subfigure[]{\includegraphics[width=0.49\textwidth,trim=0 40 0 0]{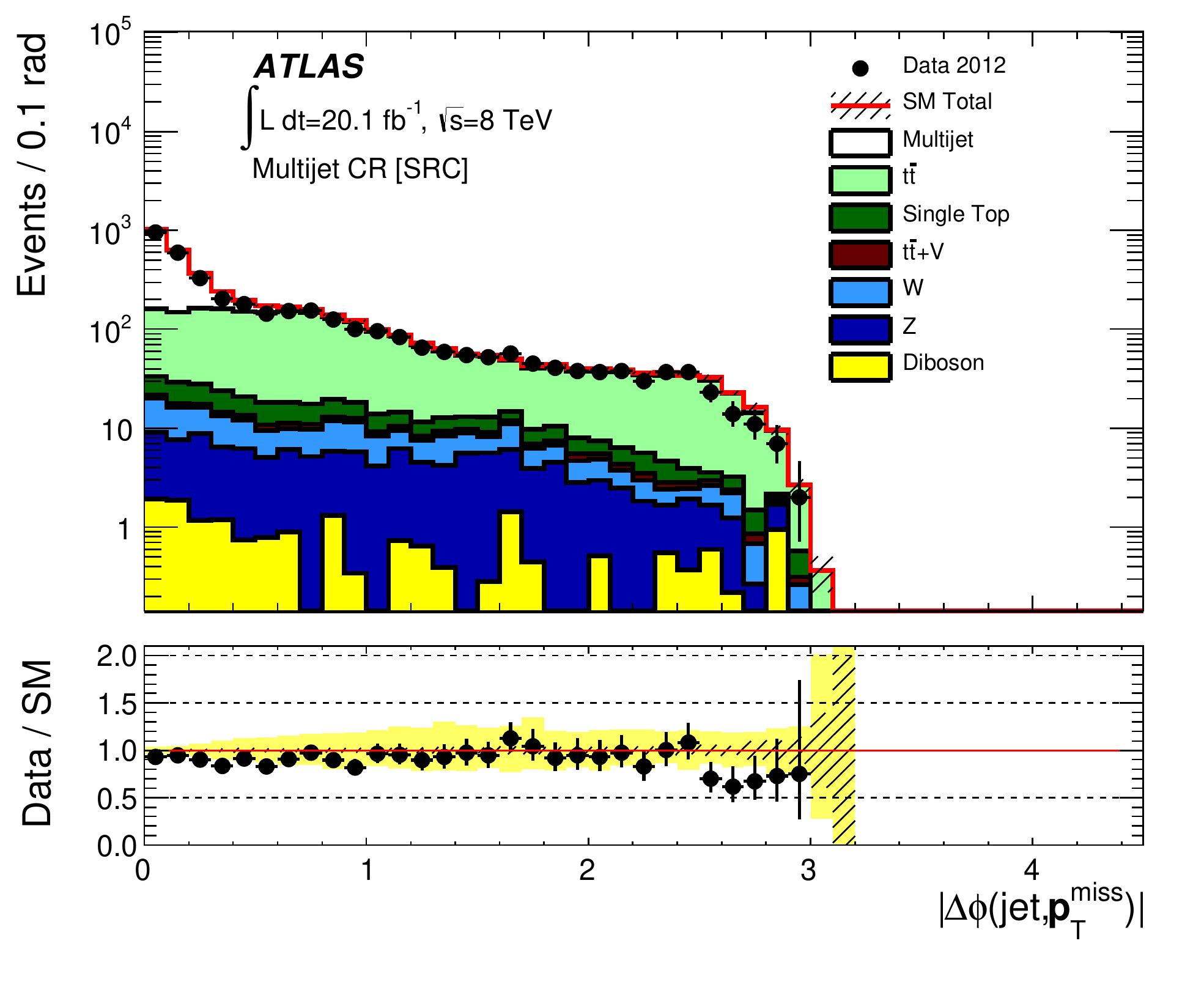}}
\subfigure[]{\includegraphics[width=0.49\textwidth,trim=0 40 0 0]{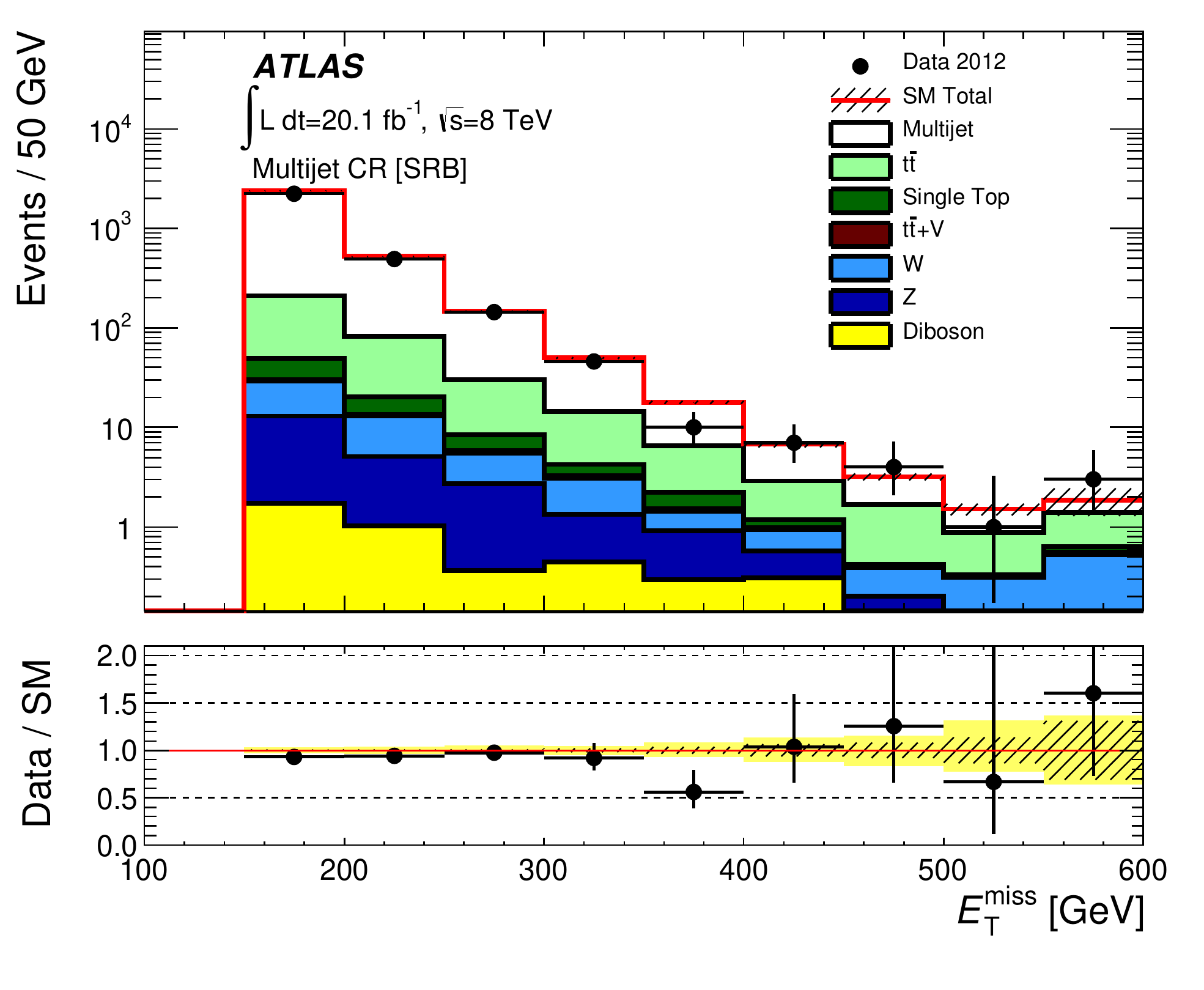}}%
\caption{ Distributions of (a) \dphijetmet\ in the multijet control region
  for SRC and (b) \met\ in the multijet control region for SRB.
  The stacked histograms show the Standard Model
  expectations, normalized using the factors summarized in table~\ref{tb:norm_after_fit}.   The ``Data/SM'' plots show the ratio of data events to the total Standard Model expectation.  The rightmost bin includes all overflows.
  The hatched uncertainty band around  the
  Standard Model expectation shows the statistical uncertainty and
  the yellow band (shown only for the ``Data/SM'' plots)
  shows the combination of statistical and detector-related systematic
  uncertainties.
}
\label{fig:QCDCR}
\end{center}
\end{figure}

The multijet (including fully hadronic \ttbar) background is evaluated
using the jet-smearing technique described in ref.~\cite{:2012rz}.
The basic concept is to take a sample of well-measured multijet events
in the data (based on low values of $\met/\sqrt{\sumet}$, where
$\sumet$ is the scalar sum of the transverse energy in the event
recorded in the calorimeter systems) and to smear the jet momentum and
$\phi$ direction with jet response functions, determined with \pythiaeight~\cite{Sjostrand:2007gs} and corrected with data,
separately for light-flavour and heavy-flavour jets.  This sample
of smeared events is normalized to the data in a control region enriched in
multijet events.  The same jet multiplicity requirements are applied
as in the signal regions, including
the requirement of two or more $b$-tagged jets.  The calorimeter-based
(track-based) \met\ is required to be greater than 150 (30)~\GeV.
The signal region
requirement on the
azimuthal separation $\dphijetmet$ is inverted to enhance the population of
mis-measured multijet events. Figure~\ref{fig:QCDCR} compares 
the \dphijetmet\ and \met\ distributions in data with expectations in the multijet control
region, after normalizing the smeared event sample to the data in the
region $\dphijetmet< 0.1$.
The multijet and fully hadronic \ttbar\ background contributions in the signal
regions are evaluated by applying all the signal region
requirements to the normalized smeared event sample.

\clearpage

\subsection{Simultaneous fit to determine SM background}
\label{sec:fit}

The observed numbers of events in the various control regions are
included in a profile likelihood fit to determine the SM background
estimates in each signal region. This procedure takes common
systematic uncertainties (discussed in detail in
section~\ref{sec:systematics}) between the control and signal regions and
their correlations into account; they are treated as nuisance
parameters in the fit and are modelled by Gaussian probability density
functions. For SRA and SRC, the free parameters in the fit are the
overall normalizations of the \ttbar, \Zjets, and multijet
background.\footnote{As the smeared events for the multijet background
  are first normalized ``by hand'' outside the fit to the data in the multijet control
  region (after correcting for non-multijet background), the
  additional normalization factor from the fit is not listed in
  table~\ref{tb:norm_after_fit}.} For SRB the normalization of the
\Wjets\ background is also a free parameter. The contributions from
all other background processes are fixed at the values expected from
the simulation, using the most accurate theoretical cross sections
available, as described in section \ref{sec:montecarlo}.  The fit to the
control regions yields normalization factors for each background
source; these are summarized in table~\ref{tb:norm_after_fit}. These
normalization factors are correlated between SRB and SRC due to the
overlap of the respective control regions; however, these correlations
do not affect the final results since SRB and SRC are never
statistically combined, as described in section \ref{sec:results}.
The resulting yields in the control regions (before and after the fit) are summarized in table~\ref{tb:CRSummary}. The contamination due to potential signal events in the control regions is negligible (at most a few per cent).

\begin{table}[!tb]
  \caption{Normalization of the $\ttbar$, \Wjets, and \Zjets\ SM
    background as obtained from the background fits for SRA, SRB and
    SRC.
  }
  \begin{center}
    \begin{tabular}{cccc}  \hline \hline
      Background Source      & SRA & SRB & SRC  \\ \hline \hline
      \\ [-2ex]
      $\ttbar$  & $1.24 \pm 0.13 $ &  $1.00^{+0.10}_{-0.05}$&  $1.07 \pm 0.11$      \\ [1ex]
      \Wjets\ &           --              &  $1.0 \pm 0.4$ &    --     \\ [1ex]
      \Zjets\ & $0.94^{+0.16}_{-0.15}$  &  $1.07 \pm 0.07$      &$1.07 \pm 0.07$     \\ [1ex] \hline \hline
    \end{tabular}
  \end{center}
  \label{tb:norm_after_fit}
\end{table}%

\input{CRSummary}

\afterpage{\clearpage}

The background estimates are validated by predicting the background in
dedicated regions and comparing to
observation. The validation regions are designed to be orthogonal to
the control and signal regions while retaining 
kinematics and event composition close to the SRs but
with little contribution from signal in any of the models considered.
For SRA, two validation regions are defined.  In the
first region (VRA1), all of the requirements for SRA1 are applied except for
those on the top mass, and the $\tau$ veto is inverted.  In the second
region (VRA2), all of the SRA1 requirements are applied except for those on
the top mass, and \mtbmetmindphi\ is required to be between 125 and
$175\,\GeV$.  For SRB, the validation region (VRB) is formed from the logical
OR of SRB1 and SRB2 except that the \met\ requirement is loosened to
$\met > 150$ \GeV, \mtbmetmindphi\ is required to be between 100 and
$175\,\GeV$, and none of the other transverse mass requirements are
applied.   For SRC, the validation regions consist of all the SRC1
requirements except that \mtbmetmindphi\ is required to be between 150
and $185\,\GeV$ and \mtbmetmaxdphi\ is required to be greater than $125\,\GeV$, and the $\tau$ veto is either inverted (VRC1) or applied (VRC2). All five validation regions are dominated by \ttbar\ background. The background yield in each
validation region, predicted from the fit to the control regions, is consistent with the observed number of events to within one standard deviation; these results are summarized in table~\ref{tb:VRSummary}.

\input{VRSummary}
\clearpage

%% file: CRSummary.tex
\begin{landscape}
\begin{table}[hp]
\begin{center}
\caption{Event yields in the control regions, before and after the profile likelihood fit. The uncertainties quoted include statistical and systematic contributions. Smaller background contributions from single-top, $\ttbar+\Wboson/\Zboson$, and diboson production are included in ``Others". \label{tb:CRSummary}}
\begin{tabular}{lcccccccccc} \hline\hline
\multicolumn{11}{l}{ } \\ [-1.5ex]
 & \multicolumn{3}{c}{CRs for SRA} & \multicolumn{4}{c}{CRs for SRB} & \multicolumn{3}{c}{CRs for SRC} \\ [-1.5ex]
\multicolumn{11}{l}{ } \\ [-1.5ex]  \cline{2-11}
\multicolumn{11}{l}{ } \\ [-1.5ex]
           &  \multicolumn{1}{c}{\ttbar\ }            & \multicolumn{1}{c}{\Zjets\ }    & \multicolumn{1}{c}{Multijets } &  \multicolumn{1}{c}{\ttbar\ }       & \multicolumn{1}{c}{\Wjets\ }    & \multicolumn{1}{c}{\Zjets\ }    & \multicolumn{1}{c}{Multijets } &  \multicolumn{1}{c}{\ttbar\ }            & \multicolumn{1}{c}{\Zjets\ }    & \multicolumn{1}{c}{Multijets } \\ [0.5ex] \hline \hline
\multicolumn{11}{l}{ } \\ [-1.3ex]
\multicolumn{11}{l}{Observed  events} \\ [0.7ex] \hline
\multicolumn{11}{l}{ } \\ [-1.5ex]
     & $247$              & $101$              & $592$      & $950$              & $440$              & $499$              & $2082$         & $313$              & $499$              & $1017$ \\ [0.5ex] \hline\hline
\multicolumn{11}{l}{ } \\ [-1.3ex]
\multicolumn{11}{l}{Fitted background events} \\ [0.7ex] \hline
\multicolumn{11}{l}{ } \\ [-1.5ex]
Total SM\phantom{hi}        & $\numRF{246.81}{3} \pm \numRF{15.84}{2}$          & $\numRF{100.98}{3} \pm \numRF{10.13}{2}$          & $\numRF{592.75}{3} \pm \numRF{26.72}{2}$    & $\numRF{949.91}{2} \pm \numRF{37.89}{1}$          & $\numRF{440.05}{3} \pm \numRF{26.86}{2}$          & $\numRF{498.98}{3} \pm \numRF{22.33}{2}$          & $\numRF{2082.08}{4} \pm \numRF{48.23}{2}$  & $\numRF{312.70}{3} \pm \numRF{17.71}{2}$          & $\numRF{499.15}{3} \pm \numRF{22.40}{2}$          & $\numRF{1017.51}{4} \pm \numRF{34.22}{2}$    \\ [0.5ex]
        $\ttbar$         & $\numRF{197.37}{3} \pm \numRF{20.69}{2}$          & $\numRF{12.62}{3} \pm \numRF{3.01}{2}$          & $\numRF{109.27}{3} \pm \numRF{23.19}{2}$  & $\numRF{801.70}{2} \pm \numRF{51.80}{1}$          & $\numRF{189.20}{3} \pm \numRF{25.43}{2}$          & $\numRF{46.10}{2} \pm \numRF{7.30}{1}$          & $\numRF{139.96}{3} \pm \numRF{13.94}{2}$  & $\numRF{239.26}{3} \pm \numRF{24.43}{2}$          & $\numRF{49.25}{2} \pm \numRF{12.16}{2}$          & $\numRF{115.18}{3} \pm \numRF{23.05}{2}$   \\
        $\Zjets$         & $\numRF{0.28}{2} \pm \numRF{0.19}{2}$          & $\numRF{73.11}{2} \pm \numRF{11.08}{2}$          & $\numRF{2.50}{2} \pm \numRF{0.61}{1}$     & $\numRF{0.59}{2} \pm \numRF{0.16}{2}$          & $\numRF{1.40}{3} \pm \numRF{0.25}{2}$          & $\numRF{422.97}{3} \pm \numRF{25.33}{2}$          & $\numRF{11.65}{3} \pm \numRF{1.55}{2}$  & $\numRF{0.18}{2} \pm \numRF{0.07}{1}$          & $\numRF{420.03}{3} \pm \numRF{25.75}{2}$          & $\numRF{6.66}{2} \pm \numRF{0.93}{1}$  \\
        $\Wjets$         & $\numRF{19.83}{2} \pm \numRF{9.20}{1}$          & \multicolumn{1}{c}{--}          & $\numRF{4.52}{2} \pm \numRF{2.15}{2}$  & $\numRF{54.43}{2} \pm \numRF{20.24}{2}$          & $\numRF{188.73}{2} \pm \numRF{40.50}{1}$          & \multicolumn{1}{c}{--}          & $\numRF{17.80}{2} \pm \numRF{7.26}{1}$   & $\numRF{28.39}{2} \pm \numRF{11.86}{2}$          & \multicolumn{1}{c}{--}          & $\numRF{8.94}{1} \pm \numRF{4.48}{1}$   \\
        Multijets         & \multicolumn{1}{c}{--}           & \multicolumn{1}{c}{--}           & $\numRF{464.70}{2} \pm \numRF{35.53}{1}$  & \multicolumn{1}{c}{--}          & \multicolumn{1}{c}{--}          & \multicolumn{1}{c}{--}          & $\numRF{1889.94}{3} \pm \numRF{50.52}{1}$  & \multicolumn{1}{c}{--}          & \multicolumn{1}{c}{--}          & $\numRF{874.12}{2} \pm \numRF{41.44}{1}$ \\ 
        Others         & $\numRF{29.33}{2} \pm \numRF{4.47}{1}$          & $\numRF{15.25}{2} \pm \numRF{3.80}{1}$          & $\numRF{11.75}{3} \pm \numRF{1.56}{2}$ & $\numRF{93.19}{2} \pm \numRF{13.06}{2}$          & $\numRF{60.72}{2} \pm \numRF{8.45}{1}$          & $\numRF{29.91}{2} \pm \numRF{9.92}{1}$          & $\numRF{22.72}{3} \pm \numRF{3.03}{2}$ & $\numRF{44.87}{2} \pm \numRF{7.37}{1}$          & $\numRF{29.87}{2} \pm \numRF{7.12}{1}$          & $\numRF{12.61}{3} \pm \numRF{1.60}{2}$  \\ \hline\hline
\multicolumn{11}{l}{ } \\ [-1.3ex]
\multicolumn{11}{l}{Expected events (before fit)} \\ [0.7ex] \hline 
\multicolumn{11}{l}{ } \\ [-1.5ex]
        $\ttbar$         & $\numRF{159.21}{3}$          & $\numRF{10.20}{3}$          & $\numRF{88.29}{2}$   & $\numRF{803.10}{2}$          & $\numRF{189.54}{3}$          & $\numRF{46.18}{2}$          & $\numRF{140.18}{3}$ & $\numRF{224.03}{3}$          & $\numRF{46.18}{2}$          & $\numRF{107.94}{3}$  \\
        $\Zjets$         & $\numRF{0.31}{2}$          & $\numRF{77.81}{2}$          & $\numRF{2.66}{2}$     & $\numRF{0.55}{2}$          & $\numRF{1.30}{3}$          & $\numRF{394.29}{3}$          & $\numRF{10.86}{3}$  & $\numRF{0.17}{2}$          & $\numRF{394.29}{3}$          & $\numRF{6.25}{2}$  \\
        $\Wjets$         & $\numRF{19.84}{2}$          & \multicolumn{1}{c}{--}          & $\numRF{4.51}{2}$     & $\numRF{52.24}{2}$          & $\numRF{181.10}{2}$          & \multicolumn{1}{c}{--}          & $\numRF{17.07}{2}$  & $\numRF{28.16}{2}$          & \multicolumn{1}{c}{--}          & $\numRF{8.88}{1}$ \\
        Multijets         & \multicolumn{1}{c}{--}          & \multicolumn{1}{c}{--}          & $\numRF{460.03}{2}$   & \multicolumn{1}{c}{--}          & \multicolumn{1}{c}{--}          & \multicolumn{1}{c}{--}          & $\numRF{2085.43}{3}$  & \multicolumn{1}{c}{--}          & \multicolumn{1}{c}{--}          & $\numRF{867.14}{2}$   \\ 
        Others         & $\numRF{29.28}{2}$          & $\numRF{15.25}{2}$          & $\numRF{11.74}{3}$   & $\numRF{93.19}{2}$          & $\numRF{60.71}{2}$          & $\numRF{29.91}{2}$          & $\numRF{22.72}{3}$  & $\numRF{44.85}{2}$          & $\numRF{29.91}{2}$          & $\numRF{12.61}{3}$    \\ \hline \hline
\end{tabular}
\end{center}
\end{table}
\end{landscape}

%% file: VRSummary.tex
\begin{table}[!b]
\begin{center}
\caption{Event yields in the validation regions compared to the background estimates obtained from the profile likelihood fit. The requirements for VRA1--C2 are described in the text. Statistical and systematic uncertainties in the number of fitted background events are shown. Smaller background contributions from multijets, single-top, $\ttbar+\Wboson/\Zboson$, and diboson production are included in ``Others". \label{tb:VRSummary}}
\begin{tabular}{lccccc} \hline\hline
\multicolumn{5}{l}{ } \\ [-1.5ex]
           &  \multicolumn{1}{c}{VRA1}     & \multicolumn{1}{c}{VRA2}    & \multicolumn{1}{c}{VRB}    &  \multicolumn{1}{c}{VRC1}            & \multicolumn{1}{c}{VRC2} \\ [0.5ex] \hline \hline
\multicolumn{5}{l}{ } \\ [-1.3ex]
\multicolumn{5}{l}{Observed  events} \\ [0.7ex] \hline
\multicolumn{5}{l}{ } \\ [-1.5ex]
          & $158$              & $51$              & $69$              & $103$              & $24$              \\ [0.5ex] \hline \hline
\multicolumn{5}{l}{ } \\ [-1.3ex]
\multicolumn{5}{l}{Fitted background events} \\ [0.7ex] \hline
\multicolumn{5}{l}{ } \\ [-1.5ex]
Total SM         & $\numRF{188.67}{3} \pm \numRF{26.41}{2}$          & $\numRF{49.51}{2} \pm \numRF{5.75}{1}$          & $\numRF{70.38}{2} \pm \numRF{19.01}{2}$          & $\numRF{110.16}{3} \pm \numRF{12.18}{2}$          & $\numRF{21.11}{3} \pm \numRF{2.93}{2}$   \\ [0.5ex]
        $\ttbar$         & $\numRF{170.11}{3} \pm \numRF{27.25}{2}$          & $\numRF{33.89}{2} \pm \numRF{6.95}{1}$          & $\numRF{59.59}{2} \pm \numRF{18.92}{2}$          & $\numRF{92.64}{2} \pm \numRF{11.95}{2}$          & $\numRF{17.34}{3} \pm \numRF{2.84}{2}$  \\
        $\Zjets$         & $\numRF{4.01}{2} \pm \numRF{1.14}{2}$          & $\numRF{1.50}{2} \pm \numRF{0.41}{1}$         & $\numRF{1.51}{2} \pm \numRF{0.46}{1}$          & $\numRF{6.94}{2} \pm \numRF{1.54}{2}$          & $\numRF{0.24}{2} \pm \numRF{0.20}{2}$   \\
        $\Wjets$         & $\numRF{2.78}{2} \pm \numRF{1.23}{2}$          & $\numRF{4.83}{2} \pm \numRF{2.18}{2}$          & $\numRF{2.06}{2} \pm \numRF{1.38}{2}$          & $\numRF{3.86}{2} \pm \numRF{1.75}{2}$          & $\numRF{1.11}{2} \pm \numRF{0.52}{1}$ \\
        Others         & $\numRF{11.76}{3} \pm \numRF{3.07}{2}$          & $\numRF{9.08}{2} \pm \numRF{2.17}{2}$          & $\numRF{7.21}{2} \pm \numRF{2.48}{2}$          & $\numRF{6.72}{2} \pm \numRF{2.04}{2}$ & $\numRF{2.41}{2} \pm \numRF{0.73}{1}$  \\ \hline\hline
\end{tabular}
\end{center}
\end{table}

%% file: systematics.tex
\label{sec:systematics}

Systematic uncertainties in the SM background estimates and signal
expectations are evaluated and included in the profile likelihood fit
described in section~\ref{sec:fit}.  The impact of each source of
systematic uncertainty is quantified as a percentage of the total
background  estimate (signal expectation) after propagating the
uncertainty from the relevant nuisance parameter, keeping all other fit
parameters fixed. All correlations with the other parameters of interest are taken into account. 

The main sources of detector-related systematic uncertainties in the SM background estimates are the jet energy resolution (JER) and jet energy scale (JES). These jet reconstruction uncertainties are propagated
to all quantities that depend on the jet energies such as the
reconstructed top mass
and the \met. Additional uncertainties in \met\ that arise from energy
deposits unassociated with reconstructed objects are also included.
The impact of these uncertainties is mitigated by the
normalization of the dominant SM background contributions in the kinematically
similar control regions. The JER uncertainty is derived from in-situ
measurements of the jet response asymmetry in dijet
events~\cite{Aad:2011he}; the effect of this uncertainty on the background estimates in the signal regions ranges from 
$6$--$15\%$ in SRA,
$16\%$ in SRB, and
$3$--$6\%$ in SRC.
The JES uncertainty is determined using the techniques described in refs.~\cite{Aad:2011he,JES:SingleHadron}; the uncertainties are determined in bins of jet $\eta$ and \pT\ and depend on jet flavour and the number of primary vertices in an event. 
The effect of the JES uncertainty on the background estimate ranges from $5$--$9\%$ in SRA, 6\% in SRB, and
$8$--$11\%$ in SRC.  Other uncertainties arising from the simulation of
$b$-tagging, pileup, the $\tau$ veto, and \mettrk\ are negligible by comparison.

A $2.8\%$ uncertainty in the luminosity determination~\cite{Aad:2013ucp}
is included for all signal and background MC simulations.

Theoretical uncertainties in the modelling of the SM background are
evaluated; their impact is mitigated by the use of control
regions to normalize the background contributions. For the \ttbar\ background,
uncertainties due to the choice of parton shower (\pythia\ vs.\ \herwig~\cite{Corcella:2002jc}+\jimmy~\cite{jimmy}), the renormalization and factorization scales
(each varied up and down by a factor of two), the amount of initial-
and final-state radiation (using \acer\ samples with differing parton
shower settings, constrained by measurement \cite{ATLAS:2012al}), and
the PDF uncertainties (derived from the
envelope of variations of the CT10 PDF summed in quadrature with the
difference with respect to the
HERA PDFs~\cite{Aaron:2009aa}) are evaluated.
The resulting uncertainties in the
total background yields are less than $10\%$ in SRA and SRC (the signal regions
with an appreciable contribution from \ttbar\ production). For
$\ttbar+\Wboson/\Zboson$ background, the theoretical uncertainty
is dominated by  the 22\% uncertainty
~\cite{Campbell:2012dh,Garzelli:2012bn}
on the NLO cross section. Additional variations considered include the choice of
renormalization and factorization scales (each varied up and down by a
factor of two), the amount of initial- and final-state radiation
(using simulated \madgraph\ samples), and the matching scale at which
additional jets from the matrix element are distinguished from those
generated by the parton shower. Finally the uncertainty arising from
the use of a finite number of additional partons in the matrix element
is assessed by comparing event yields from samples with
one versus two additional partons in the matrix element. The
resulting impact on the total background yields range from $3$--$6\%$ in SRA, $6\%$
in SRB, and is at the per cent level in SRC.

Systematic uncertainties in the modelling of the \Wjets\ background are evaluated with respect to the choice of renormalization and
factorization scales (each varied up and down by a factor of two), the
matching scale, the PDF uncertainties, and event generation with a finite
number of partons (comparing samples with four versus five additional
partons in the matrix element). An uncertainty of $38\%$ is applied to the
$\Wboson+$ heavy flavour production cross section; this is derived from
the measurement in ref.~\cite{ATLAS:2013Wb} and extrapolated to events
with at least five jets. An additional
uncertainty of $20\%$ is applied to the \Wjets\ control region for SRB
to account for  differing fractions of $\Wboson+c$
vs.\ $\Wboson+\bbbar/\ccbar$ in the control region compared to the
signal region. For the \Zjets\ background, the uncertainties due to the
choice of generator (\alpgen\ vs.\ \sherpa), the PDF
uncertainties, and event generation
with a finite number of partons are evaluated. The resulting impact on
the total background yields from all of the above-mentioned
\Wjets\ (\Zjets) theoretical uncertainties
are $1$--$2\%$ ($1$--$2\%$) in SRA, $10\%$ ($9\%$) in SRB, and $5\%$
($4$--$5\%$) in SRC. An additional systematic uncertainty of $17\%$ is
assigned to the background yield from the uncertainty in the linear fit to the $\bbbar$-fraction in
the \Zjets\ control region described in section~\ref{sec:Zjets}.

The single-top background is dominated by the $Wt$ subprocess; the
cross-section uncertainty is taken from ref.~\cite{Kidonakis:2010ux}.
Additional uncertainties are evaluated for the
selection of generator (\mcatnlo\ vs.\ \powhegbox), parton shower,
initial- and final-state radiation,
and PDF choices. Finally, the effect of the interference between
single-top and \ttbar\ production is evaluated from a comparison of
the sum of  \ttbar\ and single-top background from \powhegbox\ with the
background from an \acer\ sample of the inclusive
$\Wboson b \Wboson b$ final state.  The resulting
uncertainty in the total background estimate ranges between 1\% and 5\%,
depending on the signal region.
An uncertainty of $50\%$ is assigned to diboson production, resulting
in uncertainties in the total background yield of $< 1$\% in all
signal regions. An uncertainty of $100\%$ is assigned to the small
multijet background, with negligible impact on the total background
uncertainty. 

The theoretical uncertainties in the top squark production cross
section include uncertainties due to the chosen PDF, factorization and
renormalization scales, and strong coupling constant variations. These
uncertainties are not included as nuisance parameters in the fit;
instead, their impact is shown explicitly in the results.  In
contrast, systematic
uncertainties in the signal acceptance are included as nuisance
parameters in the fit.  The impact on the signal acceptance
of variations in the PDF, factorization and
renormalization scales, and strong coupling constant  is found to be
negligible.
The systematic uncertainty in the signal acceptance due to the
modelling of initial-state radiation is negligible in the region where
this analysis has sensitivity to top squark production.
Detector-related systematic uncertainties in
the signal acceptance are dominated by the JES ($4$--$16\%$ effect on the
signal yield in SRA, $3\%$
in SRB, $4$--$10\%$ in SRC), $b$-tagging ($7$--$8\%$ in all signal
regions) uncertainties and JER ($2$--$10\%$ in SRA, $10\%$ in SRB and $2$--$3\%$
in SRC).
All detector-related uncertainties are assumed to be fully
correlated with those of the background.

%% file: results.tex
\label{sec:results}

\begin{figure}[p] 
\begin{center}
\subfigure[]{\includegraphics[width=0.45\textwidth,trim=0 20 0 0]{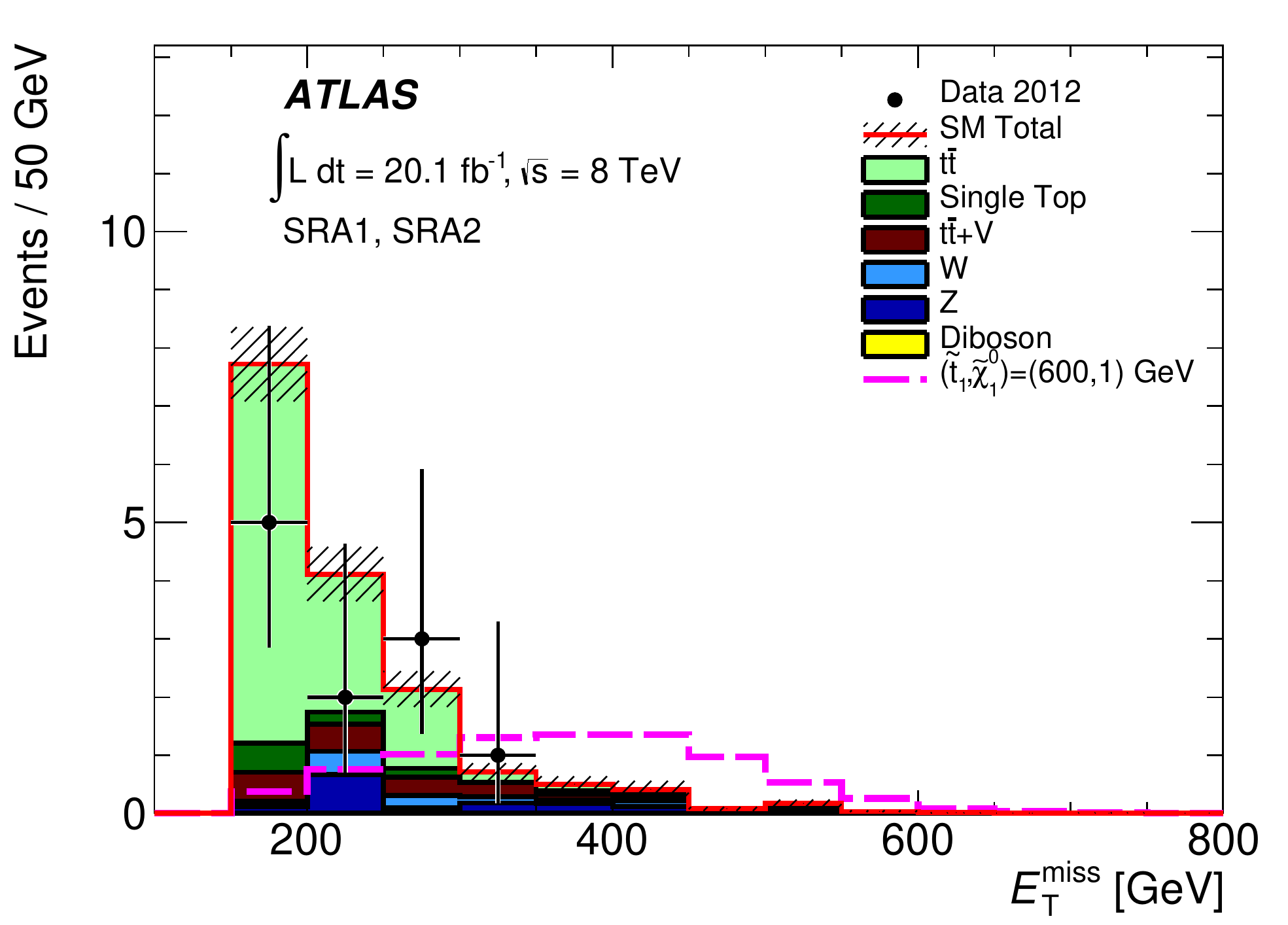}}%
\subfigure[]{\includegraphics[width=0.45\textwidth,trim=0 20 0 0]{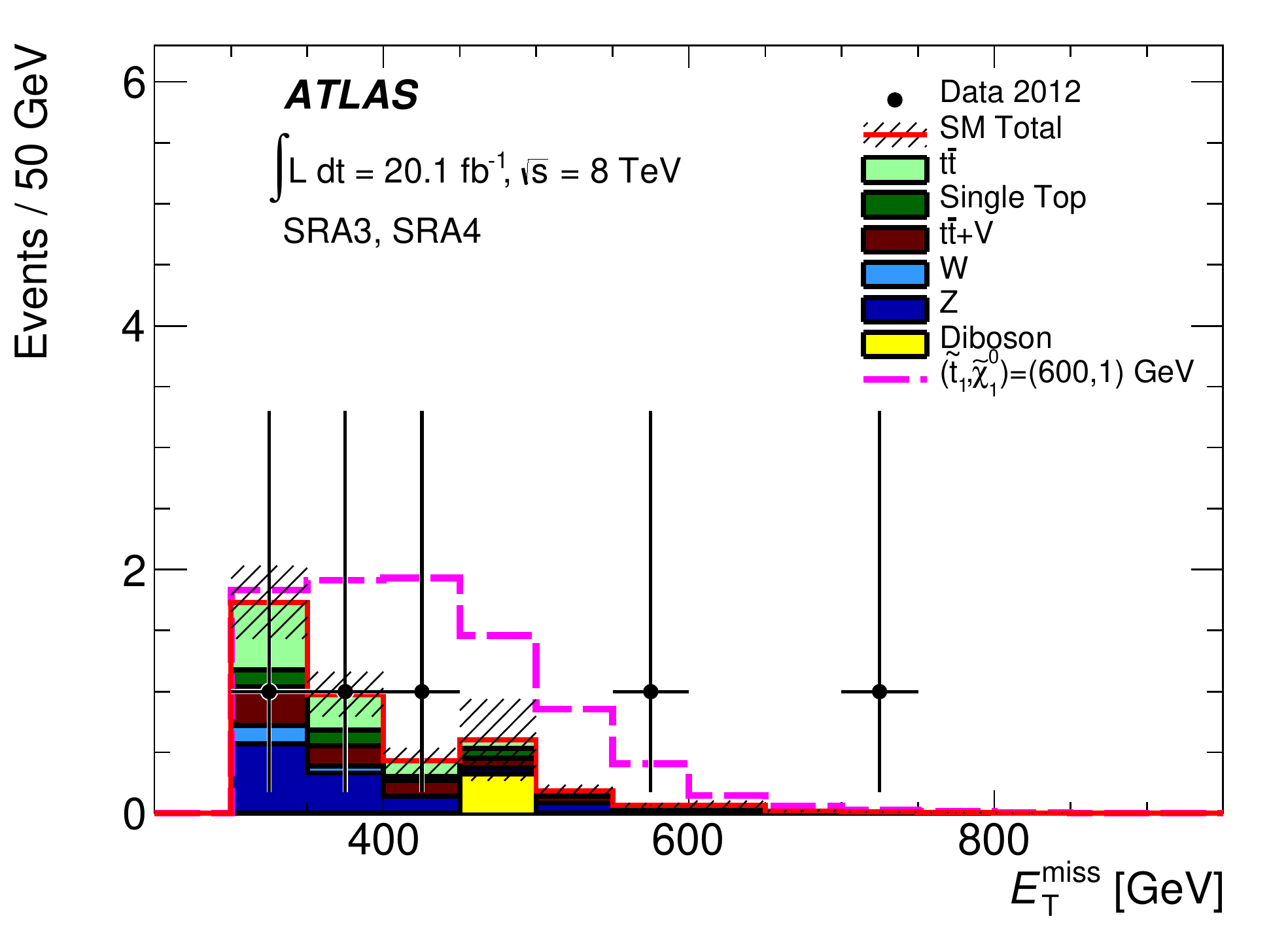}}
\subfigure[]{\includegraphics[width=0.45\textwidth,trim=0 20 0 0]{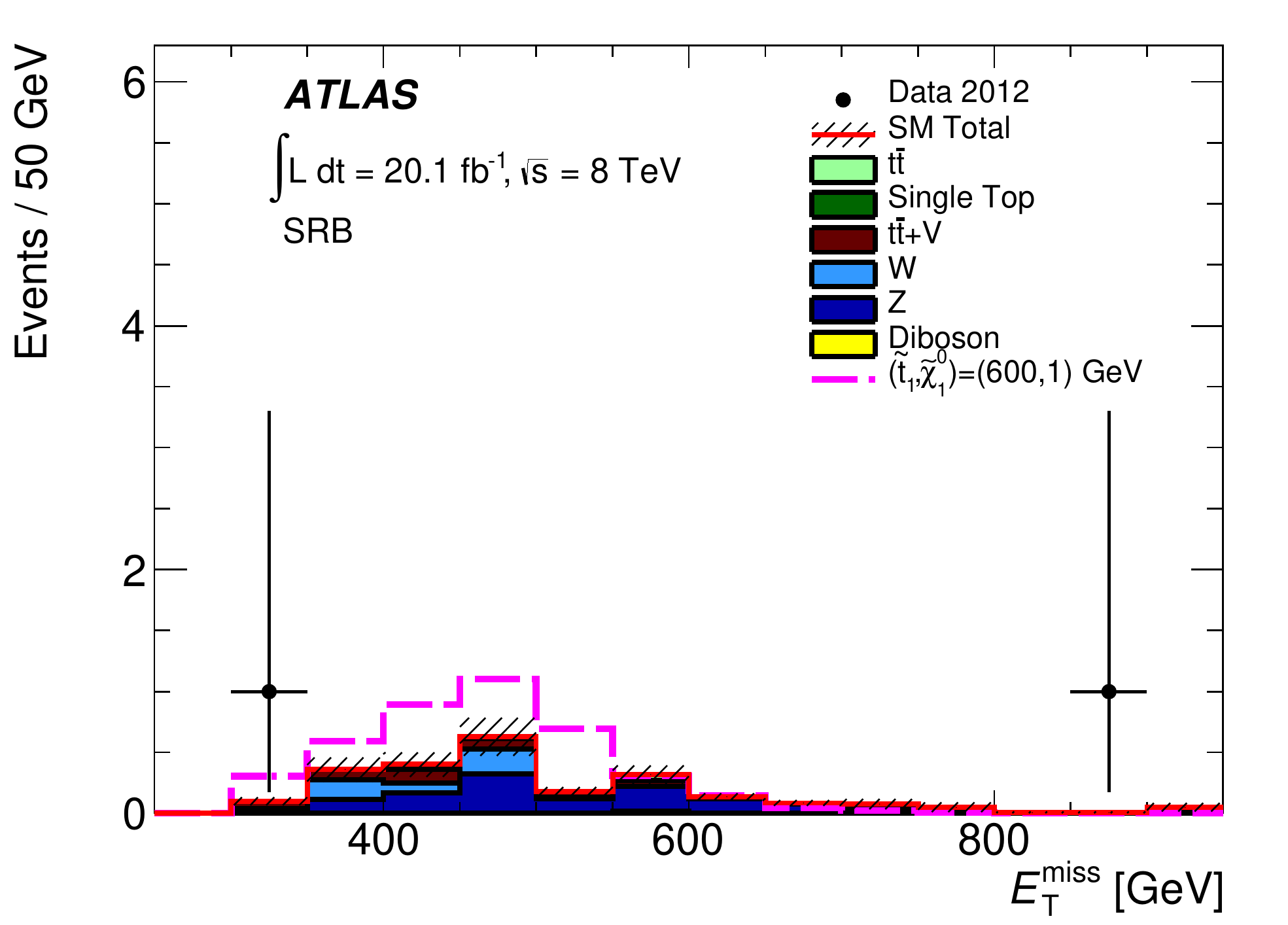}}%
\subfigure[]{\includegraphics[width=0.45\textwidth,trim=0 20 0 0]{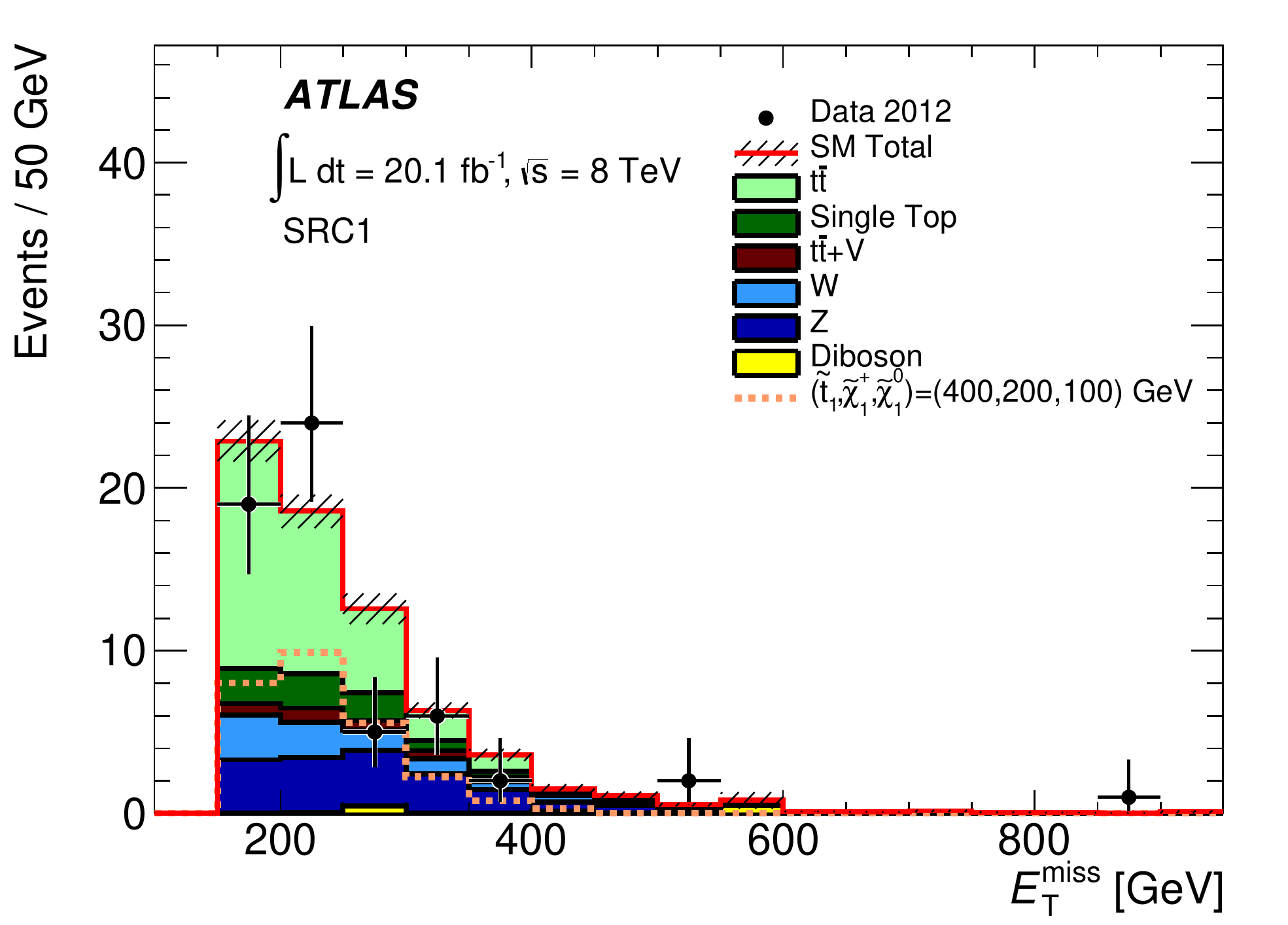}}
\subfigure[]{\includegraphics[width=0.45\textwidth,trim=0 20 0 0]{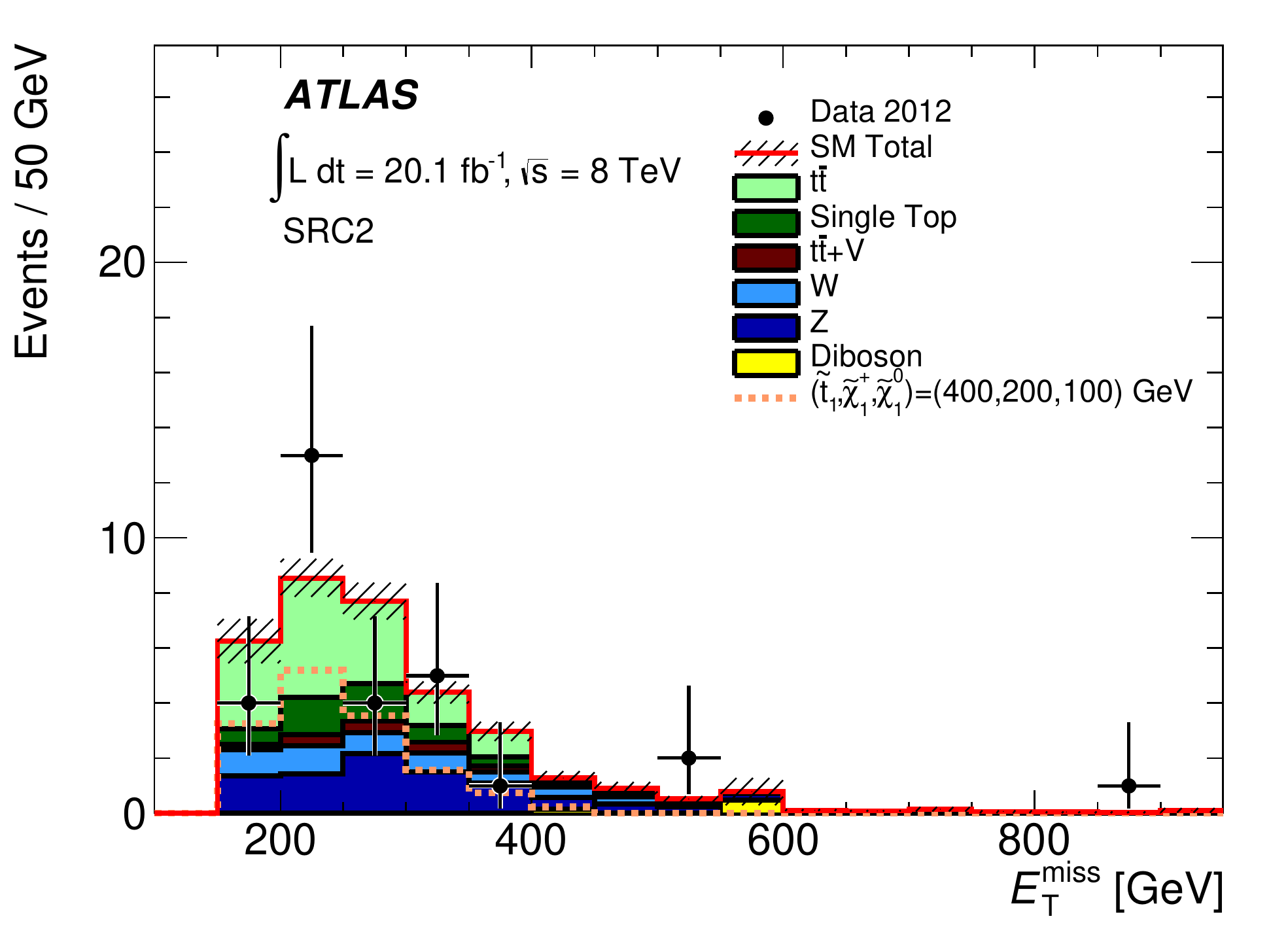}}%
\subfigure[]{\includegraphics[width=0.45\textwidth,trim=0 20 0 0]{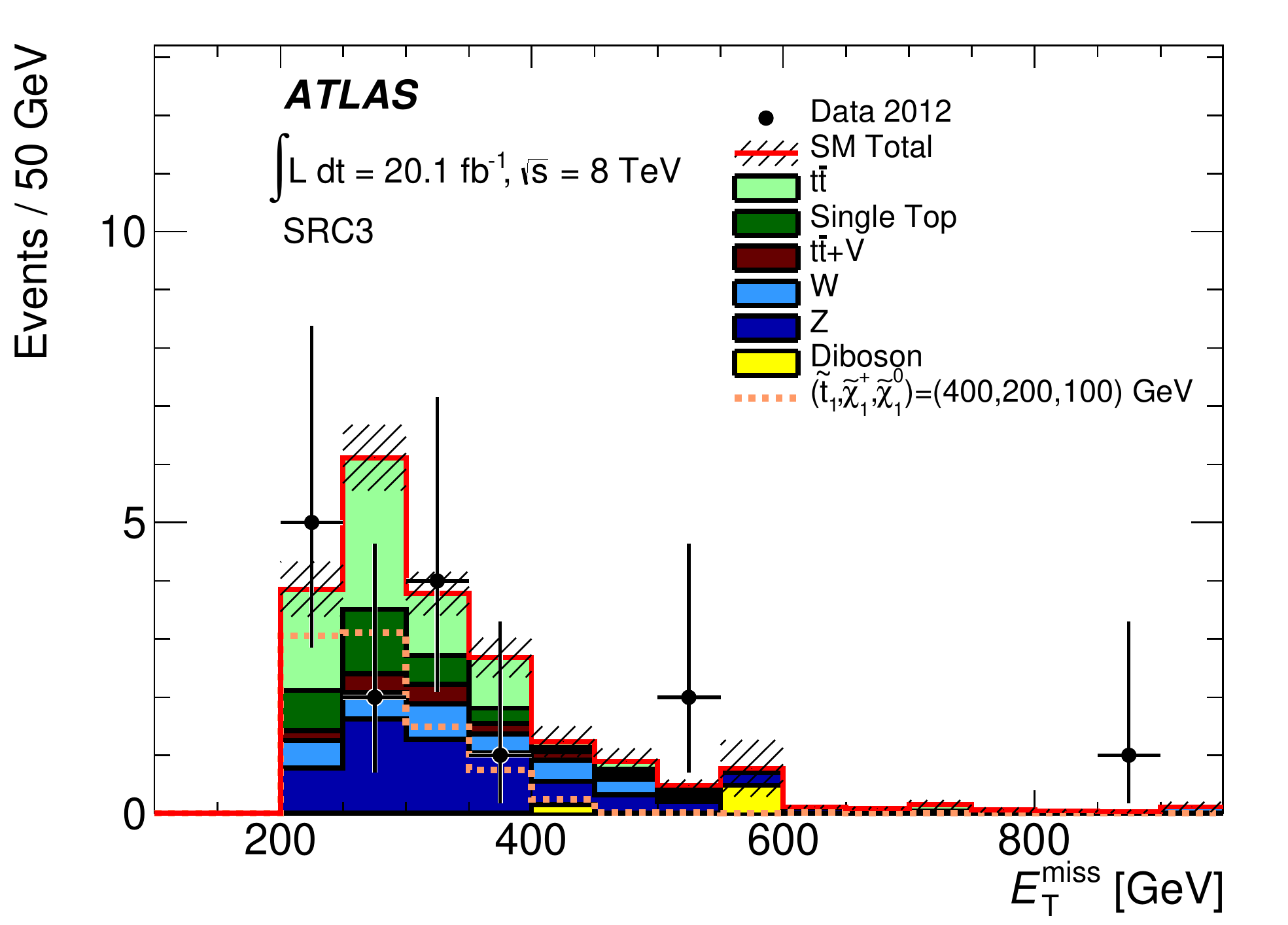}}
\caption{The \met\ distributions for SRA, SRB, and SRC. SRA1 and SRA2 (SRA3 and SRA4) differ only by the \met\ requirement. The background expectation (data) are represented by the stacked histogram (black points). For SRA and SRB, the simulated signal distribution for $m_{\stop} = 600\GeV$, $m_{\ninoone}=1\GeV$ is overlaid (pink dashed line), while for SRC the simulated signal distribution for $m_{\stop} = 400\GeV$, $m_{\chinoonepm} = 200\GeV$, and $m_{\ninoone} = 100\GeV$ is overlaid (orange dotted line). The hatched band on the SM total histogram represents the MC statistical uncertainty only.}
\label{fig:SRresults}
\end{center}
\end{figure}

The numbers of events observed in data in each of the eight signal regions are presented in table~\ref{tb:SRresults}. These results are compared to the total number of expected background events in each signal region. The total background estimate is determined from the simultaneous fit based on the profile likelihood method~\cite{Cowan:2010js} using a procedure similar to that described in section~\ref{sec:fit} but including the corresponding signal regions as well as control regions. The \met\ distributions for each signal region are displayed in figure~\ref{fig:SRresults}; the distributions for SRA1 and SRA2 as well as SRA3 and SRA4 are combined since they only differ by the \met\ requirement. In these figures, the background expectations are normalized by the factors given in table~\ref{tb:norm_after_fit}.

No significant excess above the SM expectation is observed in any of the signal regions. The probabilities are all consistent with the background-only hypothesis; the smallest $p_0$ value is $0.19$ for SRA4. The $95\%$ confidence level (CL) upper limits on the number of beyond-the-SM (BSM) events in each signal region are derived using the CL${}_s$ prescription~\cite{Read:2002hq} and calculated from asymptotic formulae~\cite{Cowan:2010js}. Any possible signal contamination in the control regions is neglected. The BSM signal strength is included as a free parameter but constrained to be non-negative. Normalizing the upper limits on the numbers of events by the integrated luminosity of the data sample, they can be interpreted as model-independent limits on the visible BSM cross sections, defined as $\sigma_{\textrm{vis}} = \sigma \cdot A \cdot \epsilon$, where $\sigma$ is the production cross section, $A$ is the acceptance, and $\epsilon$ is the selection efficiency for a BSM signal. Table~\ref{tb:SRresults} summarizes these upper limits for each signal region. A comparison between results obtained using pseudo-experiments and the asymptotic approximation was performed; the two methods are found to be in good agreement.

The results from the simultaneous fit to the signal and control regions are used to set limits on direct top squark pair production except that a fixed signal component is used here and any signal contamination in the CRs is taken into account. Again, limits are derived using the CL${}_s$ prescription and calculated from asymptotic formulae. By design, SRA is orthogonal to SRB and SRC. However, SRB and SRC are not independent (each considers five-jet events). Therefore each of SRA$1$--$4$ is statistically combined both with SRB (SRA+SRB) and with each of SRC$1$--$3$ (SRA+SRC). The SRA+SRB or SRA+SRC combination with the smallest expected $95\%$ CL${}_s$ value is chosen for each $\stopone$ and $\ninoone$ mass. In these combinations, the signal and detector-related systematic uncertainties are treated as correlated while the theoretical uncertainties and those due to the background normalizations in the independent control regions are treated as uncorrelated. `Expected' limits are calculated by setting the nominal event yield in each SR to the mean background expectation; contours that correspond to $\pm1\sigma$ uncertainties in the background estimates ($\sigma_{\textrm{exp}}$) are also evaluated. `Observed' limits for each channel are calculated from the observed event yields in the signal regions for the nominal signal cross sections and $\pm1\sigma$ theory uncertainties ($\sigma^{\textrm{SUSY}}_{\textrm{theory}}$). Numbers quoted in the text are evaluated from the observed exclusion limit based on the nominal cross section minus $1\sigma^{\textrm{SUSY}}_{\textrm{theory}}$.

\begin{landscape}
\begin{table}[p]
\begin{center}
\input{SRResults}
\end{center}
\end{table}
\end{landscape}

\clearpage

\begin{figure}[tb]
\begin{center}
\includegraphics[width=0.71\textwidth]{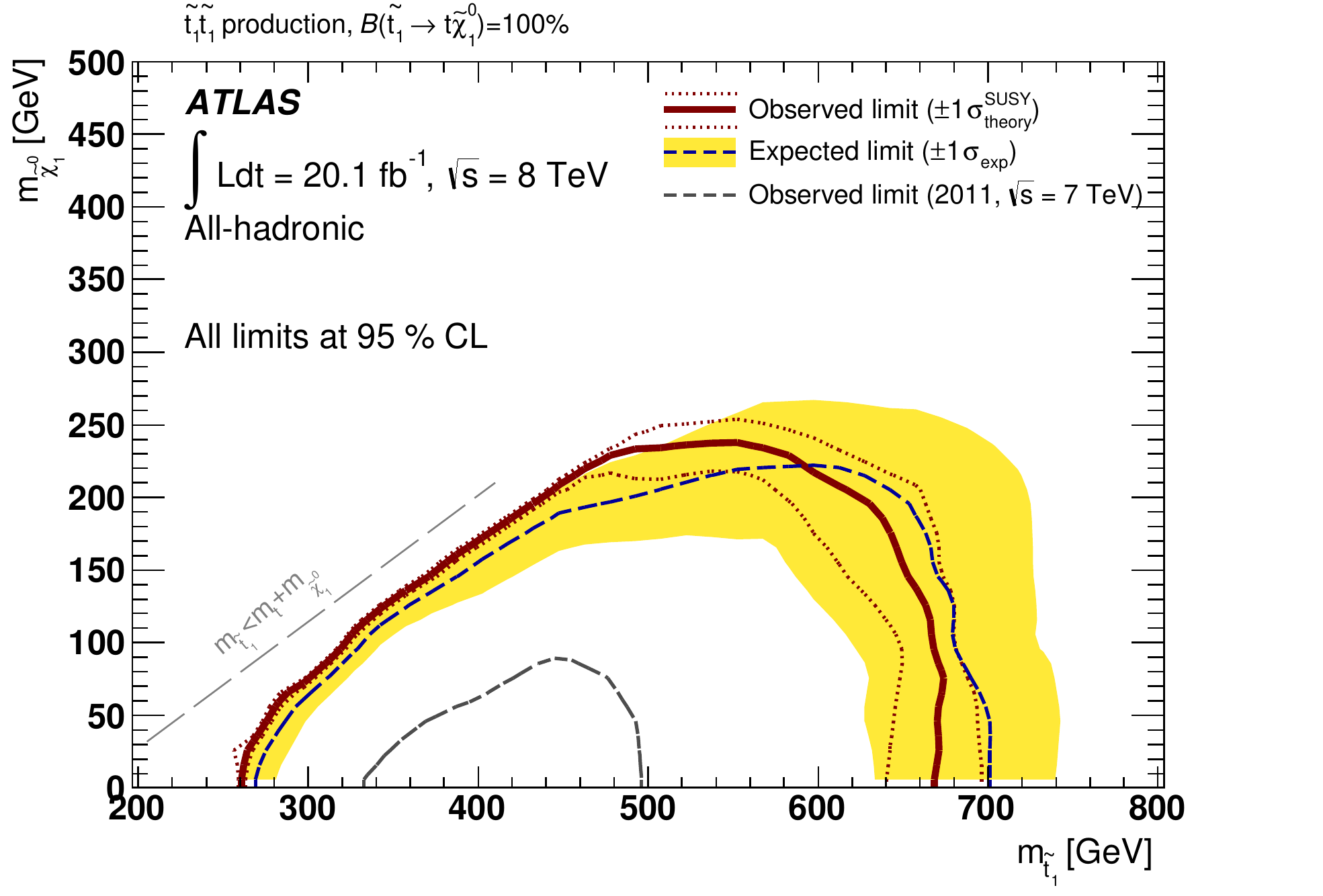}
\caption{Exclusion contours at $95\%$ CL in the scenario where both top squarks decay exclusively via $\stop \to t \ninoone$ and the top quark decays hadronically. The blue dashed line indicates the expected limit, and the yellow band indicates the $\pm1\sigma$ uncertainties, which include all uncertainties except the theoretical uncertainties in the signal. The red solid line indicates the observed limit, and the red dotted lines indicate the sensitivity to $\pm1\sigma$ variations of the signal theoretical uncertainties. The observed limit from the all-hadronic $\rts=7\TeV$ search~\cite{:2012si} is overlaid for comparison.}
\label{fig:SRABCexclusion}
\end{center}
\end{figure}

The resulting exclusion contours for the scenario where both top squarks decay via $\stop \to t \ninoone$ are shown in figure~\ref{fig:SRABCexclusion}, demonstrating an expected sensitivity to potential top squark signals of $275 < m_{\stop} < 700\GeV$ for $m_{\ninoone} < 30\GeV$. The combination of SRA1 or SRA2 with SRC1 tends to be most sensitive for smaller $\stop$--$\ninoone$ mass differences, while the combination of SRA3 or SRA4 with SRB is most sensitive at larger mass differences. Assuming $B\left(\tone \to t \ninoone\right) = 100\%$, top squark masses in the range $270$--$645\GeV$ are excluded for $m_{\ninoone} < 30\GeV$.

\begin{figure}[p]
\begin{center}
\subfigure[]{\includegraphics[width=0.74\textwidth,trim=-20 20 20 0]{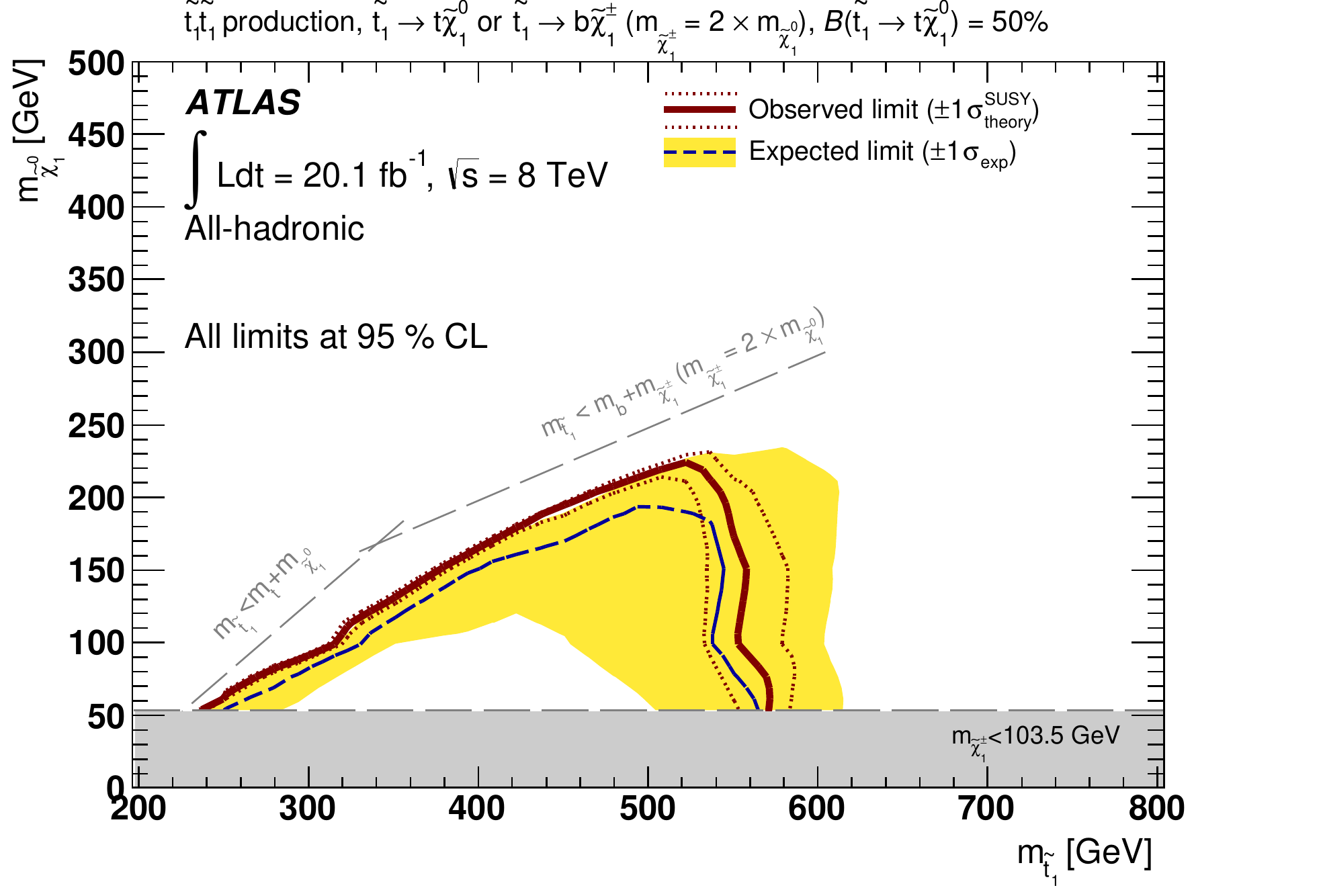}}
\subfigure[]{\includegraphics[width=0.70\textwidth,trim=10 20 0 10]{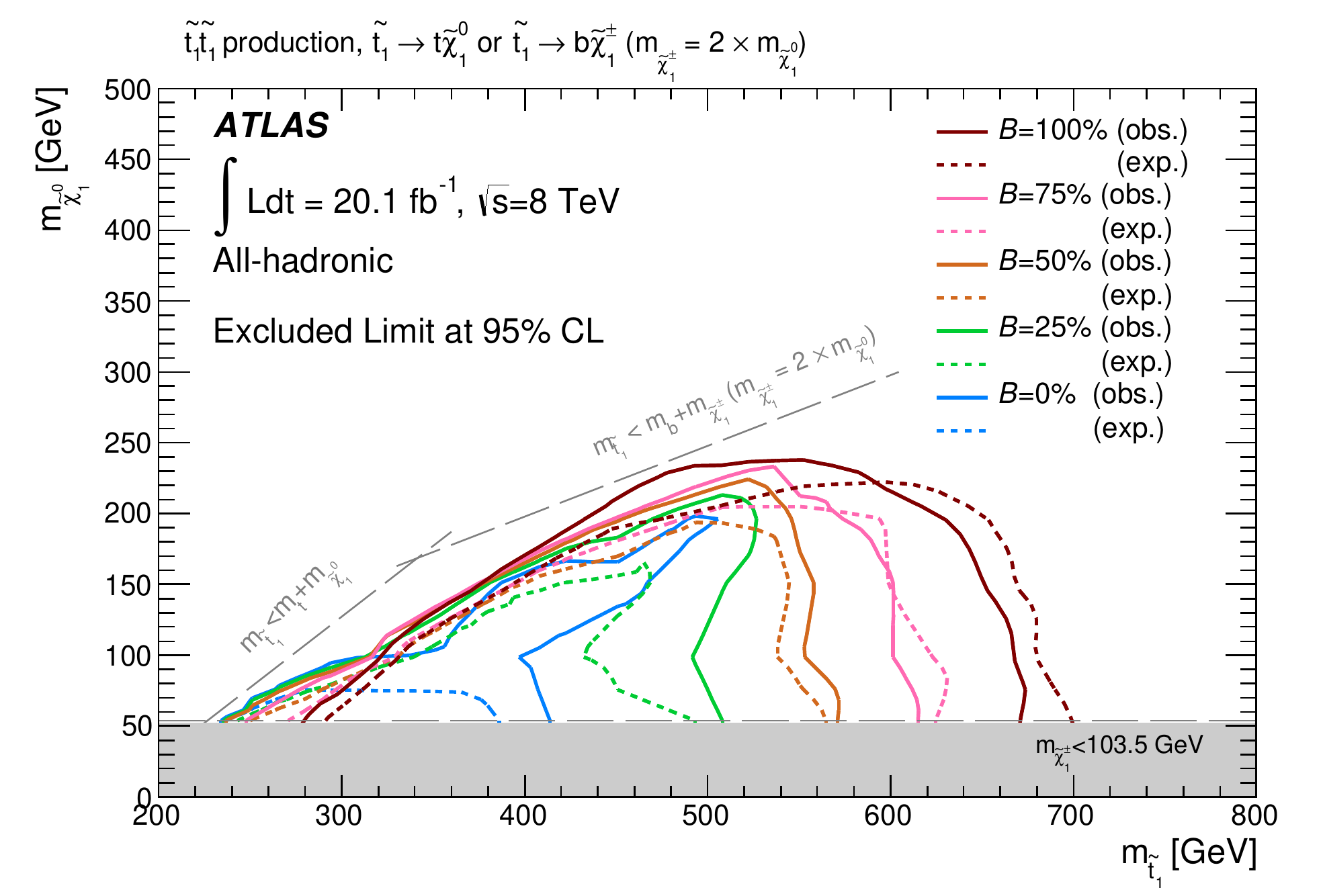}}
\caption{Exclusion contours at 95 $\%$ CL in the scenario where the top squarks are allowed to decay via $\stop \to b \chinoonepm, \chinoonepm \to\Wboson^{\left(*\right)}\ninoone$. The $\chinoonepm$ mass is fixed to twice the $\ninoone$ mass, and the grey filled areas correspond to the LEP limit of $103.5\,\GeV$ on the lightest chargino mass~\cite{LEPCharginoLimit,Heister:2003zk,Abdallah:2003xe,Acciarri:1999km,Abbiendi:2003sc}. (a) Expected and observed limits for $B\left(\tone \to t \ninoone\right) = 50\%$. The blue dashed line indicates the expected limit, and the yellow band indicates the $\pm1\sigma$ uncertainties, which include all uncertainties except the theoretical uncertainties in the signal. The red solid line indicates the observed limit, and the red dotted lines indicate the sensitivity to $\pm1\sigma$ variations of the signal theoretical uncertainties. (b)~The observed and expected exclusion contours are shown for $B\left(\tone \to t \ninoone\right)$ values from $0\%$ (inner contours) to $100\%$ (outer contours). For each branching fraction, the observed (solid line) and expected (dashed line) limits are displayed.}
\label{fig:SRBCexclusion}
\end{center}
\end{figure}

Since the top squark is assumed to decay via either $\tone \rightarrow t \ninoone$ or $\tone \rightarrow b \chinoonepm
\rightarrow b W^{\left(*\right)} \ninoone$, the results are also presented for different values of the branching fraction of $\tone \to t \ninoone$. The mass of the chargino is assumed to be twice the mass of the neutralino. The resulting exclusion contours are shown in figure~\ref{fig:SRBCexclusion}~(a), demonstrating an expected sensitivity to potential top squark signals of $260 < m_{\stop} < 565\GeV$ for $m_{\ninoone} < 60\GeV$ and $B\left(\tone \to t \ninoone\right) = 50\%$. The grey filled area corresponds to the $\ninoone$ mass region excluded by the LEP limit on the lightest chargino mass, taking into account $m_{\chinoonepm} = 2 m_{\ninoone}$~\cite{LEPCharginoLimit,Heister:2003zk,Abdallah:2003xe,Acciarri:1999km,Abbiendi:2003sc}. For $B\left(\tone \to t \ninoone\right) = 50\%$, top squark masses in the range $250$--$550\GeV$ are excluded for $m_{\ninoone} < 60\GeV$. Figure~\ref{fig:SRBCexclusion}~(b) shows the expected and observed contours for a range of $B\left(\tone \to t \ninoone\right)$ values: $100\%$, $75\%$, $50\%$, $25\%$, and $0\%$, where $0\%$ indicates that both top squarks decay exclusively via $\stop \to b \chinoonepm, \chinoonepm \to\Wboson^{\left(*\right)}\ninoone$. The excluded top squark mass ranges are summarized as a function of $B\left(\tone \to t \ninoone\right)$ in table~\ref{tb:excludedmasses}.

\clearpage

\begin{table}[tb]
  \caption{Excluded top squark masses for a range of $B\left(\tone \to t \ninoone\right)$ values, assuming $m_{\chinoonepm} = 2 m_{\ninoone}$. The excluded mass ranges correspond to the observed limit minus one standard deviation of the uncertainty in the signal cross section (the inner red dotted contour in figure~\ref{fig:SRABCexclusion}, for example).}
  \begin{center}
    \begin{tabular}{ccc}  \hline \hline
      $B\left(\tone \to t \ninoone\right)$ & $m_{\stop}$ & $m_{\ninoone}$ \\ \hline
      \\ [-2ex]
      $0\%$  & $245$--$400\GeV$ & $< 60\GeV$ \\ [1ex]
      $25\%$ & $245$--$485\GeV$ & $< 60\GeV$ \\ [1ex]
      $50\%$ & $250$--$550\GeV$ & $< 60\GeV$ \\ [1ex]
      $75\%$ & $265$--$595\GeV$ & $< 60\GeV$ \\ [1ex] 
      $100\%$ & $270$--$645\GeV$ & $< 30\GeV$ \\ \hline \hline
    \end{tabular}
  \end{center}
  \label{tb:excludedmasses}
\end{table}%

%% file: SRResults.tex
\caption{Event yields in each signal region (SRA, SRB, and SRC) are compared to the background estimate from the profile likelihood fit. Statistical, detector, and theoretical systematic uncertainties are included; the total systematic uncertainty in the background estimate includes all correlations. For each signal region, the $95\%$ CL upper limits on the expected (observed) visible cross sections $\sigma_{\textrm{vis}}\left(\textrm{exp}\right)$ ($\sigma_{\textrm{vis}}\left(\textrm{obs}\right)$) and the expected (observed) event yields $N^{95}_{\rm exp}$ ($N^{95}_{\rm obs}$) are summarized. 
}
\begin{tabular}{lcccccccc} \hline\hline
 & \multicolumn{1}{c}{SRA1} & \multicolumn{1}{c}{SRA2} & \multicolumn{1}{c}{SRA3} & \multicolumn{1}{c}{SRA4}    & \multicolumn{1}{c}{SRB} & \multicolumn{1}{c}{SRC1} & \multicolumn{1}{c}{SRC2} & \multicolumn{1}{c}{SRC3}    \\ \hline
Observed events & $11$  & $4$& $5$& $4$                    & $2$     & $59$& $30$ & $15$ \\ \hline \hline
Total SM & $\numRF{15.81}{3} \pm \numRF{1.90}{2}$ & $\numRF{4.05}{2} \pm \numRF{0.76}{1}$& $\numRF{4.07}{2} \pm \numRF{0.92}{1}$ & $\numRF{2.35}{2} \pm \numRF{0.66}{1}$ & $\numRF{2.38}{2} \pm \numRF{0.65}{1}$ & $\numRF{68.22}{2} \pm \numRF{7.34}{1}$& $\numRF{33.78}{2} \pm \numRF{4.56}{1}$& $\numRF{20.30}{3} \pm \numRF{3.00}{2}$\\ \hline
        $\ttbar$ & $\numRF{10.64}{3} \pm \numRF{1.90}{2}$ & $\numRF{1.81}{2} \pm \numRF{0.48}{1}$& $\numRF{1.05}{2} \pm \numRF{0.56}{1}$ & $\numRF{0.49}{2} \pm \numRF{0.34}{2}$   & \multicolumn{1}{l}{$0.10\;_{-\;0.10}^{+\;0.14}$}   & $\numRF{32.20}{2} \pm \numRF{4.42}{1}$& $\numRF{12.94}{3} \pm \numRF{1.95}{2}$& $\numRF{6.66}{2} \pm \numRF{1.16}{2}$         \\
        $\ttbar+\Wboson/\Zboson$ & $\numRF{1.80}{2} \pm \numRF{0.59}{1}$ & $\numRF{0.85}{2} \pm \numRF{0.29}{2}$ & $\numRF{0.82}{2} \pm \numRF{0.29}{2}$ & $\numRF{0.50}{2} \pm \numRF{0.17}{2}$     & $\numRF{0.47}{2} \pm \numRF{0.17}{2}$    & $\numRF{3.24}{2} \pm \numRF{0.82}{1}$& $\numRF{1.90}{2} \pm \numRF{0.53}{1}$& $\numRF{1.33}{2} \pm \numRF{0.36}{1}$         \\
        $\Zjets$ & $\numRF{1.42}{2} \pm \numRF{0.53}{1}$ & $\numRF{0.63}{2} \pm \numRF{0.22}{2}$& $\numRF{1.24}{2} \pm \numRF{0.39}{1}$ & $\numRF{0.68}{2} \pm \numRF{0.27}{2}$    & $\numRF{1.23}{3} \pm \numRF{0.31}{2}$   & $\numRF{15.70}{3} \pm \numRF{3.45}{2}$& $\numRF{8.98}{2} \pm \numRF{1.91}{2}$& $\numRF{6.05}{2} \pm \numRF{1.30}{2}$        \\
        $\Wjets$ & $\numRF{0.95}{1} \pm \numRF{0.45}{1}$ & $\numRF{0.46}{2} \pm \numRF{0.21}{2}$ & $\numRF{0.21}{2} \pm \numRF{0.19}{2}$ & \multicolumn{1}{l}{$0.06\;_{-\;0.06}^{+\;0.10}$}   & $\numRF{0.49}{2} \pm \numRF{0.33}{2}$    & $\numRF{8.48}{1} \pm \numRF{3.70}{1}$& $\numRF{4.78}{2} \pm \numRF{2.18}{2}$& $\numRF{2.77}{2} \pm \numRF{1.22}{2}$       \\
        Single top & $\numRF{1.00}{2} \pm \numRF{0.35}{1}$& $\numRF{0.30}{2} \pm \numRF{0.17}{2}$& $\numRF{0.44}{2} \pm \numRF{0.14}{2}$ & $\numRF{0.31}{2} \pm \numRF{0.16}{2}$    & $\numRF{0.08}{1} \pm \numRF{0.06}{1}$    & $\numRF{7.23}{2} \pm \numRF{2.92}{2}$& $\numRF{4.48}{2} \pm \numRF{1.81}{2}$& $\numRF{2.85}{2} \pm \numRF{1.39}{2}$        \\
        Diboson & \multicolumn{1}{c}{$<$ 0.4} & \multicolumn{1}{c}{$<$ 0.13}  & $\numRF{0.32}{2} \pm \numRF{0.17}{2}$ & $\numRF{0.32}{2} \pm \numRF{0.18}{2}$   & $\numRF{0.02}{1} \pm \numRF{0.01}{1}$    & $\numRF{1.14}{2} \pm \numRF{0.77}{1}$& \multicolumn{1}{l}{$\phantom{3}0.6\;_{-\;0.6}^{+\;0.7}$} & \multicolumn{1}{l}{$\phantom{2}0.6\;_{-\;0.6}^{+\;0.7}$}       \\
        Multijets & \multicolumn{1}{c}{$<0.001$} & \multicolumn{1}{c}{$<0.001$} &  \multicolumn{1}{c}{$<0.001$} & \multicolumn{1}{c}{$<0.001$}    & \multicolumn{1}{c}{$<0.001$}    & $\numRF{0.24}{2} \pm \numRF{0.24}{2}$ & $\numRF{0.06}{1} \pm \numRF{0.06}{1}$   & $\numRF{0.01}{1} \pm \numRF{0.01}{1}$      \\ \hline
$\sigma_{\textrm{vis}}\left(\textrm{obs}\right)$ [fb] &$0.33$&$0.29$&$0.33$ &$0.32$ &$0.21$ &$0.78$ &$0.62$&$0.40$   \\ 
$\sigma_{\textrm{vis}}\left(\textrm{exp}\right)$ [fb]&${0.48}^{\;+\;0.21}_{\;-\;0.14}$ &  ${0.29}^{\;+\;0.13}_{\;-\;0.09}$& ${0.29}^{\;+\;0.14}_{\;-\;0.09}$ &${0.25}^{\;+\;0.13}_{\;-\;0.07}$ &  ${0.24}^{\;+\;0.13}_{\;-\;0.06}$ &$\phantom{2}{1.03}^{\;+\;0.42}_{\;-\;0.29}$&${0.73}^{\;+\;0.31}_{\;-\;0.21}$ &${0.55}^{\;+\;0.24}_{\;-\;0.15}$ \\ 
$N^{95}_{\rm obs}$          & $6.6$  & $5.7$ & $6.7$  &$6.5$ & $4.2$  & $15.7$ & $12.4$ & $8.0$  \\                                                    
$N^{95}_{\rm exp}$          & ${9.7}^{\;+\;4.3}_{\;-\;3.0}$&${5.8}^{\;+\;2.6}_{\;-\;1.8}$ & ${5.9}^{\;+\;2.8}_{\;-\;1.9}$&${5.0}^{\;+\;2.6}_{\;-\;1.4}$ & ${4.7}^{\;+\;2.6}_{\;-\;1.2}$&${20.7}^{\;+\;8.4}_{\;-\;5.8}$ & ${14.7}^{\;+\;6.2}_{\;-\;4.2}$ &${11.0}^{\;+\;4.9}_{\;-\;3.1}$\\  \hline  \hline  
\end{tabular}
\label{tb:SRresults}

%% file: conclusions.tex
\label{sec:conclusions}

The results of a search for direct top squark production with an all-hadronic experimental signature of jets and missing transverse momentum are presented, using an integrated luminosity of \totalluminumnoerr\ of proton--proton collision data at $\rts=8\TeV$ collected by the ATLAS detector at the LHC. In this search, the top squark is assumed to decay via $\stop \to t \ninoone$ or $\stop\to b\chinoonepm$. In addition to the nominal fully resolved topology that requires at least six jets, the sensitivity of the analysis is increased by including a partially resolved topology (four or five jets). Furthermore, a dedicated signal region requiring exactly five jets augments the sensitivity to top squark decays via $\stop \to b \chinoonepm$. These three categories of events are statistically combined to provide improved sensitivity to direct top squark production.

No excess over the SM background prediction is observed, and exclusion limits are reported as a function of the top squark and neutralino masses for a range of the branching fractions of $\tone \to t \ninoone$ from $0$--$100\%$. For $B\left(\tone \to t \ninoone\right) = 100\%$, top squark masses in the range $270$--$645\GeV$ are excluded for a $m_{\ninoone} < 30\GeV$, while for $B\left(\tone \to t \ninoone\right) = 50\%$ and $m_{\chinoonepm} = 2 m_{\ninoone}$, top squark masses in the range $250$--$550\GeV$ are excluded for a $m_{\ninoone} < 60\GeV$. These limits significantly extend previous results.

%% file: acknowledgments.tex
We thank CERN for the very successful operation of the LHC, as well as the
support staff from our institutions without whom ATLAS could not be
operated efficiently.

We acknowledge the support of ANPCyT, Argentina; YerPhI, Armenia; ARC,
Australia; BMWF and FWF, Austria; ANAS, Azerbaijan; SSTC, Belarus; CNPq and FAPE
SP,
Brazil; NSERC, NRC and CFI, Canada; CERN; CONICYT, Chile; CAS, MOST and NSFC,
China; COLCIENCIAS, Colombia; MSMT CR, MPO CR and VSC CR, Czech Republic;
DNRF, DNSRC and Lundbeck Foundation, Denmark; EPLANET, ERC and NSRF, European Un
ion;
IN2P3-CNRS, CEA-DSM/IRFU, France; GNSF, Georgia; BMBF, DFG, HGF, MPG and AvH
Foundation, Germany; GSRT and NSRF, Greece; ISF, MINERVA, GIF, I-CORE and Benozi
yo Center,
Israel; INFN, Italy; MEXT and JSPS, Japan; CNRST, Morocco; FOM and NWO,
Netherlands; BRF and RCN, Norway; MNiSW and NCN, Poland; GRICES and FCT, Portuga
l; MNE/IFA, Romania; MES of Russia and ROSATOM, Russian Federation; JINR; MSTD,
Serbia; MSSR, Slovakia; ARRS and MIZ\v{S}, Slovenia; DST/NRF, South Africa;
MINECO, Spain; SRC and Wallenberg Foundation, Sweden; SER, SNSF and Cantons of
Bern and Geneva, Switzerland; NSC, Taiwan; TAEK, Turkey; STFC, the Royal
Society and Leverhulme Trust, United Kingdom; DOE and NSF, United States of
America.

The crucial computing support from all WLCG partners is acknowledged
gratefully, in particular from CERN and the ATLAS Tier-1 facilities at
TRIUMF (Canada), NDGF (Denmark, Norway, Sweden), CC-IN2P3 (France),
KIT/GridKA (Germany), INFN-CNAF (Italy), NL-T1 (Netherlands), PIC (Spain),
ASGC (Taiwan), RAL (UK) and BNL (USA) and in the Tier-2 facilities
worldwide.

%% file: atlas_authlist.tex
\begin{flushleft}
{\Large The ATLAS Collaboration}

\bigskip

G.~Aad$^{\rm 84}$,
B.~Abbott$^{\rm 112}$,
J.~Abdallah$^{\rm 152}$,
S.~Abdel~Khalek$^{\rm 116}$,
O.~Abdinov$^{\rm 11}$,
R.~Aben$^{\rm 106}$,
B.~Abi$^{\rm 113}$,
M.~Abolins$^{\rm 89}$,
O.S.~AbouZeid$^{\rm 159}$,
H.~Abramowicz$^{\rm 154}$
% ...
\bigskip
\\
% ...
$^{11}$ Institute of Physics, Azerbaijan Academy of Sciences, Baku, Azerbaijan\\
% ...
$^{84}$ CPPM, Aix-Marseille Universit{\'e} and CNRS/IN2P3, Marseille, France\\
% ...
$^{89}$ Department of Physics and Astronomy, Michigan State University, East Lansing MI, United States of America\\
% ...
$^{106}$ Nikhef National Institute for Subatomic Physics and University of Amsterdam, Amsterdam, Netherlands\\
% ...
$^{112}$ Homer L. Dodge Department of Physics and Astronomy, University of Oklahoma, Norman OK, United States of America\\
$^{113}$ Department of Physics, Oklahoma State University, Stillwater OK, United States of America\\
% ...
$^{116}$ LAL, Universit{\'e} Paris-Sud and CNRS/IN2P3, Orsay, France\\
% ...
$^{152}$ Institute of Physics, Academia Sinica, Taipei, Taiwan\\
% ...
$^{154}$ Raymond and Beverly Sackler School of Physics and Astronomy, Tel Aviv University, Tel Aviv, Israel\\
% ...
$^{159}$ Department of Physics, University of Toronto, Toronto ON, Canada\\
% ...
\end{flushleft}


%% file: paper.bbl
\providecommand{\href}[2]{#2}\begingroup\raggedright\begin{thebibliography}{100}

\bibitem{:2012gk}
{\bf ATLAS} Collaboration, {\it {Observation of a new particle in the search
  for the Standard Model Higgs boson with the ATLAS detector at the LHC}},
  {\em Phys. Lett.} {\bf B 716} (2012) 1--29,
  [\href{http://xxx.lanl.gov/abs/1207.7214}{{\tt arXiv:1207.7214}}].

\bibitem{:2012gu}
{\bf CMS} Collaboration, {\it {Observation of a new boson at a mass of 125 GeV
  with the CMS experiment at the LHC}},  {\em Phys. Lett.} {\bf B 716} (2012)
  30--61, [\href{http://xxx.lanl.gov/abs/1207.7235}{{\tt arXiv:1207.7235}}].

\bibitem{Weinberg:1975gm}
S.~Weinberg, {\it {Implications of Dynamical Symmetry Breaking}},  {\em Phys.
  Rev.} {\bf D 13} (1976) 974--996.

\bibitem{Gildener:1976ai}
E.~Gildener, {\it {Gauge Symmetry Hierarchies}},  {\em Phys. Rev.} {\bf D 14}
  (1976) 1667--1672.

\bibitem{Weinberg:1979bn}
S.~Weinberg, {\it {Implications of Dynamical Symmetry Breaking: An Addendum}},
  {\em Phys. Rev.} {\bf D 19} (1979) 1277--1280.

\bibitem{Susskind:1978ms}
L.~Susskind, {\it {Dynamics of Spontaneous Symmetry Breaking in the Weinberg-
  Salam Theory}},  {\em Phys. Rev.} {\bf D 20} (1979) 2619--2625.

\bibitem{Miyazawa:1966}
H.~Miyazawa, {\it {Baryon Number Changing Currents}},  {\em Prog. Theor. Phys.}
  {\bf 36 (6)} (1966) 1266--1276.

\bibitem{Ramond:1971gb}
P.~Ramond, {\it {Dual Theory for Free Fermions}},  {\em Phys. Rev.} {\bf D 3}
  (1971) 2415--2418.

\bibitem{Golfand:1971iw}
Y.~Golfand and E.~Likhtman, {\it {Extension of the Algebra of Poincare Group
  Generators and Violation of p Invariance}},  {\em JETP Lett.} {\bf 13} (1971)
  323--326.

\bibitem{Neveu:1971rx}
A.~Neveu and J.~H. Schwarz, {\it {Factorizable dual model of pions}},  {\em
  Nucl. Phys.} {\bf B 31} (1971) 86--112.

\bibitem{Neveu:1971iv}
A.~Neveu and J.~H. Schwarz, {\it {Quark Model of Dual Pions}},  {\em Phys.
  Rev.} {\bf D 4} (1971) 1109--1111.

\bibitem{Gervais:1971ji}
J.~Gervais and B.~Sakita, {\it {Field theory interpretation of supergauges in
  dual models}},  {\em Nucl. Phys.} {\bf B 34} (1971) 632--639.

\bibitem{Volkov:1973ix}
D.~Volkov and V.~Akulov, {\it {Is the Neutrino a Goldstone Particle?}},  {\em
  Phys. Lett.} {\bf B 46} (1973) 109--110.

\bibitem{Wess:1973kz}
J.~Wess and B.~Zumino, {\it {A Lagrangian Model Invariant Under Supergauge
  Transformations}},  {\em Phys. Lett.} {\bf B 49} (1974) 52--54.

\bibitem{Wess:1974tw}
J.~Wess and B.~Zumino, {\it {Supergauge Transformations in Four-Dimensions}},
  {\em Nucl. Phys.} {\bf B 70} (1974) 39--50.

\bibitem{Dimopoulos:1981zb}
S.~Dimopoulos and H.~Georgi, {\it {Softly Broken Supersymmetry and SU(5)}},
  {\em Nucl. Phys.} {\bf B 193} (1981) 150--162.

\bibitem{Witten:1981nf}
E.~Witten, {\it {Dynamical Breaking of Supersymmetry}},  {\em Nucl. Phys.} {\bf
  B 188} (1981) 513--554.

\bibitem{Dine:1981za}
M.~Dine, W.~Fischler, and M.~Srednicki, {\it {Supersymmetric Technicolor}},
  {\em Nucl. Phys.} {\bf B 189} (1981) 575--593.

\bibitem{Dimopoulos:1981au}
S.~Dimopoulos and S.~Raby, {\it {Supercolor}},  {\em Nucl. Phys.} {\bf B 192}
  (1981) 353--368.

\bibitem{Sakai:1981gr}
N.~Sakai, {\it {Naturalness in Supersymmetric GUTs}},  {\em Z. Phys.} {\bf C
  11} (1981) 153--157.

\bibitem{Kaul:1981hi}
R.~Kaul and P.~Majumdar, {\it {Cancellation of quadratrically divergent mass
  corrections in globally supersymmetric spontaneously broken gauge theories}},
   {\em Nucl. Phys.} {\bf B 199} (1982) 36--58.

\bibitem{Barbieri:1987fn}
R.~Barbieri and G.~Giudice, {\it {Upper Bounds on Supersymmetric Particle
  Masses}},  {\em Nucl. Phys.} {\bf B 306} (1988) 63--76.

\bibitem{deCarlos:1993yy}
B.~de~Carlos and J.~Casas, {\it {One loop analysis of the electroweak breaking
  in supersymmetric models and the fine tuning problem}},  {\em Phys. Lett.}
  {\bf B 309} (1993) 320--328,
  [\href{http://xxx.lanl.gov/abs/hep-ph/9303291}{{\tt hep-ph/9303291}}].

\bibitem{Fayet:1976et}
P.~Fayet, {\it {Supersymmetry and Weak, Electromagnetic and Strong
  Interactions}},  {\em Phys. Lett.} {\bf B 64} (1976) 159--162.

\bibitem{Fayet:1977yc}
P.~Fayet, {\it {Spontaneously Broken Supersymmetric Theories of Weak,
  Electromagnetic and Strong Interactions}},  {\em Phys. Lett.} {\bf B 69}
  (1977) 489--494.

\bibitem{Farrar:1978xj}
G.~R. Farrar and P.~Fayet, {\it {Phenomenology of the Production, Decay, and
  Detection of New Hadronic States Associated with Supersymmetry}},  {\em Phys.
  Lett.} {\bf B 76} (1978) 575--579.

\bibitem{Fayet:1979sa}
P.~Fayet, {\it {Relations Between the Masses of the Superpartners of Leptons
  and Quarks, the Goldstino Couplings and the Neutral Currents}},  {\em Phys.
  Lett.} {\bf B 84} (1979) 416--420.

\bibitem{Beenakker:1997ut}
W.~Beenakker et~al., {\it Stop production at hadron colliders},  {\em Nucl.
  Phys.} {\bf B 515} (1998) 3--14.

\bibitem{Beenakker:2010nq}
W.~Beenakker et~al., {\it {Supersymmetric top and bottom squark production at
  hadron colliders}},  {\em JHEP} {\bf 08} (2010) 098,
  [\href{http://xxx.lanl.gov/abs/1006.4771}{{\tt arXiv:1006.4771}}].

\bibitem{Beenakker:2011fu}
W.~Beenakker et~al., {\it {Squark and gluino hadroproduction}},  {\em Int. J.
  Mod. Phys.} {\bf A 26} (2011) 2637--2664.

\bibitem{Aad:2012cz}
{\bf ATLAS} Collaboration, {\it {Search for scalar top quark pair production in
  natural gauge mediated supersymmetry models with the ATLAS detector in $pp$
  collisions at $\sqrt{s}=7$ TeV}},  {\em Phys. Lett.} {\bf B 715} (2012)
  44--60, [\href{http://xxx.lanl.gov/abs/1204.6736}{{\tt arXiv:1204.6736}}].

\bibitem{:2012si}
{\bf ATLAS} Collaboration, {\it {Search for a supersymmetric partner to the top
  quark in final states with jets and missing transverse momentum at
  $\sqrt{s}=7$ TeV with the ATLAS detector}},  {\em Phys. Rev. Lett.} {\bf 109}
  (2012) 211802, [\href{http://xxx.lanl.gov/abs/1208.1447}{{\tt
  arXiv:1208.1447}}].

\bibitem{Aad:2012tx}
{\bf ATLAS} Collaboration, {\it {Search for light scalar top quark pair
  production in final states with two leptons with the ATLAS detector in
  $\sqrt{s}=7$ TeV proton-proton collisions}},  {\em Eur. Phys. J.} {\bf C 72}
  (2012) 2237, [\href{http://xxx.lanl.gov/abs/1208.4305}{{\tt
  arXiv:1208.4305}}].

\bibitem{Aad:2012uu}
{\bf ATLAS} Collaboration, {\it {Search for a heavy top-quark partner in final
  states with two leptons with the ATLAS detector at the LHC}},  {\em JHEP}
  {\bf 11} (2012) 094, [\href{http://xxx.lanl.gov/abs/1209.4186}{{\tt
  arXiv:1209.4186}}].

\bibitem{Aad:2012yr}
{\bf ATLAS} Collaboration, {\it {Search for light top squark pair production in
  final states with leptons and $b$-jets with the ATLAS detector in
  $\sqrt{s}=7$ TeV proton-proton collisions}},  {\em Phys. Lett.} {\bf B 720}
  (2013) 13--31, [\href{http://xxx.lanl.gov/abs/1209.2102}{{\tt
  arXiv:1209.2102}}].

\bibitem{Aad:2013ija}
{\bf ATLAS} Collaboration, {\it {Search for direct third-generation squark pair
  production in final states with missing transverse momentum and two $b$-jets
  in $\sqrt{s} =$ 8 TeV $pp$ collisions with the ATLAS detector}},  {\em JHEP}
  {\bf 10} (2013) 189, [\href{http://xxx.lanl.gov/abs/1308.2631}{{\tt
  arXiv:1308.2631}}].

\bibitem{Aad:2014mha}
{\bf ATLAS} Collaboration, {\it {Search for direct top squark pair production
  in events with a Z boson, b-jets and missing transverse momentum in
  $\sqrt{s}$=8 TeV pp collisions with the ATLAS detector}},
  \href{http://xxx.lanl.gov/abs/1403.5222}{{\tt arXiv:1403.5222}}.

\bibitem{Chatrchyan:2012uea}
{\bf CMS} Collaboration, {\it {Inclusive search for supersymmetry using the
  razor variables in $pp$ collisions at $\sqrt{s}=7$ TeV}},  {\em Phys. Rev.
  Lett.} {\bf 111} (2013), no.~8 081802,
  [\href{http://xxx.lanl.gov/abs/1212.6961}{{\tt arXiv:1212.6961}}].

\bibitem{Chatrchyan:2012wa}
{\bf CMS} Collaboration, {\it {Search for supersymmetry in final states with
  missing transverse energy and 0, 1, 2, or at least 3 b-quark jets in 7 TeV pp
  collisions using the variable $\alpha_{\rm{T}}$}},  {\em JHEP} {\bf 01}
  (2013) 077, [\href{http://xxx.lanl.gov/abs/1210.8115}{{\tt
  arXiv:1210.8115}}].

\bibitem{Chatrchyan:2013lya}
{\bf CMS} Collaboration, {\it {Search for supersymmetry in hadronic final
  states with missing transverse energy using the variables $\alpha_{\rm{T}}$
  and b-quark multiplicity in pp collisions at 8 TeV}},  {\em Eur. Phys. J.}
  {\bf C 73} (2013) 2568, [\href{http://xxx.lanl.gov/abs/1303.2985}{{\tt
  arXiv:1303.2985}}].

\bibitem{Chatrchyan:2013xna}
{\bf CMS} Collaboration, {\it {Search for top-squark pair production in the
  single-lepton final state in pp collisions at $\sqrt{s}$ = 8 TeV}},  {\em
  Eur. Phys. J.} {\bf C 73} (2013) 2677,
  [\href{http://xxx.lanl.gov/abs/1308.1586}{{\tt arXiv:1308.1586}}].

\bibitem{Chatrchyan:2013mya}
{\bf CMS} Collaboration, {\it {Search for top squark and higgsino production
  using diphoton Higgs boson decays}},  {\em Phys. Rev. Lett.} {\bf 112} (2014)
  161802, [\href{http://xxx.lanl.gov/abs/1312.3310}{{\tt arXiv:1312.3310}}].

\bibitem{Khachatryan:2014doa}
{\bf CMS} Collaboration, {\it {Search for top-squark pairs decaying into Higgs
  or Z bosons in pp collisions at $\sqrt{s}$ = 8 TeV}},
  \href{http://xxx.lanl.gov/abs/1405.3886}{{\tt arXiv:1405.3886}}.

\bibitem{DetectorPaper:2008}
{\bf ATLAS} Collaboration, {\it {The ATLAS Experiment at the CERN Large Hadron
  Collider}},  {\em JINST} {\bf 3} (2008) S08003.

\bibitem{Aad:2013ucp}
{\bf ATLAS} Collaboration, {\it {Improved luminosity determination in $pp$
  collisions at $\sqrt{s}$ = 7 TeV using the ATLAS detector at the LHC}},  {\em
  Eur. Phys. J.} {\bf C 73} (2013) 2518,
  [\href{http://xxx.lanl.gov/abs/1302.4393}{{\tt arXiv:1302.4393}}].

\bibitem{Frixione:2007vw}
S.~Frixione, P.~Nason, and C.~Oleari, {\it {Matching NLO QCD computations with
  Parton Shower simulations: the POWHEG method}},  {\em JHEP} {\bf 11} (2007)
  070, [\href{http://xxx.lanl.gov/abs/0709.2092}{{\tt arXiv:0709.2092}}]. {For
  \ttbar\ (single top) production, POWHEG-BOX version 1.0 r2129 is interfaced
  with \pythia\ 6.427 (6.426).}

\bibitem{Aad:2012hg}
{\bf ATLAS} Collaboration, {\it {Measurements of top quark pair relative
  differential cross-sections with ATLAS in $pp$ collisions at $\sqrt{s}=7$
  TeV}},  {\em Eur. Phys. J.} {\bf C 73} (2013) 2261,
  [\href{http://xxx.lanl.gov/abs/1207.5644}{{\tt arXiv:1207.5644}}].

\bibitem{kersevan:2004yg}
B.~P. Kersevan and E.~Richter-Was, {\it {The Monte Carlo event generator AcerMC
  version 2.0 with interfaces to PYTHIA 6.2 and HERWIG 6.5}},
  \href{http://xxx.lanl.gov/abs/hep-ph/0405247}{{\tt hep-ph/0405247}}. {AcerMC
  version 3.8 is interfaced with \pythia\ 6.426.}

\bibitem{Gleisberg:2008ta}
T.~Gleisberg et~al., {\it {Event generation with SHERPA 1.1}},  {\em JHEP} {\bf
  02} (2009) 007, [\href{http://xxx.lanl.gov/abs/0811.4622}{{\tt
  arXiv:0811.4622}}]. {SHERPA version 1.4.1.}

\bibitem{Alwall:2011uj}
J.~Alwall et~al., {\it {MadGraph 5 : Going Beyond}},  {\em JHEP} {\bf 06}
  (2011) 128, [\href{http://xxx.lanl.gov/abs/1106.0522}{{\tt
  arXiv:1106.0522}}]. {MadGraph 5 version 1.3.33 is interfaced with \pythia\
  6.426.}

\bibitem{ATL-PHYS-PUB-2011-009}
{\bf ATLAS} Collaboration, {\it {ATLAS tunes of PYTHIA 6 and Pythia 8 for
  MC11}},  {\em ATL-PHYS-PUB-2011-009} (2011).
  http://cdsweb.cern.ch/record/1363300.

\bibitem{Sjostrand:2006za}
T.~Sj{\"o}strand, S.~Mrenna, and P.~Z. Skands, {\it {PYTHIA 6.4 Physics and
  Manual}},  {\em JHEP} {\bf 05} (2006) 026,
  [\href{http://xxx.lanl.gov/abs/hep-ph/0603175}{{\tt hep-ph/0603175}}].

\bibitem{PhysRevD.82.074018}
P.~Z. Skands, {\it {Tuning Monte Carlo generators: The Perugia tunes}},  {\em
  Phys. Rev.} {\bf D 82} (2010) 074018,
  [\href{http://xxx.lanl.gov/abs/1005.3457}{{\tt arXiv:1005.3457}}].

\bibitem{Lai:2010vv}
H.-L. Lai et~al., {\it {New parton distributions for collider physics}},  {\em
  Phys. Rev.} {\bf D 82} (2010) 074024,
  [\href{http://xxx.lanl.gov/abs/1007.2241}{{\tt arXiv:1007.2241}}].

\bibitem{Pumplin:2002vw}
J.~Pumplin et~al., {\it {New generation of parton distributions with
  uncertainties from global QCD analysis}},  {\em JHEP} {\bf 07} (2002) 012,
  [\href{http://xxx.lanl.gov/abs/hep-ph/0201195}{{\tt hep-ph/0201195}}].

\bibitem{Catani:2009sm}
S.~Catani et~al., {\it {Vector boson production at hadron colliders: A Fully
  exclusive QCD calculation at NNLO}},  {\em Phys. Rev. Lett.} {\bf 103} (2009)
  082001, [\href{http://xxx.lanl.gov/abs/0903.2120}{{\tt arXiv:0903.2120}}].

\bibitem{Martin:2009iq}
A.~Martin, W.~Stirling, R.~Thorne, and G.~Watt, {\it {Parton distributions for
  the LHC}},  {\em Eur. Phys. J.} {\bf C 63} (2009) 189--285,
  [\href{http://xxx.lanl.gov/abs/0901.0002}{{\tt arXiv:0901.0002}}].

\bibitem{Cacciari:2011hy}
M.~Cacciari, M.~Czakon, M.~Mangano, A.~Mitov, and P.~Nason, {\it {Top-pair
  production at hadron colliders with next-to-next-to-leading logarithmic
  soft-gluon resummation}},  {\em Phys. Lett.} {\bf B 710} (2012) 612--622,
  [\href{http://xxx.lanl.gov/abs/1111.5869}{{\tt arXiv:1111.5869}}].

\bibitem{Czakon:2011xx}
M.~Czakon and A.~Mitov, {\it {Top++: A Program for the Calculation of the
  Top-Pair Cross-Section at Hadron Colliders}},
  \href{http://xxx.lanl.gov/abs/1112.5675}{{\tt arXiv:1112.5675}}. {Top++
  version 2.0.}

\bibitem{Campbell:2012dh}
J.~M. Campbell and R.~K. Ellis, {\it {$t \bar{t} W^{+-}$ production and decay
  at NLO}},  {\em JHEP} {\bf 07} (2012) 052,
  [\href{http://xxx.lanl.gov/abs/1204.5678}{{\tt arXiv:1204.5678}}].

\bibitem{Garzelli:2011is}
M.~Garzelli, A.~Kardos, C.~Papadopoulos, and Z.~Trocsanyi, {\it {$Z^{0}$ -
  boson production in association with a top anti-top pair at NLO accuracy with
  parton shower effects}},  {\em Phys. Rev.} {\bf D 85} (2012) 074022,
  [\href{http://xxx.lanl.gov/abs/1111.1444}{{\tt arXiv:1111.1444}}].

\bibitem{Kidonakis:2010tc}
N.~Kidonakis, {\it {NNLL resummation for s-channel single top quark
  production}},  {\em Phys. Rev.} {\bf D 81} (2010) 054028,
  [\href{http://xxx.lanl.gov/abs/1001.5034}{{\tt arXiv:1001.5034}}].

\bibitem{Kidonakis:2010ux}
N.~Kidonakis, {\it {Two-loop soft anomalous dimensions for single top quark
  associated production with a $\Wboson^-$ or $H^-$}},  {\em Phys. Rev.} {\bf D
  82} (2010) 054018, [\href{http://xxx.lanl.gov/abs/1005.4451}{{\tt
  arXiv:1005.4451}}].

\bibitem{Kidonakis:2011wy}
N.~Kidonakis, {\it {Next-to-next-to-leading-order collinear and soft gluon
  corrections for t-channel single top quark production}},  {\em Phys. Rev.}
  {\bf D 83} (2011) 091503, [\href{http://xxx.lanl.gov/abs/1103.2792}{{\tt
  arXiv:1103.2792}}].

\bibitem{Campbell:2011bn}
J.~M. Campbell, R.~K. Ellis, and C.~Williams, {\it {Vector boson pair
  production at the LHC}},  {\em JHEP} {\bf 07} (2011) 018,
  [\href{http://xxx.lanl.gov/abs/1105.0020}{{\tt arXiv:1105.0020}}].

\bibitem{LEPCharginoLimit}
{LEP SUSY Working Group (ALEPH, DELPHI L3, OPAL), Notes LEPSUSYWG/01-03.1,
  http://lepsusy.web.cern.ch/lepsusy/Welcome.html}.

\bibitem{Bahr:2008pv}
M.~B{\"a}hr et~al., {\it {Herwig++ Physics and Manual}},  {\em Eur. Phys. J.}
  {\bf C 58} (2008) 639--707, [\href{http://xxx.lanl.gov/abs/0803.0883}{{\tt
  arXiv:0803.0883}}].

\bibitem{Kramer:2012bx}
M.~Kr{\"a}mer et~al., {\it {Supersymmetry production cross sections in $pp$
  collisions at $\sqrt{s}=7$ TeV}},
  \href{http://xxx.lanl.gov/abs/1206.2892}{{\tt arXiv:1206.2892}}.

\bibitem{atlassimulation}
{\bf ATLAS} Collaboration, {\it {The ATLAS Simulation Infrastructure}},  {\em
  \EPJ} {\bf C 70} (2010) 823--874,
  [\href{http://xxx.lanl.gov/abs/1005.4568}{{\tt arXiv:1005.4568}}].

\bibitem{geant4}
{\bf GEANT4} Collaboration, S.~Agostinelli et~al., {\it {GEANT4: A simulation
  toolkit}},  {\em Nucl. Instrum. Meth.} {\bf A 506} (2003) 250--303.

\bibitem{ATL-PHYS-PUB-2010-013}
{\bf ATLAS} Collaboration, {\it {The simulation principle and performance of
  the ATLAS fast calorimeter simulation FastCaloSim}},  {\em
  ATL-PHYS-PUB-2010-013} (2010). http://cdsweb.cern.ch/record/1300517.

\bibitem{PV}
{\bf ATLAS} Collaboration, {\it {Performance of primary vertex reconstruction
  in proton-proton collisions at $\sqrt{s}$ = 7 TeV in the ATLAS experiment}},
  {\em ATLAS-CONF-2010-069} (2010). http://cdsweb.cern.ch/record/1281344.

\bibitem{topoclusters}
W.~Lampl et~al., {\it {Calorimeter Clustering Algorithms: Description and
  Performance}},  {\em ATL-LARG-PUB-2008-002} (2008).
  http://cdsweb.cern.ch/record/1099735.

\bibitem{Cacciari:2008gp}
M.~Cacciari, G.~P. Salam, and G.~Soyez, {\it {The Anti-k(t) jet clustering
  algorithm}},  {\em JHEP} {\bf 04} (2008) 063,
  [\href{http://xxx.lanl.gov/abs/0802.1189}{{\tt arXiv:0802.1189}}].

\bibitem{Cacciari:2005hq}
M.~Cacciari and G.~P. Salam, {\it {Dispelling the $N^{3}$ myth for the $k_t$
  jet-finder}},  {\em Phys. Lett.} {\bf B 641} (2006) 57--61,
  [\href{http://xxx.lanl.gov/abs/hep-ph/0512210}{{\tt hep-ph/0512210}}].

\bibitem{Cacciari:2011ma}
M.~Cacciari, G.~P. Salam, and G.~Soyez, {\it {FastJet User Manual}},  {\em Eur.
  Phys. J.} {\bf C 72} (2012) 1896,
  [\href{http://xxx.lanl.gov/abs/1111.6097}{{\tt arXiv:1111.6097}}].

\bibitem{Issever:2004qh}
C.~Issever, K.~Borras, and D.~Wegener, {\it {An Improved weighting algorithm to
  achieve software compensation in a fine grained LAr calorimeter}},  {\em
  Nucl. Instrum. Meth.} {\bf A 545} (2005) 803--812,
  [\href{http://xxx.lanl.gov/abs/physics/0408129}{{\tt physics/0408129}}].

\bibitem{Cacciari:2007fd}
M.~Cacciari and G.~P. Salam, {\it {Pileup subtraction using jet areas}},  {\em
  Phys. Lett.} {\bf B 659} (2008) 119--126,
  [\href{http://xxx.lanl.gov/abs/0707.1378}{{\tt arXiv:0707.1378}}].

\bibitem{Aad:2011he}
{\bf ATLAS} Collaboration, {\it {Jet energy measurement with the ATLAS detector
  in proton-proton collisions at $\sqrt{s}=7$ TeV}},  {\em Eur. Phys. J.} {\bf
  C 73} (2013) 2304, [\href{http://xxx.lanl.gov/abs/1112.6426}{{\tt
  arXiv:1112.6426}}].

\bibitem{ATLAS-CONF-2012-043}
{\bf ATLAS} Collaboration, {\it {Measurement of the b-tag Efficiency in a
  Sample of Jets Containing Muons with 5 \ifb~of Data from the ATLAS
  Detector}},  {\em ATLAS-CONF-2012-043} (2012).
  http://cdsweb.cern.ch/record/1435197.

\bibitem{ATLAS-CONF-2011-089}
{\bf ATLAS} Collaboration, {\it {Calibrating the b-Tag Efficiency and Mistag
  Rate in 35 $pb^{-1}$ of Data with the ATLAS Detector}},  {\em
  ATLAS-CONF-2011-089} (2011). http://cdsweb.cern.ch/record/1356198.

\bibitem{ATLAS-CONF-2011-102}
{\bf ATLAS} Collaboration, {\it {Commissioning of the ATLAS high-performance
  b-tagging algorithms in the 7 TeV collision data}},  {\em
  ATLAS-CONF-2011-102} (2011). http://cdsweb.cern.ch/record/1369219.

\bibitem{Aad:2011mk}
{\bf ATLAS} Collaboration, {\it {Electron performance measurements with the
  ATLAS detector using the 2010 LHC proton-proton collision data}},  {\em Eur.
  Phys. J.} {\bf C 72} (2012) 1909,
  [\href{http://xxx.lanl.gov/abs/1110.3174}{{\tt arXiv:1110.3174}}].

\bibitem{ATLAS-CONF-2011-021}
{\bf ATLAS} Collaboration, {\it {A measurement of the ATLAS muon reconstruction
  and trigger efficiency using $J/\psi$ decays}},  {\em ATLAS-CONF-2011-021}
  (2011). http://cdsweb.cern.ch/record/1336750.

\bibitem{ATLAS-CONF-2011-063}
{\bf ATLAS} Collaboration, {\it {Muon reconstruction efficiency in reprocessed
  2010 LHC proton-proton collision data recorded with the ATLAS detector}},
  {\em ATLAS-CONF-2011-063} (2011). http://cdsweb.cern.ch/record/1345743.

\bibitem{:2012rz}
{\bf ATLAS} Collaboration, {\it {Search for squarks and gluinos with the ATLAS
  detector in final states with jets and missing transverse momentum using 4.7
  $fb^{-1}$ of $\sqrt{s}=7$ TeV proton-proton collision data}},  {\em Phys.
  Rev.} {\bf D 87} (2013) 012008,
  [\href{http://xxx.lanl.gov/abs/1208.0949}{{\tt arXiv:1208.0949}}].

\bibitem{Sjostrand:2007gs}
T.~Sj{\"o}strand, S.~Mrenna, and P.~Z. Skands, {\it {A Brief Introduction to
  PYTHIA 8.1}},  {\em Comput. Phys. Commun.} {\bf 178} (2008) 852--867,
  [\href{http://xxx.lanl.gov/abs/0710.3820}{{\tt arXiv:0710.3820}}].

\bibitem{JES:SingleHadron}
{\bf ATLAS} Collaboration, {\it {Single hadron response measurement and
  calorimeter jet energy scale uncertainty with the ATLAS detector at the
  LHC}},  {\em Eur. Phys. J.} {\bf C 73} (2013) 2305,
  [\href{http://xxx.lanl.gov/abs/1203.1302}{{\tt arXiv:1203.1302}}].

\bibitem{Corcella:2002jc}
G.~Corcella et~al., {\it {HERWIG 6.5 release note}},
  \href{http://xxx.lanl.gov/abs/hep-ph/0210213}{{\tt hep-ph/0210213}}.

\bibitem{jimmy}
J.~Butterworth, J.~R. Forshaw, and M.~Seymour, {\it {Multiparton interactions
  in photoproduction at HERA}},  {\em \ZfP} {\bf C 72} (1996) 637--646,
  [\href{http://xxx.lanl.gov/abs/hep-ph/9601371}{{\tt hep-ph/9601371}}].

\bibitem{ATLAS:2012al}
{\bf ATLAS} Collaboration, {\it {Measurement of $t \bar{t}$ production with a
  veto on additional central jet activity in pp collisions at sqrt(s) = 7 TeV
  using the ATLAS detector}},  {\em Eur. Phys. J.} {\bf C 72} (2012) 2043,
  [\href{http://xxx.lanl.gov/abs/1203.5015}{{\tt arXiv:1203.5015}}].

\bibitem{Aaron:2009aa}
{{\bf H1} and {\bf ZEUS} Collaborations}, {\it {Combined Measurement and QCD
  Analysis of the Inclusive e$^\pm$ p Scattering Cross Sections at HERA}},
  {\em JHEP} {\bf 01} (2010) 109,
  [\href{http://xxx.lanl.gov/abs/0911.0884}{{\tt arXiv:0911.0884}}].

\bibitem{Garzelli:2012bn}
M.~Garzelli, A.~Kardos, C.~Papadopoulos, and Z.~Trocsanyi, {\it
  {$t\bar{t}W^{\pm}$ and $t\bar{t}$ Z Hadroproduction at NLO accuracy in QCD
  with Parton Shower and Hadronization effects}},  {\em JHEP} {\bf 11} (2012)
  056, [\href{http://xxx.lanl.gov/abs/1208.2665}{{\tt arXiv:1208.2665}}].

\bibitem{ATLAS:2013Wb}
{\bf ATLAS} Collaboration, {\it {Measurement of the cross-section for W boson
  production in association with b-jets in pp collisions at sqrt(s) = 7 TeV
  with the ATLAS detector}},  {\em J. High Energy Phys.} {\bf 06} (2013) 084,
  [\href{http://xxx.lanl.gov/abs/1302.2929}{{\tt arXiv:1302.2929}}].

\bibitem{Cowan:2010js}
G.~Cowan, K.~Cranmer, E.~Gross, and O.~Vitells, {\it {Asymptotic formulae for
  likelihood-based tests of new physics}},  {\em Eur. Phys. J.} {\bf C 71}
  (2011) 1554, [\href{http://xxx.lanl.gov/abs/1007.1727}{{\tt
  arXiv:1007.1727}}].

\bibitem{Read:2002hq}
A.~L. Read, {\it {Presentation of search results: The CL(s) technique}},  {\em
  J. Phys.} {\bf G 28} (2002) 2693--2704.

\bibitem{Heister:2003zk}
{\bf ALEPH} Collaboration, A.~Heister et~al., {\it {Absolute mass lower limit
  for the lightest neutralino of the MSSM from $e^{+}e^{-}$ data at $\sqrt{s}$
  up to 209~GeV}},  {\em Phys. Lett.} {\bf B 583} (2004) 247--263.

\bibitem{Abdallah:2003xe}
{\bf DELPHI} Collaboration, J.~Abdallah et~al., {\it {Searches for
  supersymmetric particles in $e^{+}e^{-}$ collisions up to 208~GeV and
  interpretation of the results within the MSSM}},  {\em Eur. Phys. J.} {\bf C
  31} (2003) 421--479, [\href{http://xxx.lanl.gov/abs/hep-ex/0311019}{{\tt
  hep-ex/0311019}}].

\bibitem{Acciarri:1999km}
{\bf L3} Collaboration, M.~Acciarri et~al., {\it {Search for charginos and
  neutralinos in $e^{+} e^{-}$ collisions at $\sqrt{s}$ = 189~GeV}},  {\em
  Phys. Lett.} {\bf B 472} (2000) 420--433,
  [\href{http://xxx.lanl.gov/abs/hep-ex/9910007}{{\tt hep-ex/9910007}}].

\bibitem{Abbiendi:2003sc}
{\bf OPAL} Collaboration, G.~Abbiendi et~al., {\it {Search for chargino and
  neutralino production at $\sqrt{s} = 192$~GeV to 209 GeV at LEP}},  {\em Eur.
  Phys. J.} {\bf C 35} (2004) 1--20,
  [\href{http://xxx.lanl.gov/abs/hep-ex/0401026}{{\tt hep-ex/0401026}}].

\end{thebibliography}\endgroup
